\documentclass[12pt]{article}
\usepackage{a4}
\usepackage{epsf,rotate}
\usepackage{subeqn}	
\usepackage{cite}
\usepackage{array,dcolumn}	

\newcolumntype{d}[1]{D{.}{.}{#1}}

\def\openone{\leavevmode\hbox{\small1\kern-3.8pt\normalsize1}}%
\newcommand{\1}{\openone}

\def\slash#1{\setbox0=\hbox{$#1$}#1\hskip-\wd0\dimen0=5pt\advance
       \dimen0 by-\ht0\advance\dimen0 by\dp0\lower0.5\dimen0\hbox
         to\wd0{\hss\sl/\/\hss}}

\newlength{\miniwidth}
\newlength{\miniwidthplot}
\newlength{\nseparation}
\newenvironment{nfigure}
	{\begin{figure}[htbp]\hrule\vspace{\nseparation}\par}
	{\vspace{\nseparation}\par \hrule \end{figure}}
\newenvironment{ntable}
	{\begin{table}[htbp]\hrule\vspace{\nseparation}\par}
	{\vspace{\nseparation}\par \hrule \end{table}}

\newcommand{\lt}{\left}
\newcommand{\rt}{\right}
\newcommand{\ov}{\overline}
\newcommand{\nn}{\nonumber \\}
\newcommand{\no}{\nonumber}
\newcommand{\diff}[1]{\frac{d}{d #1}}

\newcommand{\g}{\gamma}
\newcommand{\gL}{\lt(1-\gamma_5\rt)}

\newcommand{\lef}{\gL}

\newcommand{\bare}{\mathrm{bare}}
\newcommand{\ba}{\bare}
\newcommand{\eps}{\varepsilon}
\newcommand{\Lagr}{{\mathcal{L}}}
\newcommand{\lag}{\Lagr}
\newcommand{\eff}{{\mathrm{eff}}}
\newcommand{\eq}[1]{(\ref{#1})}
\newcommand{\fig}[1]{Fig.~\ref{#1}}
\newcommand{\tab}[1]{Table~\ref{#1}}
\newcommand{\sect}[1]{sect.~\ref{#1}}
\newcommand{\apx}[1]{appendix~\ref{#1}}

\newcommand{\li}{{\,\mathrm{Li}}}
\newcommand{\mc}{m_c^2}
\newcommand{\mt}{m_t^2}
\newcommand{\ms}{m_s^2}
\newcommand{\md}{m_d^2}
\newcommand{\mw}{M_W^2}
\newcommand{\tw}{\widetilde{\openone}} 
\newcommand{\topenone}{\widetilde{\openone}}
\newcommand{\as}{\alpha_s}
\newcommand{\T}{\,\mathrm{\mathbf{T\,}}}
\newcommand{\Tg}{\,\mathrm{\mathbf{T_g\,}}}
\newcommand{\gev}{\,\mathrm{GeV}}
\newcommand{\mev}{\,\mathrm{MeV}}
\newcommand{\laMSb}{\ensuremath{\Lambda_{\overline{\mathrm{MS}}}}}
\newcommand{\laQCD}{\ensuremath{\Lambda_{\mathrm{QCD}}}}
\newcommand{\gf}{\,G_F}
\newcommand{\gft}{\, G_F^2}
\newcommand{\kkm}{\ensuremath{\mathrm{K^0\!-\!\ov{K^0}}\,}-mixing\ }
\newcommand{\bbm}{\ensuremath{\mathrm{B^0\!-\!\ov{B^0}}\,}-mixing\ }
\newcommand{\kkmd}{\ensuremath{\mathrm{K_L\!-\!K_S}\,}-mass difference\ }
\newcommand{\dstwo}{\ensuremath{\mathrm{|\Delta S| \!=\!2}}}
\newcommand{\dsone}{\ensuremath{\mathrm{|\Delta S| \!=\!1}}}

\newcommand{\eom}{{\mathrm{EOM}}}

\newcommand{\msb}{\ensuremath{\ov{\textrm{MS}}}}
\newcommand{\dmu}{\mu \diff{\mu}}

\newcommand{\gzero}{\gamma^{(0)}}

\newcommand{\wt}[1]{\widetilde{#1}}

\newcommand{\Qeom}{Q_{\mathrm{g}\eom}}
\newcommand{\Qfeom}{Q_{\mathrm{q}\eom}}

\newcommand{\cll}[1]{\wt{C}_{S2}^{(#1)}}
\newcommand{\oll}{\ensuremath{\tilde{Q}_{S2}}}
\newcommand{\zll}{\wt{Z}_{S2}}
\newcommand{\F}{F}
\newcommand{\kctsa}{k^{ct}_{sing,a}}
\newcommand{\kcts}{k^{ct}_{sing}}
\newcommand{\kctoa}{k^{ct}_{oct,a}}
\newcommand{\kcto}{k^{ct}_{oct}}
\newcommand{\ktsa}{k^{t}_{sing,a}}
\newcommand{\kts}{k^{t}_{sing}}

\newcommand{\kto}{k^{t}_{oct}}
\newcommand{\mr}[1]{\mathrm{#1}}
\newcommand{\leso}{\lag ^\mr{|\Delta S| =  1}}
\newcommand{\leo}{\lag_\mr{eff}^\mr{|\Delta S| =  2}}
\newcommand{\oloc}{\wt{Q}_7}
\newcommand{\cloc}{\wt{C}_7}
\newcommand{\zloc}{\wt{Z}_{77}}
\newcommand{\zbi}[1]{\wt{Z}_{#1,7}}
\newcommand{\gloc}{\wt{\gamma}_{77}}
\newcommand{\gbi}[1]{\wt{\gamma}_{#1,7}}
\newcommand{\zbil}[3]{\left[\wt{Z}^{#1}_{#2}\right]_{#3,7}}
\newcommand{\zlocl}[2]{\left[\wt{Z}^{#1}_{#2}\right]_{77}}
\newcommand{\uloc}{\wt{U}_{77}}
\newcommand{\jloc}{\wt{J}_{77}}
\newcommand{\bilr}[1]{\mathcal{R}_{#1}}
\newcommand{\errorpm}[2]{
\raisebox{-0.5ex}{\shortstack[l]{$\scriptstyle+#1$\\$\scriptstyle-#2$}}
}
\newcommand{\pr}{Phys.\ Rev.\ }
\newcommand{\prd}{\pr D}
\newcommand{\np}{Nucl.\ Phys.\ }
\newcommand{\npb}{\np B}
\newcommand{\prl}{\pr Lett.\ }
\newcommand{\pl}{Phys.\ Lett.\ }

\newcommand{\zp}{Z.\ Phys.\ }
\newcommand{\zpc}{\zp C}
\newcommand{\prp}{preprint }
\newcommand{\ajb}{A.~J.~Buras}
\newcommand{\hn}{S.~Herrlich and U.~Nierste}

\newlength{\notewidth}
\begin{document}
\renewcommand{\thefootnote}{\fnsymbol{footnote}}

\textsf{
\begin{minipage}{2in}
\begin{flushleft}
	DESY 96-048 \\
	TUM-T31-86/96
\end{flushleft}
\end{minipage}
\hfill
\begin{minipage}{2in}
\begin{flushright}
	hep-ph/9604330 \\
	April 1996
\end{flushright}
\end{minipage}
\vspace{3ex} \\
}
\begin{center}
	{\LARGE
	The Complete \dstwo -Hamiltonian in the Next-To-Leading Order
	\footnote{Work supported by the German
		``Bundesministerium f\"ur Bildung, Wis\-sen\-schaft,
                  For\-schung und Tech\-no\-lo\-gie''
		under contract no.~06-TM-743.}
	}
\end{center}
\vspace{1pc}
\begin{center}
	{\large Stefan Herrlich
	\footnote{e-mail:
		\texttt{Stefan.Herrlich@feynman.t30.physik.tu-muenchen.de}}
	} \\
	\textsl{DESY-IfH, Platanenallee 6, D-15738 Zeuthen, Germany}
	\\[2pc]
	{\large Ulrich Nierste
	\footnote{e-mail:
		\texttt{Ulrich.Nierste@feynman.t30.physik.tu-muenchen.de}}
	} \\
	\textsl{Physik-Department, TU M\"unchen, D-85747 Garching, Germany}
\end{center}
\vspace{2pc}

\setcounter{footnote}{0}
\renewcommand{\thefootnote}{\arabic{footnote})}
\renewcommand{\arraystretch}{1.3}

\begin{abstract}
We present the complete next-to-leading order short-distance QCD
corrections to the effective \dstwo -hamiltonian in the Standard
Model.  The calculation of the coefficient $\eta_3$ is described in
great detail. It involves the two-loop mixing of bilocal structures
composed of two \dsone\ operators into \dstwo\ operators. The
next-to-leading order corrections enhance $\eta_3$ by 27\% to
\begin{displaymath}
\eta_3=0.47
\raisebox{-0.5ex}{\shortstack[l]{$\scriptstyle+0.03$\\
$\scriptstyle-0.04$}}
\end{displaymath}
thereby affecting the phenomenology of $\epsilon_K$ sizeably.
$\eta_3$ depends on the physical input parameters $m_t$, $m_c$ and
$\laMSb$ only weakly. The quoted error stems from renormalization
scale dependences, which have reduced compared to the old leading log
result.  The known calculation of $\eta_1$ and $\eta_2$ is repeated in
order to compare the structure of the three QCD coefficients.  We
further discuss some field theoretical aspects of the calculation such
as the renormalization group equation for Green's functions with two
operator insertions and the renormalization scheme dependence caused
by the presence of evanescent operators.
\end{abstract}
\newpage
\tableofcontents
%

\section{Introduction}
\label{sect:intro}

\dstwo\ transitions induce the mixing between the neutral Kaon states  
$\mathrm{K^0}$ and $\mathrm{\ov{K^0}}$.  The investigation of \kkm\
has revealed a lot about the short distance structure of nature: In
1970 Glashow, Iliopoulos and Maiani (GIM) postulated the existence of
the charm quark \cite{gim} from the suppression of this and other
flavour-changing neutral current (FCNC) processes. Then Gaillard and
Lee estimated the mass of the charm quark from the measured value of
the {\kkmd}\ \cite{gl}.  Further the violation of the CP symmetry in
nature has been first observed in \kkm\ \cite{ccft} in 1964, long
before the Standard Model of elementary particles has been
constructed.  The quantity $\epsilon_K$ characterizing this indirect
CP-violation is up to now the only unambiguously determined measure of
CP-violation in nature.  Well before the discovery of the $\tau$
lepton Kobayashi and Maskawa \cite{km} realized that the explanation
of CP-violation within the Standard Model requires a third fermion
family.  In the subsequent decades the analysis of $\epsilon_K$ has
clearly been indispensable in the determination of the elements of the
Cabibbo-Kobayashi-Maskawa (CKM) matrix. Here the CKM phase $\delta$,
which is the only source of CP-violation in the Standard Model, is
derived as a function of four key parameters: the magnitudes of the
CKM elements $V_{cb}$ and $V_{ub}$, the non-perturbative QCD parameter
$B_K$ and the top quark mass $m_t$.  $\epsilon_K$ depends on $m_t$,
because \kkm is a loop process with top quarks in the intermediate
state.  As a special feature one cannot find a solution for $\delta$
from the measured value of $\epsilon_K$ for too low values of the four
key quantities.  This has allowed to derive lower bounds on $m_t$ in
the time before the top discovery.  Yet now in the top era one can use
the measured value for $m_t$ to constrain the allowed region for the
CKM parameters {\cite{hn3}}.  But also the accuracy of the other three
parameters in the game has made significant progress in the last few
years.  To keep up with this progress the theorist's tools to predict
the strength of the \dstwo\ transitions must be sharpened as well, as
we will show in the following.

To be specific, let us look at the \dstwo-hamiltonian: 
\begin{eqnarray}  
H^{\dstwo} &=&  \frac{G_{F}^2}{16 \pi^2} \mw \, \lt[ \, 
                  \lambda_c^2 \, \eta_1  \, 
                  S(\frac{\mc }{\mw }) \:  + \:
                  \lambda_t^2 \, \eta_2  \, 
                  S(\frac{\mt }{\mw }) \rt. \nn 
&& \phantom{\frac{G_{F}^2}{16 \pi^2} \mw \,}   
         \lt.   + \, 2 \, \lambda_c \, \lambda_t \, \eta_3 \, 
                  S(\frac{\mc }{\mw },
                  \frac{\mt }{\mw }) \,
                   \rt]  
 b(\mu) \oll(\mu) \: + \:  \mathrm{h.c.} \label{s2}
\end{eqnarray} 
Here $G_{F}$ is the Fermi constant, $\lambda_j=V_{jd} V_{js}^{*}$
comprises the CKM-factors, and \oll\ is the local four quark
operator (see \fig{loc})
\begin{eqnarray}
\oll &=& 
( \ov{s}_j \gamma_\mu (1-\gamma_5) d_j)  
(\ov{s}_k \gamma^\mu (1-\gamma_5) d_k) \; = \; (\ov{s} d)_{V-A} 
(\ov{s} d)_{V-A}  \label{ollintro}
\end{eqnarray}
with $j$ and $k$ being colour indices.  The Inami-Lim functions $S(x)$
and $S(x,y)$ \cite{il} depend on the masses of the charm- and
top quark and describe the \dstwo\ transition amplitude in the absence
of strong interactions.  They are obtained by calculating the lowest
order box diagrams depicted in \fig{box}.
\begin{nfigure}
\begin{minipage}[t]{\miniwidth}
\centerline{\epsfxsize=3cm \epsffile{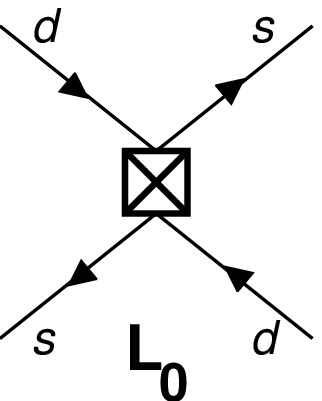}}
\caption[The \dstwo\ operator]{\slshape
        The diagram for the matrix element of \oll\
        to order  $\as^0$.
        The cross denotes the insertion of the effective
        $\mathrm{\Delta S} = -2$ operator \oll.
        }
\label{loc}
\end{minipage}
\hfill
\begin{minipage}[t]{\miniwidth}
\centerline{\epsfxsize=5cm \epsffile{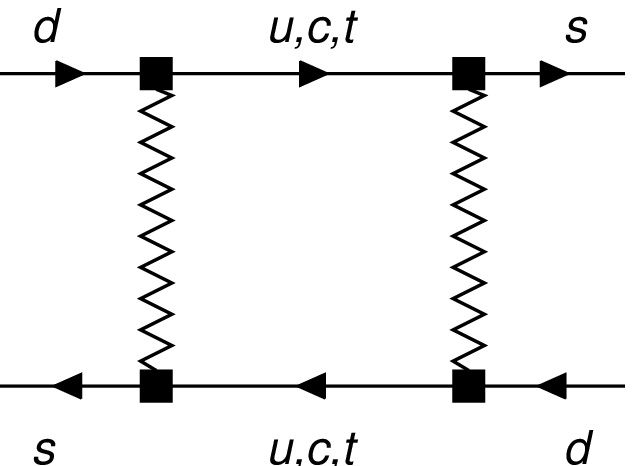}}
\caption[$\Delta S\!=\!2$ box diagram]{\slshape
        The lowest order box diagrams mediating a \dstwo\ transition.
        The zigzag lines stand for W-bosons or
        fictitious Higgs particles.  The diagrams rotated by 
        90$^\circ$ must also be considered.}
\label{box} 
\end{minipage}
\end{nfigure}

We will be interested in the short distance QCD corrections comprised
in the coefficients $\eta_1$, $\eta_2$ and $\eta_3$ with a common
factor $b(\mu)$ split off.  They describe the effect of dressing the
lowest order diagram in \fig{box} with gluons in all possible ways.
The $\eta_i$'s are functions of the charm and top quark masses and of
the QCD scale parameter \laQCD.  Further they depend on various
renormalization scales.  This dependence, however, is artificial, as
it originates from the truncation of the perturbation series, and
diminishes order-by-order in $\as$.

The hadronic matrix element of \oll\ between the neutral Kaon states
is parametrized as
\begin{eqnarray}
 \langle   \ov{K^0} \mid \oll (\mu) \mid {K^0} \rangle &=&
\frac{8}{3} \frac{B_K}{ b (\mu )} f_K^2 m_K^2 .    
\label{bk}
\end{eqnarray}
Here $m_K$ and $f_K$ are mass and decay constant of the neutral K
meson and $\mu$ is the renormalization scale at which the short
distance calculation of \eq{s2} is matched with the non-perturbative
evaluation of \eq{bk}.  $B_K$ in \eq{bk} is defined in a
renormalization group (RG) invariant way, because the $\mu$-dependent
terms cancel when the physical matrix element $\langle \ov{K^0} \mid
H^{\dstwo} \mid {K^0} \rangle$ is expressed in terms of $B_K$.

The first determination of \eq{s2} in the free quark model (i.e.\ with
$\eta_i \cdot b(\mu)=1$) is due to Va\u{\i}nste\u{\i}n and Khriplovich
\cite{vk} and Gaillard and Lee \cite{gl}.  Then the QCD factor
$\eta_1$, which is only sensitive to the first two quark families, has
been calculated in the leading-logarithmic approximation by
Va\u{\i}nste\u{\i}n, Zakharov, Novikov and Shifman \cite{vzns}.  They
have explicitly extracted the coefficient of the leading logarithm
$\as \ln (\mc/\mw)$ from the diagrams depicted in \fig{boxqcd} and
summed this logarithm to all orders in perturbation theory with the
help of the RG equation.  In the same way the coefficient $\eta_2$ has
been obtained by Vysotski\u{\i} \cite{vys} for the case of a light top
quark.  Then Gilman and Wise \cite{gw} have introduced a more
efficient method to achieve the leading log summation.  Following
Witten \cite{wit} they have applied Wilson's operator product
expansion \cite{w} consequently to the \dsone-substructure and could
reproduce the results of \cite{vzns,vys} for $\eta_1$ and $\eta_2$.
Further they have correctly determined $\eta_3$, which involves a
larger operator basis than $\eta_1$ and $\eta_2$ due to the presence
of penguin operators \cite{vzs}.  It is difficult, if not impossible,
to achieve this calculation with the older methods of \cite{vzns,vys}.
Further the leading order (LO) calculation of \cite{gw} has only
required one-loop calculations to obtain the leading logarithms of the
diagrams in \fig{boxqcd}.  The results of \cite{gw} have later been
extended to the case of a heavy top quark by Flynn and by Datta,
Fr\"ohlich and Paschos \cite{fly}.

Yet LO results suffer from certain systematic drawbacks and a
precision calculation must include the next-to-leading order (NLO)
terms.  We sketch the reasons here:
\begin{enumerate}
\renewcommand{\labelenumi}{\roman{enumi})}
\item The fundamental QCD scale parameter \laMSb\ is not
well-defined in the LO.
\item The quark mass dependence of the $\eta_i$'s is not
correctly reproduced by the LO expressions.  Especially the
$m_t$-dependent terms in $\eta_3 \cdot S(x_c,x_t)$ belong to the NLO.
\item Similarly the question of the \emph{definition} of the quark
masses (i.e.\ the renormalization scheme and scale) to be used in
\eq{s2} is a next-to-leading order issue: Hence one has to go to
the NLO to know how to use $m_t$ as determined at Fermilab in
low-energy hamiltonians like \eq{s2}.
\item The LO results for $\eta_1$ and $\eta_3$ 
show a large dependence on the renormalization scales, at which one
integrates out heavy particles.  In the NLO these uncertainties are
reduced considerably.
\item One must go to the NLO to judge whether perturbation theory 
works, i.e.\ whether the radiative corrections are small.  After all
the corrections can be sizeable.
\end{enumerate}
\begin{nfigure}
\centerline{\epsfxsize=\textwidth \epsffile{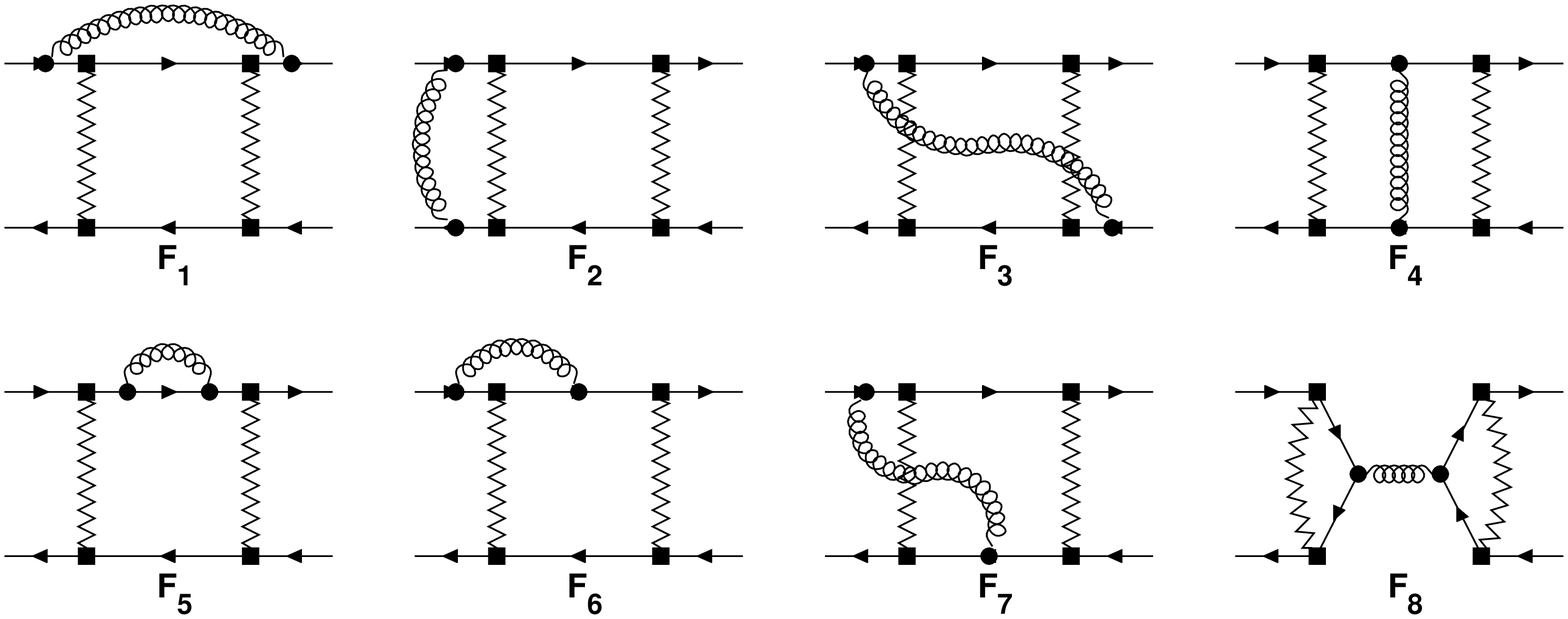}}
\caption[QCD corrections to the box diagram]{\slshape \protect
        The classes of diagrams constituting the 
        $O(\as)$-contribution 
        to \dstwo\ transitions in the SM;
        the remaining diagrams are obtained by left-right and
        up-down reflections.
        The curly lines denote gluons.
        Also QCD counterterm diagrams have to be included.
        Diagram $\mathsf{F_8}$ equals 0 for zero external momenta.
}
\label{boxqcd}
\end{nfigure}
The first step of the extension of \eq{s2} beyond the leading order
has been done by Buras, Jamin and Weisz, who have derived the NLO
expression for $\eta_2$ \cite{bjw}. Then we have calculated $\eta_1$
in the NLO \cite{hn1}, and the present work is devoted to present the
details of our NLO calculation of $\eta_3$.  For completeness we will
also list the results of \cite{bjw,hn1} for $\eta_1$ and $\eta_2$ and
illustrate the different structure of $\eta_1$, $\eta_2$ and $\eta_3$.
The numerical results and the phenomenological implications of our
findings on the analysis of $\epsilon_K$ and the \kkmd\/ have already
been given in \cite{hn3} (for an update see \cite{talk}). We assume
that the reader is familiar with the general concepts of Wilson's 
operator product expansion, the renormalization group and operator
mixing. A detailed description of these tools in the context of 
\kkm\ can be found in \cite{thu,ths}.

The paper is organized as follows: In \sect{sect:sm} we first set up
our notation and then discuss the {\dstwo}\ transition in the Standard
Model (SM). We identify the large logarithms in the transition
amplitude to orders $\as^0$ and $\as^1$ and compare their numerical
sizes.  Sect.~\ref{sect:abovec} is devoted to the effective
description of the \dstwo\ transition between the scales
$\mu_{tW}=O(m_t,M_W)$, at which the top quark and the W-boson are
integrated out, and $\mu_c=O(m_c)$, at which the charm quark is
removed as a dynamic degree of freedom.  Here we construct the
operator basis used in the effective lagrangian. We then match the SM
amplitude to the effective matrix elements at the initial scale
$\mu_{tW}$ and determine the RG evolution down to $\mu_c$.  The latter
requires the solution of RG equations for double operator insertions.
Subsequently we describe the two-loop calculations needed to obtain
the anomalous dimension tensor.  Sect.~\ref{sect:belowc} deals with
the effective theory below the scale $\mu_c$.  Here we first match the
effective four-flavour theory obtained in the last section to an
effective three-quark theory and then perform the RG evolution down to
some hadronic scale $\mu$.  In \sect{sect:final} we summarize the
analytic result and present an approximate formula having an accuracy
of approximately 1\%.  Sect.~\ref{sect:num} contains the numerical
analysis, which includes a discussion of the residual scale dependence
of our result as well as the dependence on physical parameters.  Then
we close the paper with our conclusions.  The appendices contain the
results of the two-loop diagrams, the renormalization factors and
important RG quantities appearing in the calculation.

The impatient reader may skip sects.~\ref{sect:massdef},
\ref{sect:sm-1}, \ref{sect:unphysical}, \ref{sect:rgsingle},
\ref{sect:inhom}, \ref{sect:evas} and \ref{sect:dort} and restrict
her or his attention in sects.~\ref{sect:final} and \ref{sect:num} to
the final results in \eq{SummaryResEta3}, \eq{SummaryResEta3*},
\eq{num-final}, \eq{num-final2}, \tab{tab:eta12} and the plots.  
%

\section{$\dstwo$ Transitions in the Standard Model}
\label{sect:sm}

\subsection{Notations and Conventions}
\label{sect:sm-conv}
Before writing down the result for the diagram of \fig{box}, we set up
the conventions and notations used in this work.

Throughout this paper we will use dimensional regularization and the
{\msb}\ renormalization scheme \cite{bbdm}.  Since only open fermion
lines appear during the calculation, we can safely use a naive
anticommuting $\gamma_5$ (NDR scheme) as justified in \cite{bw,krei}.
The result for $\eta_3$ will be scheme independent, the only scheme
dependence of $H^{\dstwo}$ in \eq{s2} resides in the factor $b(\mu)$.
The scheme dependences of $b(\mu)$ and the hadronic matrix element in
\eq{bk} must cancel, so that $B_K$ is scheme independent. For the 
W-propagator the 't Hooft-Feynman gauge will be used, while the QCD
gauge parameter $\xi$ is kept arbitrary.

Let $\widetilde{G}$ be the \dstwo\ Standard Model Green's function,
which is understood to be truncated, connected and Fourier-transformed
into momentum space.  In this work we are interested in the lowest
order contribution to $\wt{G}$ given by the box diagram of \fig{box}
and the QCD radiative correction to it.  The different contributions
from the internal quarks involve different CKM factors
$\lambda_j=V_{jd} V_{js}^{*}$, $j=u,c,t$.  The GIM mechanism
$\lambda_t + \lambda_c + \lambda_u = 0$ allows to eliminate
$\lambda_u$.  Now the contributions from light internal quarks must be
treated differently from those involving the heavy top quark.
We therefore split $\wt{G}$ up as
\begin{eqnarray}
\widetilde{G} &=& \lambda_c^2 \widetilde{G}^c
 +\lambda_t^2 \widetilde{G}^t
 + 2 \lambda_c \lambda_t \widetilde{G}^{ct}.
\label{smbox}
\end{eqnarray}
The upper indices in the three terms in (\ref{smbox}) denote the
internal quark flavours involved. Further each term contains
contributions with up quarks due to the GIM mechanism. When discussing
the quark mass dependence of the $\wt{G}^j$'s we will frequently use
the abbreviation $x_i=m_i^2/\mw$ for the squared ratio of some quark
mass and the W mass.  In the effective hamiltonian (\ref{s2}) the
three terms involving $\eta_1$, $\eta_2$ and $\eta_3$ emerge from
$\wt{G}^c$, $\wt{G}^t$ and $\wt{G}^{ct}$ respectively.

Frequently we will use the abbreviations $L=1-\gamma_5$ and
$R=1+\gamma_5$.  $N$ is the number of colours, the $T^{a}$'s denote
the generators of the colour group $SU(N)$ in the fundamental
representation, and the $f^{abc}$'s are the structure constants.  We
will use $\openone$ and $\topenone$ to denote colour singlet and
antisinglet, i.e.\ $(L \otimes R) \cdot \topenone$ means $\ov{s}_i
(1-\gamma_5) d_j \cdot \ov{s}_j (1+\gamma_5) d_i$ with $i,j$ being
colour indices.  The $SU(N)$ Casimir factor involved will be
$C_F=(N^2-1)/(2 N)$.  We will frequently express Green's functions in
terms of the matrix element of the local \dstwo\ operator $\oll$
defined in \eq{ollintro} and displayed in \fig{loc}.

The $\widetilde{G}^j$'s in \eq{smbox} will be expanded in $\as$ as
\begin{eqnarray}
\widetilde{G}^{j} &=& \widetilde{G}^{j,\, (0)}
      + \frac{\as}{4 \pi} \widetilde{G}^{j,\,(1)} + O(\as^2).
\label{greenex}
\end{eqnarray}
The $\widetilde{G}^{j,\,(n)}$'s, $n\ge 1$, involve infrared (mass)
singularities, which will be regularized by small quark masses $m_s$
and $m_d$.  The matrix element of some operator $Q$ between quark states
will be denoted by $\langle Q \rangle $ and expanded as
\begin{eqnarray}
\langle Q \lt( \mu \rt) \rangle &=&
\langle Q  \rangle^{(0)}  + \frac{\as (\mu) }{4 \pi}
\langle Q \lt( \mu \rt) \rangle^{(1)} + O \lt( \as^2 (\mu) \rt).
\label{qex}
\end{eqnarray}

\subsection{Zeroth Order Amplitude}
\label{sect:sm-0}
In the leading order of $m_c/m_{\mathrm{heavy}}$, where
$m_{\mathrm{heavy}}$ stands for $m_t$ or $M_W$,
one can neglect the external momenta in \eq{greenex} and \eq{qex}.

One obtains for the three terms in \eq{smbox}:
\begin{subequations}
\label{gall-0}
\begin{eqnarray}
i \widetilde{G}^{ct, \,(0)} &=& \frac{\gf^2}{16 \pi^2} \mw S(x_c,x_t)
\langle \oll \rangle^{(0)},
\label{gct}
\\
i \widetilde{G}^{j,\,(0)} &=& \frac{\gf^2}{16 \pi^2} \mw S(x_j)
\langle \oll \rangle^{(0)}, \quad j=c,t.
\label{gj}
\end{eqnarray}
\end{subequations}
Here the Inami-Lim function \cite{il} $S(x_j,x_k)$ equals
\begin{eqnarray}
S(x_j,x_k) &=& \widetilde{S}(x_j,x_k) - \widetilde{S}(x_j,0) -
\widetilde{S}(x_k,0) + \widetilde{S}(0,0),
\end{eqnarray}
where the result of the box diagram with internal quarks $j$ and $k$
is denoted by $\widetilde{S}(x_j,x_k)$ and the up-quark mass is set to
zero.  Further $S(x_j)=S(x_j, x_j)$.  Here one realizes that the
effect of the GIM mechanism is not only to forbid FCNC's at tree
level, but also to cancel the constant terms in the $\widetilde{S}$'s
and to nullify {\kkm} in the case of degenerate quark masses.

Let us look at the three contributions \eq{gct} and \eq{gj} to
\eq{smbox} in more detail:
\begin{subequations}
\label{sfkts}
\begin{eqnarray}
S(x_t) &=& x_t \lt[ \frac{1}{4} + \frac{9}{4}
\frac{1}{1-x_t} - \frac{3}{2} \frac{1}{(1-x_t)^2} \rt]
- \frac{3}{2} \lt[ \frac{x_t}{1-x_t} \rt]^3 \ln x_t,
\label{sxt} \\
S(x_c) &=& x_c + O(x_c^2),
\label{ilgim} \\
S(x_c,x_t) &=& - x_c \ln x_c + x_c \F(x_t)  + O(x_c^2 \ln x_c),
\label{sxcxt}
\end{eqnarray}
\end{subequations}
with
\begin{eqnarray}
\F(x_t)&=&  \frac{x_t^2-8 x_t+4 }{4 (1-x_t)^2} \ln x_t
          + \frac{3}{4} \frac{x_t}{x_t-1}.
\label{deff}
\end{eqnarray}
In \eq{ilgim} and \eq{sxcxt} we have only kept terms which are larger
than those of order $(m_s m_c)/\mw$ neglected by setting the external
momenta to zero.  Clearly $S(x_t)$ is much larger than $S(x_c)$ and
$S(x_c,x_t)$ reflecting the non-decoupling of the heavy top quark.
The vanishing of $S(x_c)$ and $S(x_c,x_t)$ in the limit $x_c
\rightarrow 0$ is sometimes called \emph{hard GIM suppression}.  In
the imaginary part of $i \wt{G}$, which is important for CP violation,
the size of $S(x_t)$ over-compensates the CKM suppression of the
corresponding term in (\ref{smbox}), but the three terms are roughly
of the same size.  Conversely the real part of $i \wt{G}$ relevant for
the \kkmd\ is dominated by $\wt{G}^c$ and therefore insensitive to
$m_t$ (see \cite{hn3}).

\subsection{Large Logarithms}
The Inami-Lim functions in \eq{sfkts} contain logarithms of the ratios of
internal particle masses.  As we will see in {\sect{sect:sm-1}} the
same is true for the QCD radiative corrections in (\ref{greenex}),
which in addition involve logarithms of the renormalization scale
$\mu$.  We now discuss these logarithms in order to illustrate the
effect of the forthcoming renormalization group (RG) improvement.

When the product of such a logarithm with $\as$ is large, one has to
sum it to all orders of perturbation theory by using RG methods. Let
us first investigate these logarithms in the zeroth order terms in
{\eq{sfkts}}: {\eq{sxt}} clearly contains no large logarithm because
of $\ln x_t \approx 1.5$.  Since (\ref{sxt}) contains no $x_c$, the
scale entering $\as$ in the QCD radiative corrections to
$\wt{G}^{t,\,(0)}$ is of the order of $M_W$ or $m_t$.  Now
$\as(M_W)/(4\pi)\cdot \ln x_t \approx 10^{-2}$ and needs not to be
summed by RG methods. We will come back to this point in
sect.~\ref{sect:dort}. Yet the radiative corrections in $\wt{G}^t$
contain the renormalization scale explicitly through $\ln (\mu/M_W) $.
The non-perturbative evaluation of the matrix element in (\ref{bk}) is
performed at a low hadronic scale, so that $\ln (\mu/M_W) $ will also
be large.  In \eq{sxcxt} we find a large logarithm $|\ln x_c| \approx
8$. It is 13 times larger than $\F(x_t)$. Further $\as (m_c)/\pi \cdot
|\ln x_c| = 1.0$, so that the summation of this logarithm is
indispensable.  In \eq{ilgim} we would naturally expect the large
logarithm $\ln x_c$, too.  Its absence is due to the GIM mechanism,
which we may term \emph{super-hard} in this case.  Yet the higher
order terms in $\wt{G}^c$ do contain $\ln x_c$, though with one power
less than those in $\wt{G}^{ct}$.  Of course $\wt{G}^{c, \, (n)}$ and
$\wt{G}^{ct, \, (n)}$, $n\geq 1$, also explicitly depend on $\mu$.  We
may group the logarithms such that this dependence appears as $\ln
(\mu/m_c)$. The summation of this logarithm is performed by the RG
evolution below the charm threshold described in
sect.~\ref{sect:belowc}.

The RG evolution from $\mu=M_W$ to $\mu=m_c$ summing $\ln x_c$ will be
described in sect.~\ref{sect:abovec}.
From \eq{sfkts} one can already read off the type of summed logarithms
in the three terms of \eq{smbox}, we summarize them in
\tab{tab:rglogs}.
\begin{ntable}
\begin{center}
\begin{tabular}{l|*{3}{r@{}c@{}l}}
Order &
& $\wt{G}^c$ &&& $\wt{G}^t$ &&& $\wt{G}^{ct}$ & \\
\hline
LO &
 & $(\as\ln x_c)^n$ & &
 & $(\as\ln x_c)^n$ & &
 & $(\as\ln x_c)^n$ & $\ln x_c$
\\
NLO &
$\as$ & $(\as\ln x_c)^n$ & &
$\as$ & $(\as\ln x_c)^n$ & &
 & $(\as\ln x_c)^n$ &
\\
NNLO &
$\as^2$ & $(\as\ln x_c)^n$ & &
$\as^2$ & $(\as\ln x_c)^n$ & &
$\as$ & $(\as\ln x_c)^n$ &
\\ \hline
$m_t$-dependence &
 & none &&
 & in LO &&
 & in NLO &
\end{tabular}
\end{center}
\caption{\slshape
Logarithms summed by the RG evolution from $M_W$ down to $m_c$ for the
three terms in \eq{smbox}, $n=0,1,2,\ldots$  The last line shows the
order in which the dependence on $m_t$ enters.  From the column
labeled by $\wt{G}^{ct}$ one reads off that the phenomenologically
interesting $m_t$ dependent terms in $S(x_c,x_t)$ in \eq{sxcxt}
actually belong to the NLO.  This emphasizes the importance of a
complete NLO calculation for $\eta_3$.}
\label{tab:rglogs}
\end{ntable}
Of course there is no charm quark in the calculation of $\wt{G}^{t}$,
the large logarithm $\ln x_c$ here emerges from $\ln \mu^2/\mw$
contained in $\wt{G}^{t,\,(n)}$ for $n \ge 1$.  If we now perform the
RG evolution from $\mu=M_W$ down to $\mu=m_c$, we will obtain the
quoted logarithm.

\subsection{The Definition of Quark Masses}\label{sect:massdef}
When discussing analytical expressions beyond the LO, one must specify
the \emph{definition} of the quark masses.  This point is often
handled incorrectly in phenomenological analyses, so that we discuss
it in some detail now.

Any perturbatively calculated interacting fermion propagator is
proportional to 
\begin{eqnarray}
 \frac{  i } 
     {\slash p  - m + \Sigma \left( p^2,m \right)  } .
\label{fermprop} 
\end{eqnarray}
Here $m$ is the renormalized current fermion mass, which enters the
Lagrangian, and $i \Sigma \left( p^2,m \right) $ is the 1PI
self-energy describing the dressing of the free fermion
propagator. $\Sigma $ starts at  second order in the gauge coupling
$g$ and may be calculated to some order $g^{2 n}$.  Now different
renormalization schemes may involve definitions $m$ and $m^\prime$ of
the fermion mass, which differ by a perturbative series:
\begin{eqnarray}
m^\prime &=& m \left( 1 + K g^2 + \ldots    \right) \nonumber.
\end{eqnarray}  
Yet also $\Sigma$ in (\ref{fermprop}) is different in both schemes, 
but the position of the pole in (\ref{fermprop}) is the same 
within the calculated order:
\begin{eqnarray}
m^\prime - \Sigma^\prime \left( m^{\prime\,2}, m^\prime \right)  
&=& m - \Sigma \left( m^2  , m \right) + O\left(g^{2n+2} \right)\nonumber
\end{eqnarray}
The freedom in the choice of the mass counterterms allows us to move
any desired constant term from $m$ to $\Sigma$.  If the fermion is a
lepton and therefore exists as a free particle, $m$ is commonly
defined as the \emph{pole} mass $m_{\mathrm{pole}}$ corresponding to
$\Sigma \left(m_{\mathrm{pole}}^2,m_{\mathrm{pole}} \right)=0$.  Since
the pole at $p^2=m^2_{\mathrm{pole}}$ in (\ref{fermprop}) is
observable for free fermions, $m_{\mathrm{pole}}$ is sometimes called
the \emph{physical} mass.  Yet the strong interaction confines quarks
into hadrons and the quark pole mass is not observable. In fact the
infrared structure of QCD imposes a strongly divergent perturbation
series upon observables expressed in terms of $m_{\mathrm{pole}}$,
which is most likely only a suitable parameter for very low orders of
perturbation theory \cite{bbb}.  Instead in QCD one preferably uses a
short distance mass such as the running quark mass $m(\mu)$ in the
\msb\ scheme.  It has the additional advantage to allow for a simpler
solution of the RG equations.  Its relation to the one-loop pole mass
reads:
\begin{eqnarray}
m_{\mathrm{pole}}^{(1)} &=& m (\mu) \left[ 1 +  \frac{\as (\mu)}{4
\pi} C_F \left( 4 + 3 \ln \frac{\mu^2}{m^2} \right) \right]
\label{mstopol} 
\end{eqnarray} 
Clearly the proper definition of the quark mass only matters beyond
the leading order. The one-loop relation (\ref{mstopol}) is the
appropriate one for the NLO calculation presented in this paper.  At
Fermilab the top quark pole mass is measured.
$m_{t,\mathrm{pole}}^{(1)}$ is larger than $m_t(m_t)$ by a factor of
1.045 corresponding to 7-8 GeV.

Finally, if the renormalization scale $\mu$ is much different from
$m$, one must sum the logarithm in (\ref{mstopol}) to all orders with
the help of the RG, see \eq{MassEvol} in appendix~\ref{apx:rgquant}.

\subsection{The $O(\as)$ Corrections}\label{sect:sm-1}
The $O(\as)$ corrections to the box diagram (see \fig{boxqcd}) were
first evaluated in \cite{bjw} for the case of arbitrary internal quark
masses.  These corrections have been necessary to obtain $\eta_1$ and
$\eta_2$ in the NLO \cite{hn1,bjw}.  We stress here that one does not
need them for the NLO calculation of $\eta_3$, which is the novel
issue presented in this work.  Nevertheless it is instructive to look
at $\wt{G}^{ct,\, (1)}$ as well for three reasons: First one can
identify the logarithms summed by the RG evolution, which provides a
very good check of the results presented in sects.~\ref{sect:abovec}
and \ref{sect:belowc}.  Second one can partly estimate the size of the
next-to-next-to-leading order (NNLO) terms.  Third the $O(\as)$ terms
will be useful in the discussion of the proper treatment of the
physics between the scales $\mu=m_t$ and $\mu=M_W$ presented in
sect.~\ref{sect:dort}.

Generally the $O(\as)$ terms are of the form
\begin{eqnarray}
i \widetilde{G}^{j, (1)} &=&
\frac{\gf^2}{16\pi^2} \mw \left\{
h^{j} \langle \oll \rangle^{(0)}
+ h^{j}_T \langle T \rangle^{(0)}
+ h^{j}_U \langle U \rangle^{(0)}
\right\}, \quad j=c,t,ct.
\label{gall-1}
\end{eqnarray}
Here new operators have emerged\footnote{We omit the spinors on the
external quark lines.}
\begin{subequations}
\label{tu}
\begin{eqnarray}
T &=& \left( L \otimes L + R \otimes R - \sigma_{\mu\nu} \otimes
\sigma^{\mu\nu} \right) \frac{N-1}{2N} \openone
\\
U &=& \frac{1}{2} \left(\gamma_{\mu}L \otimes \gamma^{\mu} R +
\gamma_{\mu} R \otimes \gamma^{\mu} L \right) \left(
\frac{N^2+N-1}{2N} \openone - \frac{1}{2N} \topenone \right)
\no \\
 & & - \left( L \otimes R + R \otimes L \right) \left(
\frac{N^2+N-1}{2N} \topenone - \frac{1}{2N} \openone \right),
\end{eqnarray}
\end{subequations}
which are written in a manifestly Fierz self-conjugate way.  In the
following we will discuss the coefficient functions $h$ in \eq{gall-1}
in great detail, starting with those of the new operators:
\begin{subequations}
\label{h-1}
\begin{eqnarray}
h^{ct}_T \; = \; h_T \, S(x_c,x_t); && \qquad \qquad
h^{j}_T \; = \; h_T \, S(x_j), \qquad j=c,t; \\
h^{ct}_U \; = \; h_U \, S(x_c,x_t); && \qquad \qquad
h^{j}_U \; = \; h_U \, S(x_j), \qquad j=c,t; \\
h_T \; = \; -3-\xi;  && \qquad \qquad 
h_U \; = \; \frac{3+\xi}{2} \, \frac{m_d m_s}{\ms-\md} \, \ln
\frac{\ms}{\md} .
\end{eqnarray}
\end{subequations}
We first observe that \eq{gall-1} is obviously unphysical, because the
functions in (\ref{h-1}) are gauge dependent. This is an artifact of
the use of small quark masses to regularize the infrared singularities
while at the same time using on-shell quarks with zero four-momentum
for the external states.  For the same reason we encounter the new
operators $T$ and $U$. Yet the one-loop matrix element of $\oll$
corresponding to the Feynman diagrams of \fig{fig:ds2-1}
\begin{nfigure}
\centerline{\epsfxsize=\textwidth \epsffile{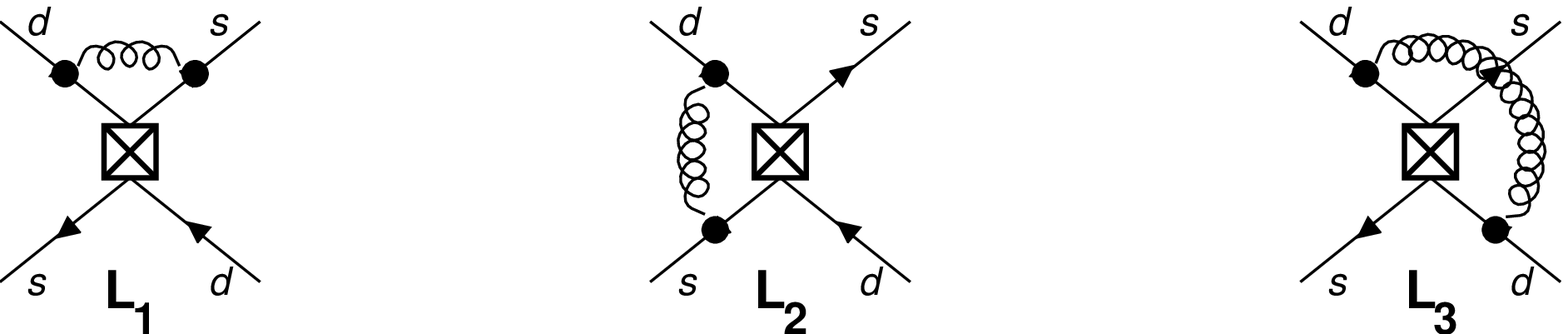}}
\caption{\slshape
The classes of diagrams constituting the $O(\as)$-contribution to the
matrix element of $\oll$; the remaining diagrams are obtained by
mirror reflections.  The cross denotes the insertion of $\oll$, the
curly lines gluons.  A QCD counterterm has to be included.}
\label{fig:ds2-1}
\end{nfigure}
involves the same operators (\ref{tu}) with the same coefficients
\cite{bjw,hn1,bw}:
\begin{eqnarray}
\hspace{-2ex} 
\langle\oll\lt(\mu\rt)\rangle \!  &=& \!
\langle\oll\lt(\mu\rt)\rangle^{(0)} + \frac{\as\left(\mu\right)}{4\pi}
  \lt[a(\mu) \langle\oll\rangle^{(0)}
      + h_T \langle\hat{T}\rangle^{(0)}
      + h_U \langle\hat{U}\rangle^{(0)} \rt].
\label{lome}
\end{eqnarray}
We will discuss the third coefficient $a(\mu)$ in conjunction with
$h^j$ in (\ref{gall-1}).  Now (\ref{gall-1}) and (\ref{lome}) allow to
express $i \widetilde{G}^{j}$ as
\begin{subequations}
\label{factg}
\begin{eqnarray}
i \widetilde{G}^j(\mu) &=& \frac{\gf^2}{16\pi^2} \mw
\lt[ S \lt( x_j (\mu) \rt) + \frac{\as(\mu)}{4\pi} 
     k^j\lt( x_j (\mu),\mu \rt) \rt]
\langle\oll(\mu)\rangle \no \\
&&+ O(\as^2), \hspace{12em} j=c,t,
\label{factgj}
\\
i \widetilde{G}^{ct} (\mu) &=& \frac{\gf^2}{16\pi^2} \mw
\lt[ S\lt(x_c (\mu) ,x_t (\mu)\rt) + 
     \frac{\as (\mu)}{4 \pi} k^{ct} \lt( x_c (\mu) ,x_t (\mu),\mu \rt) \rt]
\langle\oll(\mu)\rangle \no \\
&&+ O( \as^2 ),
\label{factgct}
\end{eqnarray}
\end{subequations}
where the new coefficient $k^j$ is related to $h^j$ in (\ref{gall-1}) and
$a(\mu)$ in (\ref{lome}) via
\begin{subequations}
\label{hka}
\begin{eqnarray}
h^j &=& k^j(x_j,\mu)  +  a(\mu) S(x_j), \hspace{2em} j=c,t, \\
h^{ct} &=& k^{ct}(x_c,x_t,\mu)  +  a(\mu) S(x_c,x_t).
\end{eqnarray}
\end{subequations}
Now the unphysical terms $h_T$ and $h_U$ containing $\xi$ and the
infrared regulators $m_d$ and $m_s$ have been absorbed into $\langle
\oll \rangle ^{(1)}$ in (\ref{factg}). Likewise these unphysical
terms in $h^j$ have gone into $a(\mu)$ in (\ref{hka}), so that $k^j$
only depends on $x_c$, $x_t$ and $\mu$.

In the coefficient $a(\mu)$ in (\ref{lome}) we also split off
the gauge and IR parts:
\begin{eqnarray}
a(\mu) &=& c  + \frac{N-1}{2 N} 6 \ln \frac{m_d m_s}{\mu ^2} \nn
&& +
  2 \xi \lt[  \lt(C_F + \frac{N-1}{2 N}  \rt)
    \lt( 1 - \frac{\ms \ln\frac{\ms}{\mu^2} -
                     \md \ln\frac{\md}{\mu^2}}{\ms-\md} \rt)
  + \frac{N-1}{2 N} \ln \frac{m_d m_s}{\mu^2}  \rt] , \nn
c  &=& -3 \, C_F -5 \, \frac{N-1}{2 N}. \label{acloc}
\end{eqnarray}
Next we write down the results for the $k^j$'s grouped according to
powers of the large logarithm $\ln x_c$:
\begin{subequations}
\label{h-1-all}
\begin{eqnarray}
k^{ct} (x_c,x_t,\mu) &=& C_F \Biggl[
-3 x_c \ln^2 x_c
+x_c \ln x_c \left(-5-6\ln\frac{\mu^2}{\mc}+ \kctsa \left(x_t\right)\right)
\no \\
& & \hspace{2em}
+x_c \kcts\left(x_t\right)
+ x_c \ln\frac{\mu^2}{\mc} \kctsa \left(x_t\right)
\Biggr]
\no \\
& &+\frac{N-1}{2N} \Biggl[
- 9 x_c \ln^2 x_c
+x_c \ln x_c \left(3+\kctoa\left(x_t\right)-6\ln\frac{\mu^2}{\mc}\right)
\no \\
& &\hspace{4em}
+ x_c \kcto\left(x_t\right)
+x_c \ln\frac{\mu^2}{\mc} \kctoa\left(x_t\right) \Biggr] \no \\
&&+ c \, x_c \ln x_c - c \, x_c \F(x_t)
+ O\left(x_c^2 \ln^2 x_c\right),
\label{h-1-ct}
\\[3mm]
k^{c} (x_c,\mu)&=& C_F \left[ -x_c + 6x_c \ln\frac{\mu^2}{\mc}
\right]
\no \\
& &+ \frac{N-1}{2N} \Biggl[ 12 x_c \ln x_c + 6 x_c \ln
\frac{\mu^2}{\mc} -11x_c + \frac{4\pi^2}{3}x_c
\Biggr]  \no \\
&& -c \, x_c + O\left(x_c^2\right),
\label{h-1-c}
\\[3mm]
k^{t} (x_t,\mu) &=& C_F \left[\kts\left(x_t\right) + 6 \ln\frac{\mu^2}{\mw}
\ktsa\left(x_t\right)
\right]
\no \\
& & + \frac{N-1}{2N} \Biggl[ \kto\left(x_t\right) + 6 \ln\frac{\mu^2}{\mw}
S\left(x_t\right)
\Biggr]  -c \, S(x_t) .
\label{h-1-t}
\end{eqnarray}
\end{subequations}
We have hidden a complicated $x_t$ dependence in the following
functions:
\begin{subequations}
\label{h-1-xt}
\begin{eqnarray}
\kcts\left(x_t\right) &=&
-3 x_t \frac{7+7x_t+2x_t^2}{4\left(x_t-1\right)^2}
+\pi^2 \frac{-8+x_t^2}{4}
\no \\
& &
+3 x_t^2 \li_2\left(1-x_t\right) \frac{-2-2x_t+x_t^2}{2\left(1-x_t\right)^2}
\no \\
& &
+\ln x_t \frac{20-12x_t-51x_t^2-11x_t^3+6x_t^4}{4\left(1-x_t\right)^3}
\no \\
& &
+3 \ln^2 x_t \frac{4-12x_t-x_t^2-x_t^3-3x_t^4+x_t^5}{4\left(x_t-1\right)^3},
\\
\kcto\left(x_t\right) &=&
\frac{-32+8x_t+3x_t^2}{4x_t\left(x_t-1\right)}
+\pi^2 \frac{-8-7x_t^2+x_t^3}{6x_t^2}
\no \\
& & +\li_2\left(1-x_t\right)
 \frac{8-16x_t+23x_t^2-29x_t^3 + 4 x_t^4 +x_t^5}{x_t^2\left(x_t-1\right)^2}
\no \\
& &
+\ln x_t \frac{-32+36x_t-12x_t^2-7x_t^3}{4x_t\left(x_t-1\right)^2}
\no \\
& &
+\ln^2 x_t \frac{12-16x_t+5x_t^2+2x_t^3}{4\left(x_t-1\right)^2},
\\
\kts\left(x_t\right) &=&
x_t \frac{4-39x_t+168x_t^2+11x_t^3}{4\left(x_t-1\right)^3}
+3x_t^3 \li_2\left(1-x_t\right) \frac{5+x_t}{\left(x_t-1\right)^3}
\no \\
& &
+3x_t \ln x_t \frac{-4+24x_t-36x_t^2-7x_t^3-x_t^4}{2\left(x_t-1\right)^4}
\no \\
& &
+3x_t^3 \ln^2 x_t \frac{13+4x_t+x_t^2}{2\left(x_t-1\right)^4},
\\
\kto\left(x_t\right) &=&
\frac{-64+68x_t+17x_t^2-11x_t^3}{4\left(1-x_t\right)^2}
+\pi^2 \frac{8}{3x_t}
\no \\
& &
+2 \li_2\left(1-x_t\right)
\frac{8-24x_t+20 x_t^2-x_t^3+7x_t^4-x_t^5}{x_t\left(x_t-1\right)^3}
\no \\
& &
+\ln x_t \frac{-32+68x_t-32x_t^2+28x_t^3-3x_t^4}{2\left(x_t-1\right)^3}
\no \\
& &
+x_t^2 \ln^2 x_t \frac{4-7x_t+7x_t^2-2x_t^3}{2\left(x_t-1\right)^4},
\\
\kctsa\left(x_t\right) &=&
-9\frac{x_t}{\left(x_t-1\right)^2}
+3\ln x_t \frac{-4+12x_t-3x_t^2+x_t^3}{2\left(x_t-1\right)^3},
\no \\
&=& 6 \lt( \F (x_t) -1 + x_t \diff{x_t} \F(x_t) \rt) ,
\\
\kctoa\left(x_t\right) &=& \frac{9}{2}  \frac{x_t}{x_t-1}
+3 \frac{4-8x_t+x_t^2}{2\left(1-x_t\right)^2} \ln x_t
\; = \; 6 \, \F (x_t)
\\
\ktsa\left(x_t\right) &=&
x_t \frac{-4+18x_t+3x_t^2+x_t^3}{4\left(x_t-1\right)^3}
- \frac{9x_t^3}{2\left(x_t-1\right)^4} \ln x_t \no  \\
& = & x_t \diff{x_t} S(x_t) .
\end{eqnarray}
\end{subequations}
Here $\li_2\left(x\right)$ denotes the dilogarithm function
\begin{eqnarray}
\li_2\left(x\right) &=& - \int_0^1 dt \frac{\ln\left(1-xt\right)}{t}
\label{dilog}
\end{eqnarray}
and $F(x_t)$ has been defined in (\ref{deff}).  
Let us now look at the ingredients of \eq{factg} in more detail:
\eq{factg} is an operator product expansion (OPE) of the Standard Model
amplitude in terms of the local \dstwo\ operator $\oll$.  The terms in
brackets are the corresponding Wilson coefficients, yet in ordinary
perturbation theory without any RG improvement.  From \eq{sfkts} and
\eq{h-1-all} one verifies that they are gauge-independent and free of
the infrared regulators $m_s$ and $m_d$. If we had used the
dimensional method also to regularize the IR singularities, the
operators $T$ and $U$ and the gauge dependence would be absent
on both sides of \eq{factg}, but the $k^j$'s would be
unchanged. Further the Wilson coefficients do not depend on the choice
of the external states used in the calculation of the matrix element.
Now $k^t (x_t,\mu)$ is simply the $O(\as)$ part of the initial
condition for the RG improved Wilson coefficient needed for the
calculation of $\eta_2$.  $k^t( x_t,\mu)$ is called $D(x_t)$ in
\cite{bjw}. The RG evolution from $\mu=M_W$ down to a low hadronic
scale sums $\as \ln (\mu/M_W)$ to all orders in perturbation
theory. The situation would be the same with $k^{ct}$ and $k^c$ in a
fictitious world in which the charm quark is so heavy that $\ln x_c $ is
small. To describe the real nature, however, we must first sum $\as
\ln x_c$ to all orders as well.  Since this is the purpose of the
subsequent sections, we discuss the powers of $\ln x_c$ term-by-term
now.  Therefore we have arranged \eq{h-1-all} such that large
logarithms can easily be distinguished from small terms.
\begin{description}
\item [$k^{ct}$:]
In \eq{h-1-ct} one immediately observes two terms $\propto \ln^2 x_c$,
which we have expected from the fact that $S(x_c,x_t)$ in \eq{sxcxt}
already contains the logarithm $\ln x_c$.  They all belong to the LO
of RG improved perturbation theory, c.f.\ \tab{tab:rglogs}.  Further
$k^{ct}$ exhibits $\ln^1 x_c$ terms.  They are linked to the $\ln^0
x_c$ term of $S(x_c,x_t)$ and constitute the first terms of the NLO
expression.  Note that the LO terms, the one $\propto\ln x_c$ in
$S(x_c,x_t)$ and the $\ln^2 x_c$ term of $k^{ct}$, are
independent of $m_t$.  Top dependence first enters through the NLO
terms $\ln^0 x_c$ in \eq{sxcxt} and the functions $\kctsa$ and
$\kctoa$.  Finally $k^{ct}$ contains a $\ln^0 x_c$ piece, which
already belongs to the next-to-next-to-leading order (NNLO).
Therefore we will not need it in our analysis.
\item [$k^{t}$:]
\eq{h-1-t} contains the logarithm  $\ln \mu^2/\mw$,
which gets large when $\mu$ gets small.  We expect the appearance of
just this logarithm, because there are no other largely separated mass
scales in $\wt{G}^t$.  The whole $S(x_t)$ in \eq{sxt} and the term
$\propto \ln \mu^2/\mw $ in \eq{h-1-t} belong to the LO.  The $\ln^0
\mu^2/\mw$  part of \eq{h-1-t} is a NLO term.
\item [$k^{c}$:]
In \eq{h-1-c} we find the large logarithm $\ln x_c$, which together
with $S(x_c)$ belongs to the LO terms in $\wt{G}^{c}$.  Note that the
bracket proportional to $C_F$ does not contain such a logarithm,
although the analogous term of $k^{t}$ contains one.  This term is
connected with the running of the corresponding quark mass.  This mass
is small in $\wt{G}^c$ but large in $\wt{G}^t$.  The non-logarithmic
part of \eq{h-1-c} again belongs to the NLO.
\end{description}

To gain an impression of  the relevance of a calculation beyond
the LO, we further look at the numerical sizes of the $S$- and
$k$-functions. For typical values of the input parameters we obtain the
numbers summarized in \tab{tab:sizes}.
\begin{ntable}
\begin{center}
\begin{tabular}{l|@{\extracolsep{1em}}*{2}{d{6}}@{\extracolsep{1em}}*{3}{d{6}}}
& \multicolumn{2}{c}{$S^{j}$} &
\multicolumn{3}{c}
	{$\displaystyle \frac{\as \left(m_c\right)}{4\pi} k^{j}$} \\
\cline{2-3} \cline{4-6} \\
$j$ & \ln^1 & \ln^0 & \ln^2 & \ln^1 & \ln^0 \\
\hline
$ct$ & 0.002176 & 0.000311 & -0.002766 & 0.000203 & -0.000235 \\
$c$ & 0 & 0.000264 & 0 & -0.000192 & 0.000029 \\
$t$ & 0 & 6.18 & 0 & -9.21 & -3.16
\end{tabular}
\end{center}
\caption{\slshape
Numerical values of different contributions to the full SM amplitude.
The listed values correspond to the case that $\as$ and $m_t$ are
evolved down to $\mu=m_c$ without integrating out heavy degrees of
freedom.  We have split the $S$- and $k$-functions according to the
power of $\ln x_c$ involved. We use $\as^{[f=6]}(m_c)=0.277$ and
$m_t^{[f=6]}(m_c)=309 \gev$ corresponding to $f=6$ active flavours
and typical values for $\as(M_Z)$ and $m_t(m_t)$.  The $\as^1 \ln^n
x_c$-parts are of the same size as the $\as^0 \ln^{n-1}x_c$-terms
illustrating the need to sum $\as \ln x_c$ to all orders in
perturbation theory as described in sect.~\ref{sect:abovec}. }
\label{tab:sizes}
\end{ntable}
Generally the $\as^1 \ln^{n+1}$, $n=0,1$, terms are linked to the
$\as^0 \ln^n$ terms via the RG equation, cf.\ \tab{tab:rglogs}.  In
\tab{tab:sizes} one observes that the size of the $\as^1 \ln^{n+1}$
contribution is about as large as the $\as^0 \ln^n$ one.  This
emphasizes the need for the summation of large logarithms to all
orders by means of an operator product expansion (OPE) and RG
techniques. The thereby improved result will contain the coefficients
of the logarithms listed in \tab{tab:sizes} evaluated for
$\mu=\mu_{tW}=O(m_t,M_W)$. For example one finds for 
$\mu=m_t (m_t)= 167 \gev$: 
\begin{subequations}
\begin{eqnarray}
\frac{1}{x_c} S(x_c,x_t) &=& - \ln x_c + 0.59
\\
\frac{1}{x_c} \frac{\as (\mu)}{4 \pi }k^{ct} (x_c,x_t,\mu) &=&   
0.026 \ln^2 x_c - 0.22 \ln x_c - 0.27 .
\end{eqnarray}
\end{subequations}
The large magnitudes of the NLO coefficients 0.59 and $-0.22$ compared
to the LO terms further emphasize the importance of the NLO
calculation.  Finally the constant term $-0.27$ enters the initial
condition of the NNLO calculation. It amounts to 46\% of the
corresponding NLO term 0.59. The discussed initial condition, however,
has a much smaller impact on the complete NLO result for $\eta_3$ than
the operator mixing worked out in the following section.

%

\section{Effective $\mathbf{\dstwo}$ Transitions above the
         Charm Threshold}\label{sect:abovec}

In this section we will sum the large logarithm $\ln x_c$ found in
(\ref{sxcxt}) and (\ref{h-1-all}) to all orders in perturbation
theory.  This is done in two steps: First one sets up an effective
lagrangian $\mathcal{L}_{\mathrm{eff}}$ in which the W-boson and the
top quark are removed as dynamic degrees of freedom. In
$\mathcal{L}_{\mathrm{eff}}$ the \dsone\ and \dstwo\ transitions are
described by local four-quark operators, which are multiplied by
Wilson coefficients.  The logarithm $\ln x_c$ is thereby split as $\ln
x_c= \ln (\mu^2/M_W^2) + \ln (m_c^2/\mu^2)$.  Here the former term
resides in the Wilson coefficients, which are functions of $m_t$,
$M_W$ and $\mu$, and the latter is contained in the matrix elements of
the four-quark operators depending only on $\mu$ and the light mass
parameters.  The second step is the application of the RG to the
Wilson coefficients. For $\mu=\mu_{tW}=O (M_W,m_t)$ there is no large
logarithm in the Wilson coefficients. The RG evolution from $\mu_{tW}$
down to $\mu_c=O(m_c)$ sums $\ln (\mu_c^2/\mu_{tW}^2)$ to all orders
in perturbation theory.  The RG improved coefficients finally multiply
matrix elements which do not contain large logarithms, because $\ln
(m_c^2/\mu_c^2)$ is small.

When passing with $\mu$ below $\mu_c$ we must also integrate out the
c-quark field. This will be  described in \sect{sect:belowc}.

\subsection{General Structure of the Effective Lagrangian}\label{sect:struct}

After integrating out the top quark and the W-boson we are left with
an effective five-flavour theory described by a lagrangian of the
generic form
\begin{eqnarray}
\mathcal{L}_{\mathrm{eff}}^{\dstwo} &=&
-\frac{\gf}{\sqrt{2}} V_{\mathrm{CKM}} \sum_{k} C_{k} Q_k
-\frac{\gf^2}{2} V_{\mathrm{CKM}} \sum_{l} \wt{C}_{l} \wt{Q}_{l}.
\label{lgeneric}
\end{eqnarray}
Here the $V_{\mathrm{CKM}}$ denotes products of CKM elements.  The
$Q_k$, $\wt{Q}_l$ represent local \dsone\ and \dstwo\ operators and
the $C_k$, $\wt{C}_l$ are the corresponding Wilson coefficient
functions with Fermi's constant factored out.  The \dsone\ operators
$Q_k$ are necessary for the proper treatment of \dstwo\ transitions,
because they contribute to the transition amplitude through Green's
functions with two operator insertions. An example is shown in
\fig{fig:double-cc-lo}, which is simply obtained from \fig{box} by
shrinking the W-boson lines to a point.
\begin{nfigure}
\centerline{\epsfxsize=4cm \epsffile{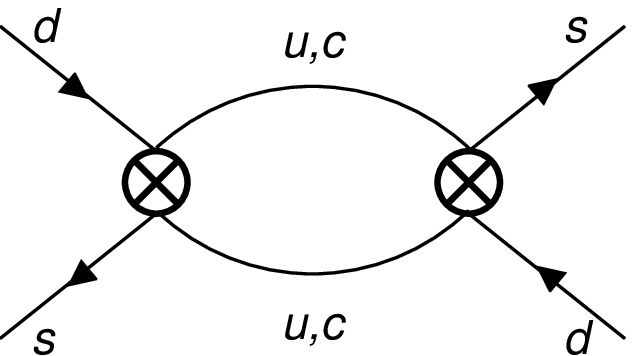}}
\caption{\slshape
Diagram $\mathsf{D_0}$ in the effective five- and four-quark theory.
The crosses denote insertions of local \dsone\ current-current
operators.}
\label{fig:double-cc-lo}
\end{nfigure}
The \dstwo\ operators $\wt{Q}_k$ (see e.g.~\fig{loc}) can likewise
be obtained by shrinking the whole box function with internal top
quarks in \fig{box} to a point. Yet the $\wt{Q}_k$'s are also needed
for the light quark contributions. Diagrams of the type in
\fig{fig:double-cc-lo} are in general divergent and require
counterterms (omitted in (\ref{lgeneric})) proportional to \dstwo\
operators.  In general both \dsone\ and \dstwo\ operators in
(\ref{lgeneric}) contribute to \dstwo\ transitions.  Yet there may be
special cases in which either the former or the latter are absent.  As
we will see later, all three possibilities are realized in $\eta_1$,
$\eta_2$ and $\eta_3$.

In the following sections the detailed structure of
${\mathcal{L}_{\mathrm{eff}}^{\dstwo}}$ will be worked out.
This requires the following steps:
\begin{enumerate}
\renewcommand{\labelenumi}{\roman{enumi})}
\item Find the minimal operator basis to be used in \eq{lgeneric}
sufficient to describe the physics of the \dstwo\ transition.  Here
one must first find a set of operators closing under
renormalization. Subsequently one can eliminate a set of unphysical
operators.
\item Match the full SM Green's function $\widetilde{G}$ of
\eq{smbox}, \eq{greenex}, \eq{gall-0} and \eq{factg} to the one
obtained in the effective theory and thereby determine the Wilson
coefficient functions $C_k$ and $\wt{C}_l$ at the initial scale
$\mu=\mu_{tW}=O(M_W,m_t)$.
\item Next prepare for the RG evolution of the Wilson coefficients
from $\mu=\mu_{tW}$ down to the final scale $\mu=\mu_c=O(m_c)$.  For
this one must derive the general RG equation for Green's functions
with double insertions (see \fig{fig:double-cc-lo}) and its solution.
The RG equation involves an \emph{anomalous dimension tensor}\/ in
addition to the familiar anomalous dimension matrices.
\item Determine the anomalous dimension tensor in the NLO for the
operator basis at hand. This requires the calculation of two-loop
diagrams.
\end{enumerate}
Finally we discuss the size of the remaining non-summed logarithm $\ln
m_t/M_W$.  We do not sum this logarithm because we simultaneously
integrate out the top quark and the W-boson.

\subsection{The Operator Basis}\label{sect:ops}
At first we restrict the set of operators in (\ref{lgeneric}) to the
lowest contributing dimension, which means dimension six for the
$Q_k$'s. As for the $\wt{Q}_l$'s, we must distinguish whether they
correspond to the SM graphs with internal top quarks or whether they
enter ${\mathcal{L}_{\mathrm{eff}}^{\dstwo}}$ as counterterms in the
light quark sector.  In the former case there is only one physical
\dstwo\ operator $\oll$, introduced in (\ref{ollintro}), with a
dimension-two Wilson coefficient $\cll{t}$ containing all information
on $m_t$ and $M_W$ \cite{bjw}.  The latter operators have the same
dimension as the diagrams they renormalize, which is eight as can be
easily seen from \fig{fig:double-cc-lo}.  Higher dimension operators
correspond to terms suppressed by powers of
$m^2_{\mathrm{light}}/m^2_{\mathrm{heavy}}$, which we already
neglected in the SM amplitude, see \eq{sfkts} and \eq{h-1-all}.

We will now establish the \dsone\ part of the operator basis, i.e.\
the $Q_k$'s in \eq{lgeneric}.  Consider first the SM \dsone\
transition of \fig{fig:ds1-full-lo} with only light quarks $k,l = u,c$
on the external legs.  Contracting the W-boson propagator to a point
yields the diagram of \fig{fig:ds1-cc-lo}, in which the cross denotes
the insertion of the \dsone\ current-current operator
\begin{eqnarray}
Q_2^{kl} &=&
\left(\bar{s} \gamma_{\mu} L k\right) \cdot
\left(\bar{l} \gamma^{\mu} L d\right) \cdot
\openone
\label{defQ2} ,
\end{eqnarray}
which we have already met in \fig{fig:double-cc-lo}.
\begin{nfigure}
\begin{minipage}[t]{\miniwidth}
\centerline{\epsfysize=3truecm \epsffile{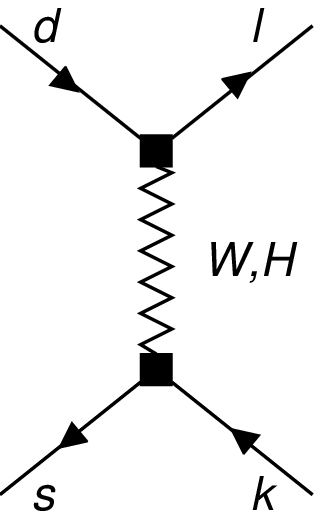}}
\caption{\slshape A \dsone\ transition in the standard model.}
\label{fig:ds1-full-lo}
\end{minipage}
\hfill
\begin{minipage}[t]{\miniwidth}
\centerline{\epsfysize=3truecm \epsffile{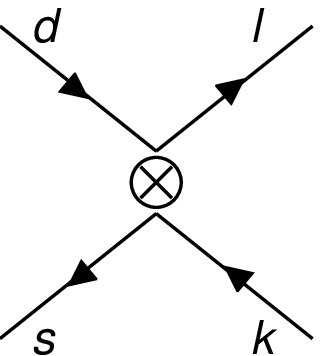}}
\caption{\slshape The matrix element of a \dsone\ current-current
operator $Q_{1,2}^{kl}$ displayed as the cross.}
\label{fig:ds1-cc-lo}
\end{minipage}
\end{nfigure}
It is well-known that QCD corrections to $Q_2^{kl}$ induce
counterterms proportional to other operators, so that $Q_2^{kl}$ mixes
with them under renormalization.

In the case of $k \neq l$ the mixing is particularly simple,
$Q_2^{kl}$ only mixes with
\begin{eqnarray}
Q_1^{kl} &=&
\left(\bar{s} \gamma_{\mu} L k\right) \cdot
\left(\bar{l} \gamma^{\mu} L d\right) \cdot
\topenone .
\label{defQ1}
\end{eqnarray}
The one-loop mixing proceeds  through the diagrams of \fig{fig:ds1-cc-nlo}.
\begin{nfigure}
\centerline{\epsfxsize=\textwidth \epsffile{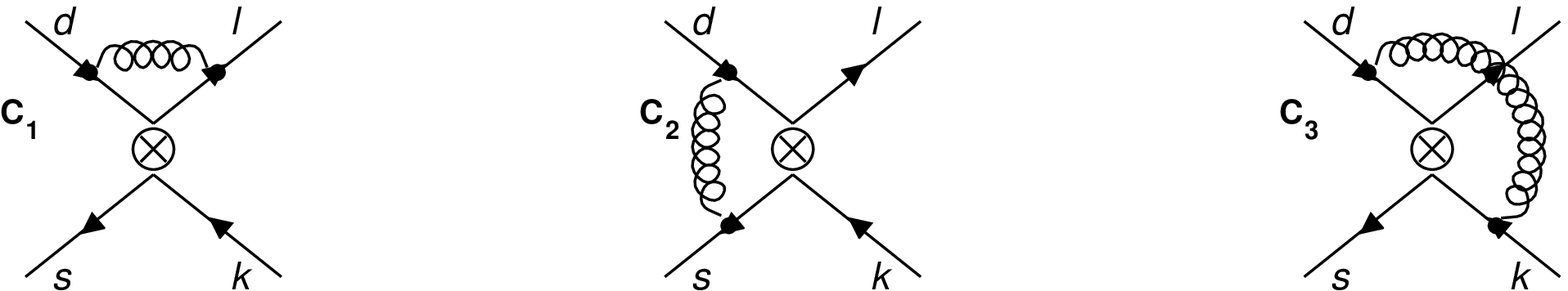}}
\caption{\slshape The classes of diagrams yielding the matrix element
of the \dsone\ current-current operators $Q_{1,2}^{kl}$ as well as the
mixing of the current-current operators among themselves to order
$\as$.}
\label{fig:ds1-cc-nlo}
\end{nfigure}
Hence the corresponding part of $\mathcal{L}_{\mathrm{eff}}^{\dstwo}$
reads
\begin{eqnarray}
- \frac{\gf}{\sqrt{2}}
 \sum_{i,j=1}^2 \sum_{k,l =u,c} V_{ks}^\ast  V_{ld} C_i Z^{-1}_{ij}
Q_j^{kl,\,\bare }.
\end{eqnarray}
Here and in the following the superscript ``bare'' denotes
unrenormalized operators, while renormalized ones do not carry an
additional superscript.  The 2$\times$2 renormalization matrix
$Z^{-1}_{ij}$ is diagonal in the basis
\begin{eqnarray}
Q_{\pm}^{kl} &=& \frac{1}{2} \left(Q_2^{kl} \pm Q_1^{kl}\right),
\label{defQpm}
\end{eqnarray}
provided one preserves Fierz symmetry in the renormalization process
\cite{bw,hn2}.

For $k=l$ additional operators enter the scene, the so-called \dsone\
penguin operators, which appear in different species.  In
\cite{bjlw,bjlw2} the NLO mixing of $Q_{1,2}$ with quark-foot
penguin operators $Q_3$ to $Q_6$ displayed in \fig{fig:ds1-4f-lo}
has been worked out.  They read
\begin{eqnarray}
Q_3 &=& \left(\ov{s} \g_\mu L d\right) \cdot \sum_{q=d,u,s,\ldots}
\left(\ov{q}\g^\mu L q\right) \cdot \1 \nn
Q_4 &=& \left(\ov{s} \g_\mu L d\right) \cdot \sum_{q=d,u,s,\ldots}
\left(\ov{q}\g^\mu L q\right) \cdot \tw \nn
Q_5 &=& \left(\ov{s} \g_\mu L d\right) \cdot \sum_{q=d,u,s,\ldots}
\left(\ov{q}\g^\mu R q\right) \cdot \1 \nn
Q_6 &=& \left(\ov{s} \g_\mu L d\right) \cdot \sum_{q=d,u,s,\ldots}
\left(\ov{q}\g^\mu R q\right) \cdot \tw
\label{pengop}
\end{eqnarray}
and enter (\ref{lgeneric}) with $V_{\mathrm{CKM}}=-\lambda_t$.
The summation runs over all active flavours, at present $q=d,u,s,c$
and $b$.

Yet we may doubt, whether $Q_1-Q_6$ are sufficient to describe the
\dsone\ substructure in \dstwo\ transition. Indeed, $Q_1-Q_6$ mix via
the diagrams of \fig{fig:ds1-cc-p-lo} into operators containing only
two quark lines such as the gluon-foot penguin operators $Q_{g1}$,
$Q_{g2}$ and $Q_{g3}$ depicted in \fig{fig:ds1-p-lo}.  Likewise loop
diagrams with $Q_{g1}-Q_{g3}$ require counterterms proportional to
$Q_3-Q_6$ (see \fig{fig:ds1-p-4f-lo}) and similarly to a ghost-foot
penguin operator
\begin{eqnarray}
Q_\mr{FP} &=& \ov{s} \g_\mu L T^a d \cdot
        \lt( \partial^\mu \ov{\eta}^b \rt) \eta^c f^{abc}.
\label{qfp}
\end{eqnarray}
Here $\eta_c$ denotes a Faddeev-Popov ghost field.  We can easily
construct a \dstwo\ diagram with one of these operators and $Q_2$, see
e.g.\ \fig{fig:ds2-g2-cc-lo}.
\begin{nfigure}
\begin{minipage}[t]{\miniwidth}
\centerline{\epsfysize=3truecm \epsffile{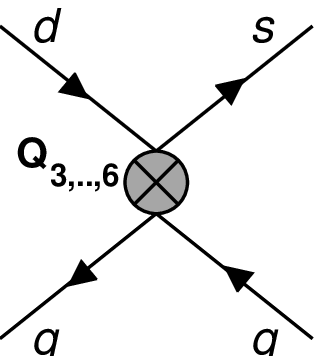}}
\caption{\slshape The four-quark penguin operators $Q_{3,\ldots,6}$}
\label{fig:ds1-4f-lo}
\end{minipage}
\hfill
\begin{minipage}[t]{\miniwidth}
\centerline{\epsfysize=3truecm \epsffile{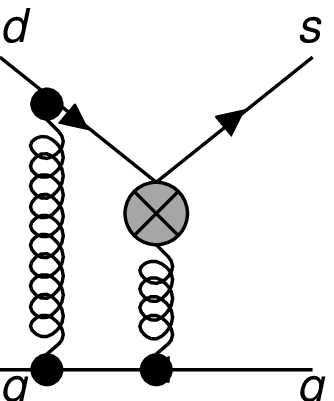}}
\caption{\slshape Diagrams for the mixing of $Q_{g1}$ denoted 
by the shaded cross into four-quark
penguin operators.  }
\label{fig:ds1-p-4f-lo}
\end{minipage}
\end{nfigure}
\begin{nfigure}
\centerline{\epsfxsize=\textwidth \epsffile{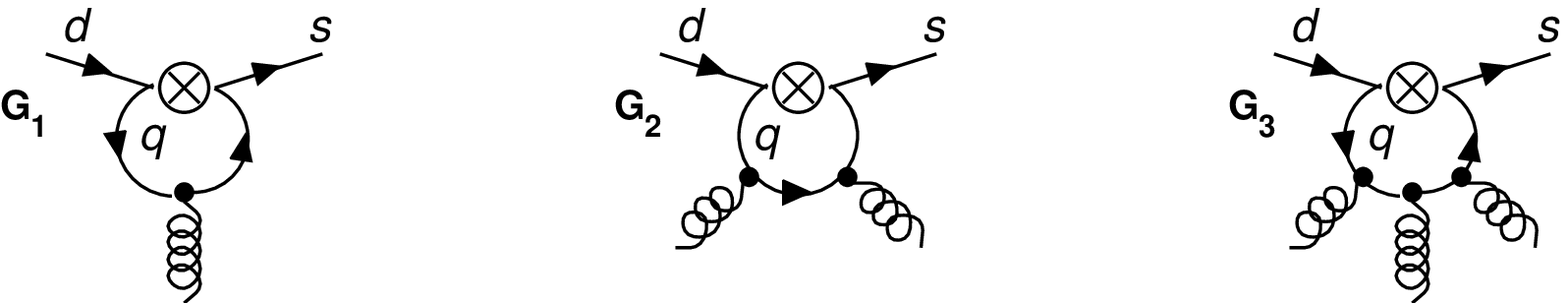}}
\caption{\slshape The mixing of a \dsone\ current-current operator,
denoted by the cross, into gluon-penguin operators.  Diagrams with the
gluon lines permuted compared to $\mathsf{G_2}$ and $\mathsf{G_3}$
must be included, too.}
\label{fig:ds1-cc-p-lo}
\end{nfigure}
\begin{nfigure}
\centerline{\epsfxsize=\textwidth \epsffile{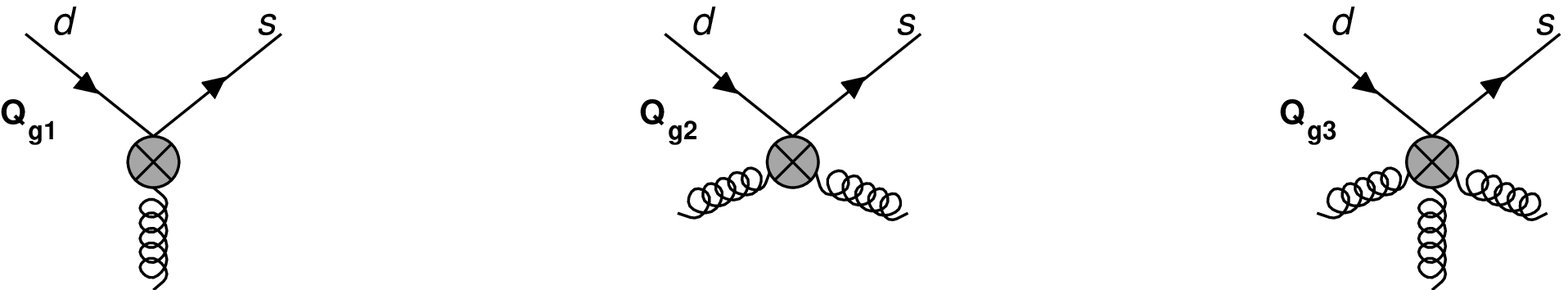}}
\caption{\slshape The \dsone\ gluon-foot penguin operators}
\label{fig:ds1-p-lo}
\end{nfigure}
\begin{nfigure}
\centerline{\epsfxsize=5cm \epsffile{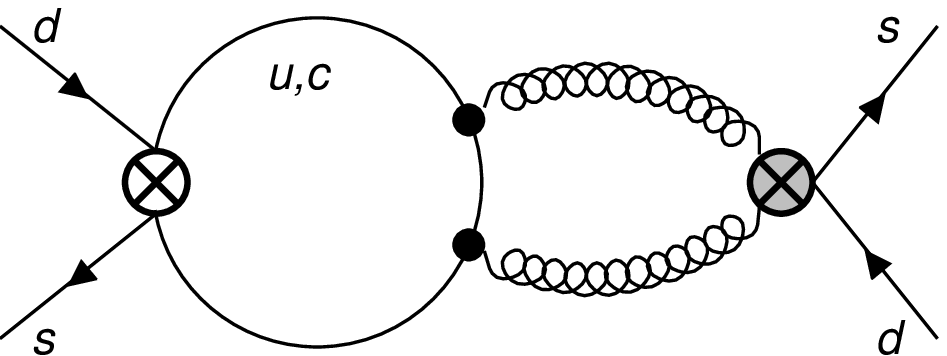}}
\caption{\slshape \dstwo\ diagram with one insertion of $Q_{1,2}$,
denoted by the light cross, and $Q_{g2}$, denoted by the shaded cross,
which could contribute to $\eta_3$.}
\label{fig:ds2-g2-cc-lo}
\end{nfigure}

Fortunately these extra operators $Q_{g1}-Q_{g3}$, $Q_\mr{FP}$ combine
to unphysical operators, which can be dropped from the renormalized
effective lagrangian \cite{jl,pol,colleom,arz,sim}.  Hence the \dsone\
operators in (\ref{lgeneric}) can be restricted to $Q_1^{kl}, \ldots,
Q_6$ and the \dsone\ part of $\mathcal{L}_{\mathrm{eff}}^{\dstwo}$
reads
\begin{eqnarray}
\leso &=& - \frac{\gf}{\sqrt{2}}
\lt[  \sum_{i=1}^2
\sum_{k,l =u,c} V_{ks}^\ast  V_{ld}  C_i Q_i^{kl} -
 \lambda_t \sum_{j=3}^6 C_j Q_j \rt].
\label{lags1}
\end{eqnarray}
We will illustrate the irrelevance of $Q_{g1}-Q_{g3},Q_\mr{FP}$ in the
following section. This, however, first requires the determination of
the local \dstwo\ operators $\wt{Q}_l$ in (\ref{lgeneric}).

Consider first the simplest case, which is realized in $\wt{G}^{t}$:
the SM result for $\wt{G}^{t}$ is proportional to the square of a
heavy mass $m_t$, $M_W$ times a function of their ratios (see \eq{gj},
{\eq{gall-1}} with $j=t$).  A diagram in the effective five-flavour
theory containing two insertions of {\dsone}\ operators can only
produce a result involving light mass parameters, which we already
neglected in \eq{gj}, \eq{gall-1}.  Therefore the diagrams with double
insertions do not contribute in this case and we are left with a
\dstwo\ contribution to $\mathcal{L}_{\mathrm{eff}}^{\dstwo}$ of the
form
\begin{eqnarray}
-\frac{G_F^2}{2} \lambda_t^2 \cll{t} \oll .
\label{clldef}
\end{eqnarray}
$\cll{t}$ has been obtained in the NLO in \cite{bjw}.

In the light quark sector the situation is a bit more involved:
Consider first the diagram of \fig{fig:double-cc-lo} with two internal
charm quarks and zero external momenta.  The only physical
dimension-eight \dstwo\ operators required to absorb its divergence
reads
\begin{eqnarray}
\oloc &=& \frac{\mc}{g^2 \mu^{2 \eps}} \oll
= \frac{\mc}{g^2 \mu^{2 \eps}} \cdot
     \ov{s} \g_\mu L d \cdot \ov{s} \g^\mu L d ,
\label{defqseven}
\end{eqnarray}
which follows from power counting and the absence of any non-zero mass
parameter apart from $m_c$.  The inverse powers of $g$ are introduced
for later convenience as in {\cite{gw}}. One may arbitrarily shift
such factors from the Wilson coefficient into the definition of the
operator. The factor $\mu^{- 2\eps}$ stems from $g_{\mr{bare}}= Z_g g
\mu^{\eps}$ and the fact that
\begin{eqnarray}
\oloc^\ba &=& \frac{ m_{c,\,\ba }^{2} }{g_{\ba }^{ 2}} \lt[
     \ov{s} \g_\mu L d \cdot \ov{s} \g^\mu L d \rt]^{\ba} .
\label{q7bare}
\end{eqnarray}
must be independent of $\mu$\footnote{Here and in the following the
quark fields like $\ov{s}$ and $d$ in (\ref{defqseven}) and
(\ref{q7bare}) are bare fields.  The wave function renormalization
constant $Z_{\psi}$ is taken into account when calculating matrix
elements}. Any other dimension-eight \dstwo\ operator contains one or
two powers of $m_c$ less than $\oloc$ and derivatives and/or gluon
fields instead.  Their on-shell matrix elements are suppressed by
powers of $m_s/m_c$ with respect to those of $\oloc$, so that they do
not contribute to the coefficient of the leading dimension-six
operator \emph{below} the charm threshold (cf.\ (\ref{s2})).  Likewise
they cannot mix with $\oloc$ under renormalization.

Next we determine the CKM factor multiplying $\cloc \oloc$: Generally
we could expect a term proportional to $\lambda_c^2$ and one
proportional to $\lambda_c \lambda_t$ corresponding to $\wt{G}^{c}$
and $\wt{G}^{ct}$ respectively.  Due to the special structure of
$\wt{G}^{c}$, which does not contain a large logarithm to order
$\as^0$ in the SM amplitude \eq{ilgim}, no term proportional to
$\lambda_c^2 \oloc$ occurs.

Hence $\mathcal{L}_{\mathrm{eff}}^{\dstwo}$ is found as
\begin{eqnarray}
\leo &=&
- \frac{\gf}{\sqrt{2}} \sum_{i=1}^6  C_i
\lt[  \sum_{j=1}^2  Z^{-1}_{ij}
\sum_{k,l =u,c} V_{ks}^\ast  V_{ld}   Q_j^{kl, \, \bare} -
 \lambda_t \sum_{j=3}^6 Z^{-1}_{ij} Q_j^{\bare}  \rt] \nonumber \\
&& - \frac{\gft}{16\pi^2} \lambda_t^2 \cll{t} \zll^{-1} \oll^\bare
   \nonumber \\
&& - \frac{\gft}{2} \lambda_c \lambda_t
    \lt[ \sum_{k=1}^{2} \sum_{l=1}^{6}
   C_k C_l \zbi{kl}^{-1} + \cloc \zloc^{-1} \rt]
   \oloc^{\bare}
   \nonumber \\
&& +  \textrm{counterterms proportional to unphysical operators} .
\label{lags2}
\end{eqnarray}
Here the first line is the pure {\dsone}\ part \eq{lags1} written in
terms of bare operators and the corresponding renormalization
factors. The second line contains everything related to a single
insertion of the local \dstwo\ operator $\oll$.  The third line
consists of two parts: the counterterm involving $\wt{Z}^{-1}_{kl,7}$
renormalizes the matrix elements of the type in
{\fig{fig:double-cc-lo}} or {\fig{fig:double-peng-lo}} with one
insertion of $Q_l$ and one of $Q_k$.  The part of {\eq{lags2}}
involving $\zloc^{-1}$ expresses the renormalization of the local
{\dstwo}\ operator $\oloc$ in single insertions to cancel the
divergences of the diagrams depicted in \fig{fig:ds2-1}.  The absence
of a local \dstwo\ operator proportional to $\lambda_c^2$ in
(\ref{lags2}) is due to the GIM mechanism (see \cite{gw,hn1}).  The
structure of \eq{lags2} becomes much more transparent, if one regards
the Z-factors as renormalization factors of the effective couplings
$C_l$, $\cloc$ rather than of the operators.  The bracket in the third
line of \eq{lags2} may then simply be interpreted as the ``bare''
Wilson coefficient of $\oloc$ in analogy to the bare coupling in QCD.
It is renormalized both by QCD and by effective \dsone\ interactions.
The last line of \eq{lags2} contains counterterms proportional to
unphysical operators, which are described in the next section.  One
automatically includes them in the calculation, if one simply
subtracts the subdivergences diagram-by-diagram in the effective
theory.

\subsubsection{Unphysical Operators}\label{sect:unphysical}

We will discuss three types of unphysical operators: \emph{BRS-exact}
operators, those which vanish by a field {\emph{equation of motion}}
(EOM) and {\emph{evanescent}} operators.  Therefore let class
$\mathcal{B}$ contain all BRS-exact operators, class $\mathcal{E}$ the
operators vanishing by a field equation of motion and class
$\mathcal{P}$ the physical and evanescent operators.

Let us first reduce the operator basis with respect to classes
$\mathcal{B}$ and $\mathcal{E}$.  The techniques to do this have been
worked out in {\cite{jl,pol,colleom,arz,sim,col}}.  They are widely
used for the treatment of Green's functions with single operator
insertions. Here the new issue is the application of the theorems
concerning double insertions \cite{colleom,sim} in a concrete
calculation.

The mixing of the classes introduced above follows the general pattern
\begin{eqnarray}
Z^{-1} = \left(
\begin{array}{ccc}
* & * & * \\
0 & * & * \\
0 & 0 & *
\end{array}
\right)
\hspace{3em}
\mbox{corresponding to class}
\hspace{3em}
\left(
\begin{array}{c}
\mathcal{P} \\ \mathcal{B} \\ \mathcal{E}
\end{array}
\right).
\label{unphMix}
\end{eqnarray}
The block-triangular form of \eq{unphMix} ensures that the Wilson
coefficients of the operators from classes $\mathcal{B}$ and
$\mathcal{E}$ do not mix into the ones of physical and evanescent
operators in $\mathcal{P}$.

Further we know that the on-shell matrix elements of operators in
class $\mathcal{B}$ or $\mathcal{E}$ vanish.  This is important for
the matching of transition amplitudes in different theories, e.g.\ of
the full SM and of an effective five-flavour theory.  The vanishing of
the on-shell matrix elements of operators in $\mathcal{B}$ and
$\mathcal{E}$ ensures that they do not contribute to physical Wilson
coefficients at the matching scale.

Since operators from $\mathcal{B}$ and $\mathcal{E}$ contribute
neither to the matching nor to the mixing of class $\mathcal{P}$, we
may neglect them in the discussion of the RG equation and evolution.

Let us now organize the penguin zoo according to the classes
$\mathcal{P}$, $\mathcal{B}$ and $\mathcal{E}$. Due to the theorems of
\cite{jl,pol,colleom,arz,sim,col} the operators
$Q_{g1}-Q_{g3},Q_\mr{FP}$ can be arranged to appear in the
combinations
\begin{eqnarray}
Q_\mathrm{BRS} = Q_\mathrm{FP} + Q_\mathrm{gf} \quad \in {\cal B}
\label{qbrs}
\end{eqnarray}
and
\begin{eqnarray}
\Qeom &=& Q_\mathrm{g} - \frac{1}{4} \left(Q_4+Q_6\right) +
\frac{1}{4N} \left(Q_3+Q_5\right) - Q_\mathrm{BRS} \quad \in {\cal E}
\label{qeom}
\end{eqnarray}
with
\begin{eqnarray}
Q_\mathrm{g} =
\frac{1}{g} \bar{s} \gamma_{\mu} L T^{a} d \cdot D_{\nu}^a F^{\mu\nu},
\qquad
Q_\mathrm{gf} \; = \;  \frac{1}{g} \frac{1}{\xi}
\bar{s} \gamma_{\mu} L T^{a} d
\cdot
\partial^{\mu} \partial^{\nu} A^{a}_{\nu}.
\label{qgf}
\end{eqnarray}
The latter is stemming from the gauge fixing part of the QCD
Lagrangian. $\Qeom$  vanishes by the equation of
motion of the gluon field. $Q_\mathrm{BRS}$ and $\Qeom$
are discussed in detail in  \cite{sim}.

Since now $Q_{g1}-Q_{g3},Q_\mr{FP}$ have been traded for
linear combination belonging to classes ${\cal E}$ and ${\cal B}$,
one may drop them from the  renormalized  operator basis
when calculating \dsone\ transitions.


In the case of double insertions the situation is more complex,
because now the operators of class $\mathcal{E}$ may give a nonzero
contribution in on-shell matrix elements and therefore their presence
cannot be ignored for the matching.  Yet it is possible to absorb the
effects of these operators into the coefficient of a {\dstwo}\
operator \cite{col,sim}.  Such non-vanishing matrix elements with two
insertions of $\Qeom$ appear in $\wt{G}^{t}$ of \eq{smbox}.  Since
the effective five- or four-flavour theory does not contain $m_t$ and
$M_W$ anymore, this contribution is suppressed by a factor of
$m^2_{\mathrm{light}}/m^2_{\mathrm{heavy}}$ compared to the term in
the second line of (\ref{lags2}) and can therefore be neglected.
Yet in the calculation of $\wt{G}^{ct}$ we will face the operator
\begin{eqnarray}
\Qfeom &=& \mc \, \ov{s} R \slash D d , \label{defqeom}
\end{eqnarray}
which vanishes by the quark equation of motion.  Its effect on \dstwo\
amplitudes can likewise be absorbed into $\cloc$.

On-shell matrix elements involving one or two operators from class
$\mathcal{B}$ still vanish, we can therefore drop them in the case of
double insertions, too.

Another important class of unphysical operators are the evanescent
operators also contained in $\mathcal{P}$, which generally appear
in theories with four-fermion interactions, if one uses dimensional
regularization.  As an example one may look at the $O(\as)$
corrections to the matrix element of $Q_{1,2}$, displayed in
{\fig{fig:ds1-cc-nlo}}.  When calculating diagram $\mathsf{C_2}$,
$\mathsf{C_3}$ one faces the structure
\begin{eqnarray}
Q^\prime &\! = \!  &
 \g_\rho \g_\sigma
\g_{\mu } \lef \otimes \g^{\mu } \g^\sigma   \g^\rho \lef  =
( 4- a \varepsilon ) Q + E_1[Q]
 +O(\varepsilon^2),
\label{DefE1}
\end{eqnarray}
where $Q=\gamma_\mu L \otimes \gamma^\mu L$. $E_1[Q]$ is evanescent,
i.e.\ it vanishes for $D=4$. $a$ is an arbitrary real parameter, its
choice belongs to the definition \eq{DefE1} of $E_1[Q]$.

When perturbative results are improved by means of the OPE and RG
techniques, subtleties arise: Evanescent operators can affect the
matching procedure \cite{bw} and the operator mixing {\cite{col,dg}}.
In {\cite{col,bw}} a finite renormalization of the evanescent
operators has been proposed to render their matrix elements zero.
Doing so the Wilson coefficients of the evanescent operators become
irrelevant at the matching scale.  If this should hold at any other
scale, one has to ensure that the Wilson coefficients of the
evanescent operators do not mix into the ones of the physical
operators.  This has been proven in \cite{dg} for a very special and
calculationally inconvenient definition of the evanescent operators.
In \cite{hn2} we achieved the following improvements:
\begin{enumerate}
\renewcommand{\labelenumi}{\roman{enumi})}
\item We have generalized the proof of \cite{dg} to an arbitrary definition
of the evanescent operators, which includes the one chosen in
{\cite{bw}}.
\item We have shown that the arbitrariness in the definition of the
evanescent operators displayed in (\ref{DefE1}) introduces a scheme
dependence into the \emph{physical} Wilson coefficients at the
matching scale as well as into the \emph{physical} anomalous dimension
matrix starting in the NLO.  This distinguishes the evanescent
operators from the operators in classes ${\cal E}$ and ${\cal B}$.  Of
course this scheme dependence cancels in the product of Wilson
coefficients and matrix elements at any scale.  We give explicit
formulae to transform Wilson coefficients or anomalous dimension
matrices from one scheme to another.  These formulae are particularly
necessary if one wants to combine Wilson coefficients and anomalous
dimension matrices calculated with different definitions of the
evanescent operators.
\item We have extended the findings from the case of single insertions to
double insertions, which is needed for this work.
\end{enumerate}
It is important to note that the first and third point above enables
us to use the results of \cite{bw,bjlw,bjlw2} for the \dsone\
hamiltonian in {\eq{lags1}}.

The physical operators in $\mathcal{P}$ needed in our calculation are
\begin{eqnarray}
\hspace{-5ex} &&
Q_{i}^{kl}, \quad i=1,2, \quad k,l=u,c; 
\qquad \qquad Q_{j}, \quad j=3, \ldots  6; 
\qquad \qquad \oloc .
\label{OpBasisS1}
\end{eqnarray}
For the evanescent operators appearing in the calculation
we use the  definition:
\begin{subequations}
\label{StdEvas}
\begin{eqnarray}
\hspace{-2em}
E_1[Q_j] &\! = \! & \left[
\gamma_\mu \gamma_\nu \gamma_\eta L \otimes \gamma^\eta \gamma^\nu \gamma^\mu L
- \left(4+a_1\eps\right) \gamma_\mu L \otimes \gamma^\mu L
\right] K_{1j} , \hspace{1em} j=1,\ldots 4,
\label{StdEvas1-12}\\
\hspace{-2em}
E_1[Q_j] &\! = \! & \left[
\gamma_\mu \gamma_\nu \gamma_\eta R \otimes \gamma^\eta \gamma^\nu \gamma^\mu L
- \left(16+a_2\eps\right) \gamma_\mu R \otimes \gamma^\mu L
\right] K_{1j} , \quad j=5,6 ,
\label{StdEvas1-56}\\
\hspace{-2em}
E_1[\oloc] &\! = \! & \frac{\mc}{g^2} \left[
\gamma_\mu \gamma_\nu \gamma_\eta L \otimes \gamma^\eta \gamma^\nu \gamma^\mu L
- \left(4+\tilde{a}_1\eps\right) \gamma_\mu L \otimes \gamma^\mu L
\right] K_{12} ,
\label{StdEvas1-loc}\\
\hspace{-2em}
E_2[\oloc] &\! = \! & \frac{\mc}{g^2} \left[
\gamma_\mu \gamma_\nu \gamma_\eta \gamma_\sigma \gamma_\tau L \otimes
\gamma^\tau \gamma^\sigma \gamma^\eta \gamma^\nu \gamma^\mu L \right.
\no \\
& & \hspace{2em} \left.
- \left[\left(4+\tilde{a}_1\eps\right)^2+\tilde{b}_1 \eps \right]
\gamma_\mu L \otimes \gamma^\mu L \right] K_{22}.
\label{StdEvas2-loc}
\end{eqnarray}
\end{subequations}
Here $E_1[Q_j]$ is the evanescent operator needed as a counterterm to
render the diagrams $\mathsf{C_2}$ and $\mathsf{C_3}$ in
{\fig{fig:ds1-cc-nlo}} with $Q_j$ inserted finite.  The colour factors
$K_{1j}$ are
\begin{eqnarray}
K_{12}=K_{13}=K_{15}&=&\frac{1}{2} \topenone - \frac{1}{2 N} \openone,
\no \\
K_{11}=K_{14}=K_{16}&=&\frac{1}{4} \openone + \frac{N^2-2}{4 N}
\topenone.
\label{DefColourEva}
\end{eqnarray}
Likewise $E_2[\oloc]$ appears in two-loop diagrams involving physical
operators or in one-loop matrix elements of $E_1[\oloc]$.

In our NLO calculation we have kept $a_1$, $a_2$, $\tilde{a}_1$ and
$\tilde{b}_1$ arbitrary for two reasons: First we want to illustrate
our findings of \cite{hn2} in \sect{sect:evas}.  Second the vanishing
of these quantities from physical results provides a non-trivial check
of our calculation.

Apart from \sect{sect:evas} we will always state the results
corresponding to
\begin{eqnarray}
a_1 = -8, \hspace{3em}
a_2 = -16, \hspace{3em}
\tilde{a}_1 = -8,
\label{StdEvasNum}
\end{eqnarray}
in order to comply with the standard choice used in
\cite{bw,hn1,bjw,bjlw,bjlw2}.  Since NLO anomalous dimensions
and matching corrections of physical operators do not depend
on $\tilde{b}_1$, we do not give a numerical value.  Likewise
we do not need the value of the colour factor $K_{22}$.

Let us finally look at the operator $\oloc$ or equivalently at $\oll$
appearing in (\ref{lags2}) and (\ref{s2}) to introduce a different
type of evanescent operator: The Dirac and flavour structure of $\oll=
(\ov{s}d)_{V-A} (\ov{s}d)_{V-A} \openone$ is Fierz self-conjugate in
four dimensions.  Hence $\oll$ differs from its Fierz transform
$(\ov{s} d)_{V-A} (\ov{s}d)_{V-A} \topenone$ by an evanescent
operator.  In general we must therefore expect the NLO anomalous
dimension to be different for $\oll$ and its Fierz transform.  Yet the
standard definition of $E_1[\oll]$ with $\wt{a}_1=-8$ ensures Fierz
symmetry to hold at the loop level as well (see \cite{bw,hn2}).  Hence
with the choice $\wt{a}_1=-8$ it does not matter whether one uses
$\oll$ or its Fierz transform or any linear combination of them in the
calculation.  This is especially gratifying for the hamiltonian
(\ref{s2}) below the charm threshold, if the non-perturbative methods
used to obtain the matrix element of $\oll$ do not distinguish between
$\oll$ and its Fierz transform.

\subsubsection{Green's Functions from 
               $\mathbf{\Lagr_{\eff}^{\dstwo}}$}\label{sect:green}

The following sections will deal with the determination of the Wilson
coefficients and renormalization $Z$-factors present in {\eq{lags2}}.
To do so we need to know the Green's function obtained from
$\Lagr_{\eff}^{\dstwo}$ to order $\gft$.  It reads:\footnote{The RHS
is looking like the Green's function of a \dstwo\ hamiltonian, the
notation is a little bit sloppy.}
\begin{eqnarray}
\!\!
\left\langle \T
\exp\left[i \int d^D x \Lagr^{\dstwo}_{\eff}\left(x\right)\right]
\right\rangle_{\dstwo} &=&
-i \left\langle H^{c}+H^{t}+H^{ct}\right\rangle
+ O\left( G_F^3 \right) ,
\label{GreenMix}
\end{eqnarray}
where
\begin{subequations}
\label{GreenMixH}
\begin{eqnarray}
H^{c}\left(x\right) &=& \lambda_c^2 \frac{\gft}{2}
\sum_{i,i',j,j'=+,-} C_i C_j
\underbrace{Z^{-1}_{ii'} Z^{-1}_{jj'}
	\mathcal{O}^{\bare}_{i'j'}\left(x\right)}
	_{\displaystyle \equiv \mathcal{O}_{ij}\left(x\right)},
\label{GreenMixHc}
\\
H^{t}\left(x\right) &=& \lambda_t^2 \frac{\gft}{16\pi^2} \cll{t}
\zll^{-1} \oll^{\bare}\left(x\right),
\label{GreenMixHt}
\\
H^{ct}\left(x\right) &=& \lambda_c \lambda_t\!\Biggl[\frac{\gft}{2}\!\!
\sum_{i=+,-}\!\sum_{j=1}^{6} C_i C_j\!
\underbrace{\left(
	\sum_{i'=+,-} \sum_{j'=1}^{6}
	Z^{-1}_{ii'} Z^{-1}_{jj'}
	\mathcal{R}^{\bare}_{i'j'}\left(x\right)
	+
	\zbi{ij}^{-1} \oloc^{\bare}\left(x\right)
	\right)}
	_{\displaystyle \equiv \mathcal{R}_{ij}\left(x\right)}
\no \\
& & \hspace{3em}
+ \frac{\gft}{2} \cloc \zloc^{-1} \oloc^{\bare}\left(x\right) \Biggr].
\label{GreenMixHct}
\end{eqnarray}
\end{subequations}
Here $\mathcal{O}_{ij}$ and $\mathcal{R}_{ij}$
denote the bilocal structures composed of two \dsone\ operators
reading
\begin{subequations}
\label{GreenMixDouble}
\begin{eqnarray}
\hspace{-2pt}
\mathcal{O}_{ij}^{\bare}\left(x\right) &=&
\frac{-i}{2} \int d^D y \T \left[
Q_i^{cc,\bare}\left(x\right) Q_j^{cc,\bare}\left(y\right)
-Q_i^{cu,\bare}\left(x\right) Q_j^{uc,\bare}\left(y\right)
\right. \no \\
& &\hspace{6em} \left.
-Q_i^{uc,\bare}\left(x\right) Q_j^{cu,\bare}\left(y\right)
+Q_i^{uu,\bare}\left(x\right) Q_j^{uu,\bare}\left(y\right)
\right],
\no \\
& &
\label{GreenMixDoubleO}
\\
\hspace{-2pt}
\mathcal{R}_{ij}^{\bare}\left(x\right) &=&
\left\{
\begin{array}{l}
\displaystyle
\frac{-i}{2} \int d^D y \T \left[
2Q_i^{uu,\bare}\left(x\right) Q_j^{uu,\bare}\left(y\right)
-Q_i^{uc,\bare}\left(x\right) Q_j^{cu,\bare}\left(y\right)
\right. \\
\displaystyle
\hspace{6em}\left.
-Q_i^{cu,\bare}\left(x\right) Q_j^{uc,\bare}\left(y\right)
\right]
\\
\hfill\mbox{for $j=1,2$,}
\\
\displaystyle
\frac{-i}{2} \int d^D y \T \left[
\left(Q_i^{uu,\bare}-Q_i^{cc,\bare}\right)\left(x\right)
Q_j^{\bare}\left(y\right)
\right. \\
\displaystyle
\hspace{6em}\left.
+ Q_j^{\bare}\left(x\right)
\left(Q_i^{uu,\bare}-Q_i^{cc,\bare}\right)\left(y\right)
\right]
\\
\hfill\mbox{for $j=3,\ldots,6$.}
\end{array}
\right.
\no \\
& &
\label{GreenMixDoubleR}
\end{eqnarray}
\end{subequations}
In (\ref{GreenMixDoubleO}) the term $-Q_i^{cc,\bare} (x)
Q_j^{uu,\bare}(y) -Q_i^{uu,\bare} (x) Q_j^{cc,\bare}(y)$ has been
omitted, because its matrix element does not contribute to the
leading power of $m_c^2/M_W^2$ below the charm threshold. Yet this
term is necessary to annihilate the mixing of the current-current
operators into penguin operators in $H^{c}$. The absence of penguins
in $H^c$ is an effect of the GIM mechanism. In contrast GIM is
broken in $H^{ct}$ due to the large mass of the top quark.  Further
the counterterms proportional to unphysical operators are not
displayed in (\ref{GreenMixDoubleO}) and (\ref{GreenMixDoubleR}).

Here the LO matrix elements of \eq{GreenMixDoubleO} and \eq{GreenMixDoubleR}
with $j=1,2$ correspond to diagrams of the type shown in
{\fig{fig:double-cc-lo}}, \eq{GreenMixDoubleR} with $j=3,\ldots,6$ to
{\fig{fig:double-peng-lo}}.
\begin{nfigure}
\centerline{\epsfxsize=4cm \epsffile{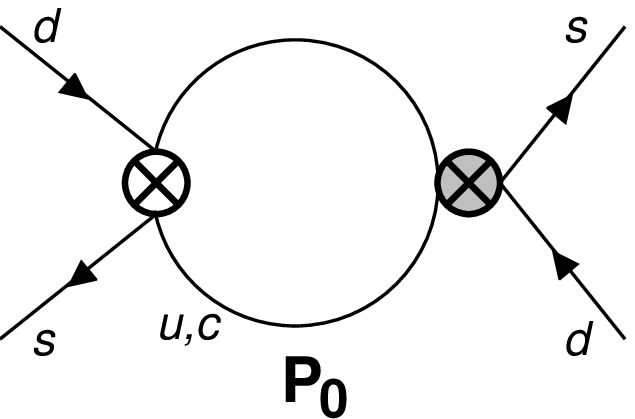}}
\caption{\slshape
Diagram $\mathsf{P_0}$ in the effective five- and four-quark theory.
The white cross denotes the insertion of a local \dsone\
current-current operator, the shaded cross the insertion of a local
four-quark penguin operator.  The penguin operator only contributes
via its up-type-quark foot, i.e.\ through the couplings
$(\ov{s}d)_{V-A} (\ov{c}c)_{V\pm A}$ and $(\ov{s}d)_{V-A}
(\ov{u}u)_{V\pm A}$.}
\label{fig:double-peng-lo}
\end{nfigure}

\subsection{Matching of the Standard Model Amplitudes to the
            Effective Theory}\label{sect:match-tW}

Let us now give the initial conditions for the Wilson coefficient
functions at the scale $\mu_{tW}$, at which the top quark and the
W-boson are integrated out. We start with the initial conditions
needed for the calculation of $\wt{G}^{ct}$. It involves the
coefficients of all \dsone\ operators comprised in $\vec{Q} = (Q_1,
Q_2, {\ldots}, Q_6)^{T}$. In the NLO the initial condition reads
{\cite{bjlw}}:
\begin{eqnarray}
\vec{C}\left(\mu_{tW}\right) &=& \left(
\begin{array}{c}
0 \\ 1 \\ 0 \\ 0 \\ 0 \\ 0
\end{array}
\right)
+ \frac{\as\left(\mu_{tW}\right)}{4\pi} \left[
\ln \frac{\mu_{tW}}{M_W} \left(
\begin{array}{c}
\gamma^{(0)}_{21} \\
\gamma^{(0)}_{22} \\
\gamma^{(0)}_{23} \\
\gamma^{(0)}_{24} \\
\gamma^{(0)}_{25} \\
\gamma^{(0)}_{26}
\end{array}
\right)
+ \left(
\begin{array}{c}
B_1 \\ B_2 \\
-\frac{1}{2N} \tilde{E}\left(x_t\left(\mu_{tW}\right)\right) \\
\frac{1}{2} \tilde{E}\left(x_t\left(\mu_{tW}\right)\right) \\
-\frac{1}{2N} \tilde{E}\left(x_t\left(\mu_{tW}\right)\right) \\
\frac{1}{2} \tilde{E}\left(x_t\left(\mu_{tW}\right)\right)
\end{array}
\right)
\right]
\no \\
& & + O(\as^2).
\label{ds1initial}
\end{eqnarray}
To obtain $\vec{C}\left(\mu_{tW}\right)$ in the LO one simply drops
the $O(\as)$ term.  In \eq{ds1initial}
\begin{subequations}
\label{ds1initials}
\begin{eqnarray}
B_1 &=& \frac{11}{2},
\hspace{3em}
B_2 = - \frac{11}{2N},
\label{ds1initialB} \\
\tilde{E}\left(x\right) &=&
- \frac{2}{3} \ln x
+ \frac{x^2 \left(15-16x+4x^2\right)}{6 \left(1-x\right)^4} \ln x
+ \frac{x \left(18-11x-x^2\right)}{12 \left(1-x\right)^3}
- \frac{2}{3}.
\label{ds1initialE}
\end{eqnarray}
\end{subequations}
$B_1$, $B_2$ and $\wt{E}(x)$ are scheme dependent, the expressions in
(\ref{ds1initialB}) and (\ref{ds1initialE}) are specific to the NDR
scheme and the definition of the evanescent operators given in
{\eq{StdEvasNum}}.  The $\ln (\mu_{tW}/M_W)$ term in {\eq{ds1initial}}
allows for $\mu_{tW} {\neq} M_W$, here the $\gamma^{(0)}_{ij}$ denote
the elements of the anomalous dimension matrix of the \dsone\
operators summarized in {\apx{apx:ds1mix}}.  It is important to note
that $Q_1$, $Q_2$ collectively denote the operators $Q_1^{kl}$,
$Q_2^{kl}$ with different flavour quantum numbers $k,l = u,c$.  Note
that $\vec{C}\left(\mu_{tW}\right)$ in \eq{ds1initial} does not depend
on the number $f$ of active flavours, so there is no difference
whether we match to an effective five-flavour theory or directly to an
effective four-flavour theory.

For the forthcoming solution of the RG equations we will also need the
diagonal basis for the current-current subset of the operator basis,
which we have already introduced in \eq{defQpm}.  The first two rows
of {\eq{ds1initial}} then translate into
\begin{eqnarray}
C_{\pm}\left(\mu_{tW}\right) &=& 1 +
\frac{\as\left(\mu_{tW}\right)}{4\pi} \left[ \ln \frac{\mu_{tW}}{M_W}
\gamma^{(0)}_{\pm} + B_{\pm} \right] + O(\as^2)
\label{ds1diaginit}
\end{eqnarray}
with
\begin{eqnarray}
B_{\pm} = B_2 \pm B_1,
\label{ds1diaginitB}
\end{eqnarray}
and the anomalous dimensions $\gamma^{(0)}_{\pm}$ of the operators
$Q_{\pm}^{kl}$ can be found in \eq{anomQpm}.

We are now in the position to calculate the initial values for the
\dstwo\ Wilson coefficient $\cloc$.  This is done by
comparing the Green's function for the \dstwo\ transition obtained in
the full SM with the same quantity obtained in the effective
five-flavour theory in \eq{GreenMix}.  In the SM expression of
$\wt{G}^{ct}$ to order $\as^0$ (cf.\ \eq{gct} and {\eq{sxcxt}}) there
is a large logarithm.  Therefore the LO matching can be done solely
with the large logarithm, the NLO matching then requires the $\ln^0$
part.  This can be seen from {\tab{tab:rglogs}}, where one simply has
to set $n=0$ to see the term which is used for the matching in a
specific order.  It is therefore sufficient for both the LO and NLO
matching to consider the effective theory only to order $\as^0$.
Since the initial value of the \dsone\ coefficients $C_k$ are of order
$\as$ for $k=1,3,4,5,6$ and $C_k= 1+O(\as)$ for $k=2,+,-$ (cf.\
{\eq{ds1initial} and {\eq{ds1diaginit}}}), the matching at the scale
$\mu=\mu_{tW}$ reads
\begin{eqnarray}
\lefteqn{ 2 i \wt{G}^{ct\, (0)} (\mu_{tW})+ O(x_c^2 \ln x_c)
    \; = \; \frac{1}{\lambda_c \lambda_t}   
            \langle H^{ct} (\mu_{tW}) \rangle ^{(0)} }  \nn
& = &  \frac{\gft}{2} \lt[
      \langle \bilr{+2}  (\mu_{tW}) \rangle ^{(0)}
      +   \langle \bilr{-2} (\mu_{tW})  \rangle ^{(0)}
       + \cloc (\mu_{tW})
         \langle \oloc (\mu_{tW})  \rangle^{(0)} \rt]   .
\label{matchctup}
\end{eqnarray}
Here \eq{GreenMixHct} has been used.
The diagram of {\fig{fig:double-cc-lo}} yields
\begin{eqnarray}
\langle \bilr{+2}(\mu_{tW}) \rangle^{(0)} &=&
\frac{\mc\left(\mu_{tW}\right)}{16 \pi^2}
\left[-8 \ln\frac{m_c}{\mu_{tW}}-2 \right] 
\langle \oll \rangle^{(0)} ,
\label{rmtw}
\end{eqnarray}
while $\langle \bilr{-2} \rangle ^{(0)}=0$.
With \eq{gct} and \eq{sxcxt} one easily finds
\begin{eqnarray}
\cloc(\mu_{tW}) &=& \left\{
\begin{array}{ll}
0 & \mbox{in LO} \\
\displaystyle \frac{\as(\mu_{tW})}{4\pi} \left(
-8\ln\frac{\mu_{tW}}{M_W}+4F\left(x_t\left(\mu_{tW}\right)\right)
+ 2
\right) & \mbox{in NLO}
\end{array}
\right. ,
\label{clocInit}
\end{eqnarray}
where $F(x_t)$ is the top dependent part of $S(x_c,x_t)$ defined in
{\eq{deff}}.  The factor $\as$ originates from the special definition
of $\oloc$ in \eq{defqseven}.  Note how the large logarithm $\ln x_c$
in \eq{sxcxt} is split between the Wilson coefficient $\cloc$ and the
matrix element in \eq{matchctup}.  Again the NLO result in
\eq{rmtw} and \eq{clocInit} is specific to the NDR scheme with
\eq{StdEvasNum}.

Next we discuss the other two flavour structures described by
$\wt{G}^{t}$ and $\wt{G}^{c}$.  For $\wt{G}^{t}$ the situation is
quite different.  Here the SM amplitude immediately has to be matched
to an effective theory containing only $\oll$.  The terms needed for
this matching can again be read off from {\tab{tab:rglogs}} if one
sets $n=0$.  In the LO the $O(\as^0)$ term of the SM amplitude and of
the effective theory matrix element is sufficient, i.e.\ {\fig{box}}
and {\fig{loc}}, while in NLO one needs the $O(\as^1)$ parts, i.e.\
{\fig{boxqcd}} and {\fig{fig:ds2-1}}.  One then easily extracts the
initial condition of the Wilson coefficient $\cll{t}$ in NLO as
\cite{bjw}
\begin{eqnarray}
\cll{t}(\mu_{tW}) &=&
\mw \left[
S\left(x_t\left(\mu_{tW}\right)\right)
+ \frac{\as\left(\mu_{tW}\right)}{4\pi}
	k^{t}\left(x_t\left(\mu_{tW}\right),\mu_{tW}\right)
\right],
\label{cllInit}
\end{eqnarray}
where $S$ and $k^{t}$ have been defined in \eq{sxt} and \eq{h-1-t}
respectively.  The LO expression is simply obtained by dropping the
$\as$ term.  Note that the term in square brackets in \eq{cllInit} 
precisely equals the one in \eq{factgj}.

The simplest case is $\wt{G}^{c}$.  Since there is no large logarithm
in the SM amplitude to order $\as^0$ due to GIM suppression, we expect
the Wilson coefficient of $\oloc$ to vanish.  We can check this
statement explicitly by performing the matching \cite{hn1}.  From
{\tab{tab:rglogs}} one again reads off the terms required for this by
setting $n=0$.  They are the $\as^0$ terms for LO and the $\as^1
\ln^0$ terms for NLO.  One therefore in LO has to calculate the finite
parts of the diagram in {\fig{fig:double-cc-lo}} with both insertions
being $Q_2$'s.  In NLO one has to do the same with the diagrams in
{\fig{fig:double-cc-nlo}}. One immediately obtains the result that the
double insertions fully account for the SM amplitude, there is no room
left for $\cloc \neq 0$ in $\wt{G}^{c}$.
\begin{nfigure}
\centerline{\epsfxsize=\textwidth \epsffile{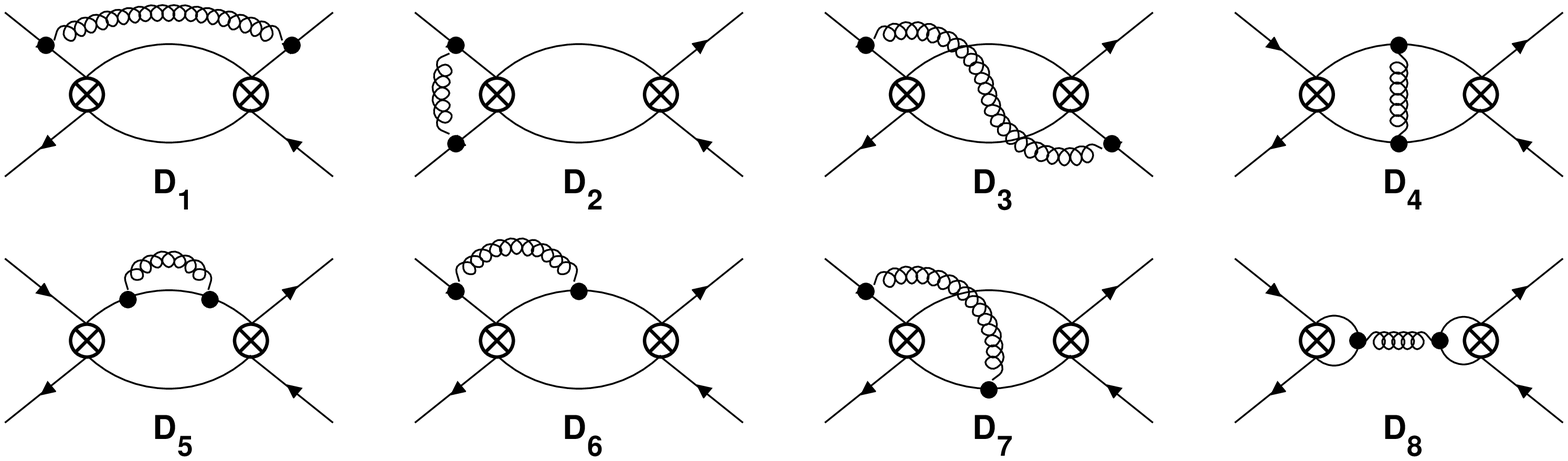}}
\caption{\slshape
The classes of diagrams yielding the $O(\as)$ contribution to
$\mathcal{O}_{ij}$; $i,j=+,-$ \eq{GreenMixDoubleO} and
$\mathcal{R}_{ij}$; $i=+,-$; $j=1,2$, \eq{GreenMixDoubleR} in the
effective five- and four-flavour theory, the remaining diagrams are
obtained by mirror reflections.  The crosses denote insertions of
{\dsone}\ current-current operators.  Additional QCD counterterms have
to be included.  The result for the divergent parts of $\mathsf{D_i}$,
$i=1,\ldots,7$ are summarized in \apx{apx:int} (\eq{div-d} and
{\tab{tab:div-d}}).  Diagram $\mathsf{D_8}$ equals 0 for zero external
momenta.}
\label{fig:double-cc-nlo}
\end{nfigure}

\subsection{Renormalization Group for Double Operator Insertions}
\label{sect:rg}

So far we have determined the Wilson coefficients taking part in the
game at the renormalization scale $\mu_{tW} = O(M_W,m_t)$.  To obtain
them at a scale $\mu < \mu_{tW}$ we need to know the RG equation,
which governs this evolution.  The new feature in the calculation
presented here is the RG equation for Green's functions with two
operator insertions.

The QCD beta function $\beta\left(g\right)$, the anomalous mass
dimension $\gamma_m$ and the anomalous dimensions for the wave
function $\gamma_\psi$ are summarized in {\apx{apx:rgquant}}.

\subsubsection{RG for Single Insertions: 
               A Short Review}\label{sect:rgsingle}
Let us first shortly review the case of single insertions adopting the
notation of {\cite{bjlw}}.  From $\dmu \Lagr^{\dsone} = 0$ in
{\eq{lags1}} one obtains the RG equation
\begin{eqnarray}
\sum_{j=1}^6 \left[ \delta_{jk}\, \dmu - \gamma_{jk} \right] C_j = 0
\label{RGsingle}
\end{eqnarray}
for the Wilson coefficient functions $C_j$, where
\begin{eqnarray}
\gamma_{ij}\left(g\left(\mu\right)\right) &=&
\sum_{k=1}^6  Z^{-1}_{ik} \dmu Z_{kj}
\label{defAnom}
\end{eqnarray}
is the anomalous dimension matrix of the \dsone\ operators $Q_k$.  The
solution of \eq{RGsingle} is given by
\begin{eqnarray}
C_j\left(\mu\right) &=& \sum_{k=1}^6 
\left[U\left(\mu,\mu_0\right)\right]_{jk} C_k\left(\mu_0\right)
\label{RGsolSingle}
\end{eqnarray}
with the evolution matrix
\begin{eqnarray}
U\left(\mu,\mu_0\right) &=&
\Tg \exp\left[\int_{g\left(\mu_0\right)}^{g\left(\mu\right)} dg'
	\frac{\gamma^{T}\left(g'\right)}{\beta\left(g'\right)}
	\right].
\label{EvolSingle}
\end{eqnarray}
Here $\Tg$ means that the matrices $\gamma(g')$, $\gamma(g'')$, \ldots
in the expanded exponential are ordered such that the couplings
increase (decrease) from right to left for $g(\mu_0) < g(\mu)$
($g(\mu_0) > g(\mu)$).

We expand the renormalization matrix $Z^{-1}$ as
\begin{eqnarray}
Z^{-1} &=& 1
+\frac{\as}{4\pi} Z^{-1,(1)}
+\left(\frac{\as}{4\pi}\right)^2 Z^{-1,(2)}
+\ldots,
\no \\
Z^{-1,(n)} &=&
\sum_{r=0}^{n} \frac{1}{\eps^r} Z_r^{-1,(n)}
+O(\eps).
\label{ExpZSingle}
\end{eqnarray}
To deal with the evanescent operators $Z^{-1}$ contains a finite
renormalization term.  From \eq{ExpZSingle} the coefficients of the
perturbative expansion of
\begin{eqnarray}
\gamma &=&
\frac{\as}{4\pi} \gamma^{(0)}
+\left(\frac{\as}{4\pi}\right)^2 \gamma^{(1)}
+\ldots,
\label{ExpAnomSingle}
\end{eqnarray}
are obtained as
\begin{subequations}
\label{AnomSingleCoeff}
\begin{eqnarray}
\gamma^{(0)} &=& 2 Z^{-1,(1)}_1 + 2\eps Z^{-1,(1)}_0 \\
\gamma^{(1)} &=& 4 Z^{-1,(2)}_1 + 2\left\{Z^{-1,(1)}_0,Z^{-1,(1)}_1\right\} +
2 \beta_0 Z^{-1,(1)}_0 + O(\eps).
\end{eqnarray}
\end{subequations}
The LO approximation of \eq{EvolSingle} reads
\begin{eqnarray}
U^{\left(0\right)}\left(\mu,\mu_0\right) &=&
\exp\left[d \cdot \ln \frac{\as\left(\mu_0\right)}{\as\left(\mu\right)}\right]
\hspace{3em}\mbox{with}\hspace{3em}
d = \frac{{\gamma^{(0)}}^{T}}{2 \beta_0}.
\label{EvolSingleLO}
\end{eqnarray}
Here $\as\left(\mu\right)$ is the running QCD coupling constant
defined in \apx{apx:rgquant}, \eq{alphas}.  The NLO expression of
{\eq{EvolSingle}} can be written as
\begin{eqnarray}
U\left(\mu,\mu_0\right) &=&
\left(1+\frac{\as\left(\mu\right)}{4\pi} J\right)
U^{\left(0\right)}\left(\mu,\mu_0\right)
\left(1-\frac{\as\left(\mu_0\right)}{4\pi} J\right),
\label{EvolSingleNLO}
\end{eqnarray}
where $J$ is a solution of the matrix equation (see e.g.\ \cite{bjlw})
\begin{eqnarray}
J + \left[ d, J\right] &=&
-\frac{{\gamma^{\left(1\right)}}^{T}}{2 \beta_0}
+\frac{\beta_1}{\beta_0} d.
\label{DefJ}
\end{eqnarray}
We remark here that it is not necessary to diagonalize (\ref{DefJ}) in
order to solve for $J$. (\ref{DefJ}) simply represents a set of 36
linear equations for the 36 elements of $J$, which are therefore
rational numbers.  $d=d^{[f]}$ and $J=J^{[f]}$ depend on the number
$f$ of active flavours through $\beta_0$, $\beta_1$, $\gamma^{(0)}$
and $\gamma^{(1)}$, which can be found in appendix~\ref{apx:rgquant}.

For an operator like $Q_+,Q_-,\oloc$ or $\oll$ which does not mix
with other operators, the matrices in (\ref{EvolSingleLO}-\ref{DefJ})
reduce to numbers.  In the following we will need
\begin{subequations}
\begin{eqnarray}
\hspace{-2ex}
\uloc^{(0)} (\mu,\mu_0) \; = \;
 \left( \frac{\alpha (\mu_0)}{\alpha (\mu )}\right)^{\wt{d}_{77}},
          &\hspace{-1ex}&\quad  
      \wt{d}_{77}\;=\; \frac{\gloc^{(0)}}{2 \beta_0}, \qquad
\jloc \; = \; - \frac{\gloc^{(1)}}{2 \beta_0}
              + \frac{ \gloc^{(0)}\, \beta_1}{2 \beta_0^2} , 
\label{def77} \\ 
          &\hspace{-1ex}&\quad  
         d_{\pm} \;=\; \frac{\g_{\pm}^{(0)}}{2 \beta_0}, \qquad
J_{\pm} \; = \; - \frac{\g_{\pm}^{(1)}}{2 \beta_0}
              + \frac{\g_{\pm}^{(0)}\, \beta_1}{2 \beta_0^2} .
\label{defpm}
\end{eqnarray}
\end{subequations}
The anomalous dimensions of $Q_{\pm}$, $\oloc$ and $\oll$ are summarized in
appendices~\ref{apx:ds1mix} and \ref{apx:ds2anom}.

During the evolution from $\mu_{tW}$ down to $\mu_c$ we have to pass
the scale $\mu_b$, at which we integrate out the bottom quark.  Since
the penguin operators $Q_{3,\ldots,6}$ explicitly depend on the number
of active flavours, we have to take into account a matching correction
$r=r^{[f]}$ \cite{bjlw}:
\begin{eqnarray}
\left\langle\vec{Q}\right\rangle\left(\mu\right) &=&
\left\langle\vec{Q}\right\rangle^{(0)}
+ \frac{\as}{4\pi} \left\langle\vec{Q}\right\rangle^{(1)}\left(\mu\right)
+ O\left(\as^2\right)
\no \\
&=&
\left[1+\frac{\as}{4\pi}r\left(\mu\right)+O\left(\as^2\right)\right]
\left\langle\vec{Q}\right\rangle^{(0)}.
\label{MatchCorrS1}
\end{eqnarray}
The matching for the Wilson coefficients $\vec{C}$ from the effective
five-flavour theory to the effective four-quark theory therefore reads
\begin{eqnarray}
\vec{C}^{[4]}\left(\mu_b\right) &=&
\left[1
+\frac{\as\left(\mu_b\right)}{4\pi}
\delta r^T \lt( \mu_b \rt)
+O\left(\as^2\right)\right]
\vec{C}^{[5]}\left(\mu_b\right)
\label{MatchWilsonS1}
\end{eqnarray}
with 
\begin{eqnarray}
\delta r \lt( \mu_b \rt) &=& 
      r^{[5]} \left( \mu_b \right) - r^{[4]}\left( \mu_b \right)
\label{defdr} .
\end{eqnarray}
For our \dsone\ operator basis the matrix $r\equiv r\left(m_b\right)$
can be found in {\apx{apx:ds1mix}}, {\eq{ds1r}}.  The required
$r\left(\mu_b\right)$ then reads
\begin{eqnarray}
r^{[f]}\left(\mu_b\right) &=&
r^{[f]}\left(m_b\right) - \ln\frac{\mu_b}{m_b}\gamma^{(0),[f]}.
\label{ds1rmu}
\end{eqnarray}
Since $\gamma^{(0)}$ and $r\left(m_b\right)$ are independent of $f$
for the current-current subspace, there is no matching correction for
$Q_1$, $Q_2$.

\subsubsection{An Inhomogeneous RG Equation}\label{sect:inhom}

The local operator counterterms proportional to
$\wt{Z}_{kl,7}^{-1}\left(\mu\right)$ in
$\Lagr_{\eff}^{\dstwo}$ of \eq{lags2} do not influence the RG
evolution of the coefficients $C_l$, but they modify the running of
$\cloc$.  We will discuss this in the following.  From $\dmu
\Lagr_{\eff}^{\dstwo} = 0$ in {\eq{lags2}} one finds
\begin{eqnarray}
\dmu \left[ \sum_{k=1}^2 \sum_{k^\prime =1}^6 
C_k\left(\mu\right) C_{k'}\left(\mu\right)
\zbi{kk'}^{-1}\left(\mu\right) + \cloc\left(\mu\right)
\zloc^{-1}\left(\mu\right) \right]
&=& 0.
\label{RGdouble}
\end{eqnarray}
This can be compactly rewritten as
\begin{eqnarray}
\dmu \cloc\left(\mu\right) &=&
\cloc\left(\mu\right) \gloc
+ \sum_{k=1}^2 \sum_{k^\prime =1}^6 
   C_{k}\left(\mu\right) C_{k'}\left(\mu\right) \gbi{kk'}
\label{RGdouble1}
\end{eqnarray}
with the \emph{anomalous dimension tensor}
\begin{eqnarray}
\gbi{kn} &=&
\frac{\as}{4\pi} \gbi{kn}^{\left(0\right)}
+ \left(\frac{\as}{4\pi}\right)^2 \gbi{kn}^{\left(1\right)}
+ \ldots
\no \\
&=&
- \sum_{k^\prime=1}^2 \sum_{n^\prime =1}^6 
\left[\gamma_{kk'} \delta_{nn'} + \delta_{kk'} \gamma_{nn'}\right]
\zbi{k'n'}^{-1} \zloc
-\left[\dmu \zbi{kn}^{-1}\right] \zloc .
\label{AnomDimTens}
\end{eqnarray}
Its perturbative coefficients analogous to \eq{AnomSingleCoeff} are
found as
\begin{subequations}
\label{AnomDoubleCoeff}
\begin{eqnarray}
\gbi{kn}^{(0)} &=&
2 \zbil{-1,(1)}{1}{kn} + 2\eps \zbil{-1,(1)}{0}{kn},
\\
\gbi{kn}^{(1)} &=&
4 \zbil{-1,(2)}{1}{kn} + 2\beta_0 \zbil{-1,(1)}{0}{kn}
\no \\
& & -2 \zbil{-1,(1)}{0}{kn} \zlocl{-1,(1)}{1}
	-2 \zbil{-1,(1)}{1}{kn} \zlocl{-1,(1)}{0}
\no \\
& & -2 \sum_{k^\prime=1}^2 \sum_{n^\prime =1}^6 \left\{
\left( \left[Z^{-1,(1)}_{0}\right]_{kk'} \delta_{nn'}
	+ \delta_{kk'} \left[Z^{-1,(1)}_{0}\right]_{nn'} \right)
	\zbil{-1,(1)}{1}{k'n'}    \right.
\no \\
& & \left. \phantom{2 \sum_{k^\prime=1}^2 \sum_{n^\prime =1}^6}
+ \left( \left[Z^{-1,(1)}_{1}\right]_{kk'} \delta_{nn'}
	+ \delta_{kk'} \left[Z^{-1,(1)}_{1}\right]_{nn'} \right)
	\zbil{-1,(1)}{0}{k'n'} \right\}
\no \\
& & + O\left(\eps\right)\label{AnomDoubleCoeff2}.
\end{eqnarray}
\end{subequations}
In (\ref{AnomDoubleCoeff}) we have also included finite
renormalization constants with subscript 0. Such finite
renormalizations appear in general when counterterms proportional to
evanescent operators must be included such as in our calculation.  For
a detailed discussion see \cite{bw,hn2}. The extra terms in
(\ref{AnomDoubleCoeff2}) involving the finite renormalization
constants can be simply included into the calculation by multiplying
all one-loop diagrams containing a finite counterterm by a factor of
1/2.\footnote{Throughout this paper we implement the $\ov{\rm MS}$ scheme
by absorbing $\gamma_{\rm E} - \ln (4 \pi)$ into the measure of the
loop integrals. Hence this trivial finite part of the counterterms
never appears explicitly in any formula.\label{foot}}

Here and in the following section \ref{sect:compact} we will present
two different ways to solve the inhomogeneous RG equation
{\eq{RGdouble1}}.
With standard methods to solve coupled differential equations one
obtains
\begin{eqnarray}
\cloc(\mu) &=&
\uloc^{(0)}\left(g\left(\mu\right),g_0\right) \cloc\left(g_0\right)
\no \\
& & + \left[1+\frac{g^2\left(\mu\right)}{16\pi^2}\jloc\right] \cdot
\int_{g_0}^{g\left(\mu\right)} dg'
\uloc^{(0)}\left(g\left(\mu\right),g'\right)
\left[1-\frac{g'^2}{16\pi^2}\jloc\right] \cdot
\no \\
& &  \phantom{+}
     \sum_{n,n^\prime,t,t^\prime =1}^2 \,
     \sum_{m,m^\prime,s,s^\prime =1}^6 
\left\{-\frac{\gbi{nm}^{(0)}}{\beta_0} \frac{1}{g'} +
\left[\frac{\beta_1}{\beta_0^2} \gbi{nm}^{(0)} -
\frac{\gbi{nm}^{(1)}}{\beta_0}\right] \frac{g'}{16\pi^2}\right\}  
 \nn 
& & \hspace{2em} \cdot
\left[\delta_{mm'}+\frac{g'^2}{16\pi^2}J_{mm'}\right]
U^{(0)}_{m's}\left(g',g_0\right)
\left[\delta_{ss'}-\frac{g_0^2}{16\pi^2}J_{ss'}\right]
C_{s'}\left(g_0\right) \no \\ 
& & \hspace{2em} \cdot    
\left[\delta_{nn'}+\frac{g'^2}{16\pi^2}J_{nn'}\right]
U^{(0)}_{n't}\left(g',g_0\right)
\left[\delta_{tt'}-\frac{g_0^2}{16\pi^2}J_{tt'}\right]
C_{t'}\left(g_0\right) .
\label{RGdouble1sol}
\end{eqnarray}
Here $\uloc^{(0)}$ and $\jloc$ are the RG quantities related to single
insertions of $\oloc$ defined in (\ref{def77}).  The QCD coupling
constant at scale $\mu_0$ has been labeled $g_0$ and here the
arguments of the evolution matrices are not the scales but the
corresponding couplings. 

The first term in \eq{RGdouble1sol} is solely related to matrix
elements with single insertions of $\oloc$.  There are no factors
involving $\jloc$ here, because the initial coefficient
$\cloc\left(g_0\right)$ starts at order $g^2$.

\eq{RGdouble1sol} nicely reveals the structure of double insertions:
First the two Wilson coefficient functions $C_{t'}$ and $C_{s'}$
independently run down from the initial scale $\mu_0$ to the
intermediate scale $\mu'$ with $g(\mu')=g'$.  Then they are linked by
the anomalous dimension tensor to the single insertion coefficient
$\cloc$, which then runs further down to the final scale $\mu$.  The
integral then performs a summation over all intermediate scales
$\mu'$.  If one wants to solve the integral in (\ref{RGdouble1sol}),
one must diagonalize at least one of the two \dsone\ evolution
matrices yielding quite cumbersome expressions.

\subsubsection{A Compact Mixing Matrix}\label{sect:compact}

For formal analyses like those in \cite{hn2} the form of
{\eq{RGdouble1sol}} is well suited.  In a practical calculation,
however, this solution of the inhomogeneous RG equation is difficult to
implement.  Here we present a simpler way to solve {\eq{RGdouble1}}.

The key to observation is that in the double insertion diagrams at
least one of the two operators always stems from the current-current
subspace of the full \dsone\ operator basis.  For this subspace we
switch to $(Q_+,Q_-)$, which has the advantage that the Wilson
coefficient functions $C_+$ and $C_-$ do not mix with each other as
long as we preserve the Fierz-symmetry of $Q_\pm$ during the
renormalization process.  This is the case for our choice of
evanescent operators in {\eq{StdEvas1-12}} \cite{bw,hn2}.  The problem
then splits into two independent inhomogeneous RG equations
\begin{eqnarray}
\dmu \cloc^{\pm}\left(\mu\right) &=&
	\gloc \cloc^{\pm}\left(\mu\right)
	+ \gbi{\pm k}  C_\pm\left(\mu\right) C_k\left(\mu\right).
\label{RGdouble1sep}
\end{eqnarray}
Here the decomposition of $\cloc(\mu_{tW})$ into $\cloc^{\pm}(\mu_{tW})$
is completely arbitrary provided one satisfies
\begin{eqnarray}
\cloc\left(\mu_{tW}\right) &=&
\cloc^{+}\left(\mu_{tW}\right) + \cloc^{-}\left(\mu_{tW}\right).
\label{clocdec}
\end{eqnarray}
This decomposition is then automatically  preserved at any
renormalization scale.
We may now cast the inhomogeneous RG equation \eq{RGdouble1sep}
together with the RG equations of the \dsone\ coefficients into two
7$\times$7 matrix equations:
\begin{eqnarray}
\dmu
\left(
\begin{array}{c}
C_\pm \vec{C} \\
\cloc^{\pm}
\end{array}
\right)
&=&
\left(
\begin{array}{cc}
\gamma^T + \gamma_{\pm} \mathbf{1} & 0 \\
\gbi{\pm}^{T} & \gloc
\end{array}
\right)
\left(
\begin{array}{c}
C_\pm \vec{C} \\
\cloc^{\pm}
\end{array}
\right)
\label{RGdouble1matrix7}
\end{eqnarray}
where $\gamma$ is the 6$\times$6 \dsone\ anomalous dimension matrix
and $\vec{C} = \left(C_1,\ldots,C_6\right)^{T}$.  Further $\gbi{\pm}$
comprises elements of the anomalous dimension tensor defined 
in \eq{RGdouble1} and \eq{AnomDimTens}:
\begin{eqnarray}
\gbi{\pm}^{T} &=&
\left(
\gbi{\pm1}, \gbi{\pm2}, \gbi{\pm3}, \gbi{\pm4}, \gbi{\pm5}, \gbi{\pm6}
\right).
\end{eqnarray}
{\eq{RGdouble1matrix7}} and its solution essentially represent the
method used by Gilman and Wise in their LO analysis \cite{gw}.  Yet
they have used an inconvenient operator basis, which contains an
operator being linearly dependent on the others.  The authors of
{\cite{gw}} therefore involve 8$\times$8 matrices with a double
eigenvalue rather than 7$\times$7 matrices as in
{\eq{RGdouble1matrix7}}.  Further their bilocal structures are defined
differently, so that they had to solve four RG matrix equations, while
we only encounter two of them (corresponding to ``$+$'' and ``$-$'' in
{\eq{RGdouble1matrix7}}).

Yet in \eq{RGdouble1matrix7} these two equations still encode a lot of
redundant information, both evolutions contain the full 6$\times$6
{\dsone}\ evolution matrix.  We can do even better and collapse them
into a single 8$\times$8 RG equation:
\begin{eqnarray}
\dmu \vec{D} &=&
\widehat{\gamma}^T \cdot \vec{D}
\label{RGdouble1matrix8}
\end{eqnarray}
with
\begin{eqnarray}
\widehat{\gamma}^T =
\left(
\begin{array}{ccc}
\gamma^T & 0 & 0 \\
\gbi{+}^{T} & \gloc - \gamma_{+} & 0 \\
\gbi{-}^{T} & 0 & \gloc - \gamma_{-}
\end{array}
\right),
&\hspace{1em}&
\vec{D}\left(\mu\right) =
\left(
\begin{array}{c}
\vec{C}\left(\mu\right) \\
\cloc^{+}\left(\mu\right) / C_{+}\left(\mu\right) \\
\cloc^{-}\left(\mu\right) / C_{-}\left(\mu\right)
\end{array}
\right).
\label{RGdouble1matrix8vec}
\end{eqnarray}
Here $\dmu \left( \cloc^{\pm}/C_\pm \right)$ appearing on the LHS of
(\ref{RGdouble1matrix8}) directly evaluates to (\ref{RGdouble1sep})
and the RG equation for $C_\pm$ involving $\g_\pm$.  Both
\eq{RGdouble1matrix7} and \eq{RGdouble1matrix8} can be solved by the
standard techniques already introduced for single insertions in
{\sect{sect:rgsingle}}, see \eq{EvolSingleLO} and \eq{EvolSingleNLO}.

Since the Wilson coefficient $\vec{D}$ contains the Wilson
coefficients of the \dsone\ operators, it receives a matching
correction when passing from the effective five to the effective four
quark theory analogously to \eq{MatchWilsonS1}:
\begin{eqnarray}
\vec{D}^{[4]}\left(\mu_b\right) &=&
\left[1
+\frac{\as\left(\mu_b\right)}{4\pi}
	\delta \widehat{r}^T \lt( \mu_b \rt) +O\left(\as^2\right)\right]
\vec{D}^{[5]}\left(\mu_b\right),
\label{MatchWilsonS2D}
\end{eqnarray}
where the 8$\times$8 matrix $\delta \widehat{r}$ is defined as
\begin{eqnarray}
\delta \widehat{r} \left(\mu_b\right) &=&
\left(
\begin{array}{*{3}{c}}
\delta r\left(\mu_b\right) & 0 & 0 \\
0 & 0 & 0 \\
0 & 0 & 0
\end{array}
\right)
\label{Defr8}
\end{eqnarray}
and $\delta r$ denotes the matching correction in the \dsone\ Wilson
coefficients introduced in \eq{defdr}.

\subsection{The $\mathbf{\dstwo}$ NLO Anomalous Dimension 
            Tensor}\label{sect:anom}

In order to calculate the solution of the RG equation
{\eq{RGdouble1matrix8}} we need to know the value of the anomalous
dimension tensor $\gbi{\pm j}$, which governs the mixing from double
insertions to $\cloc$.  This tensor is determined from the
renormalization factor $\zbi{ij}^{-1}$, see {\eq{AnomDoubleCoeff}}.

$\zbi{ij}^{-1}$ is determined from the finiteness 
of the Green's function $-i \left\langle H^{ct} \right\rangle$ in
{\eq{GreenMixHct}}.  Inserting all the required renormalization
factors including the wave-function renormalization constant
$Z_{\psi}$ we find for the $O(\as^0)$, $O(\as^1)$ term of
$\zbi{ij}^{-1}$:
\begin{subequations}
\label{bareR}
\begin{eqnarray}
\zbil{-1,(1)}{}{ij} \left\langle \oloc \right\rangle^{(0)}
\hspace{-0.6em}
&\stackrel{\mathsf{div.}}{=}&
\hspace{-0.6em}
- \left\langle \mathcal{R}_{ij} \right\rangle^{(0),\bare},
\label{bareR0}
\\
\zbil{-1,(2)}{}{ij} \left\langle \oloc \right\rangle^{(0)}
\hspace{-0.6em}
&\stackrel{\mathsf{div.}}{=}&
\hspace{-0.6em}
- \left\langle \mathcal{R}_{ij} \right\rangle^{(1),\bare}
- \sum_{i^\prime =+,-} \sum_{j^\prime=1}^6 \, 
\frac{1}{\eps} \left[2 Z_{\psi,1}^{(1)} \delta_{ii'} \delta_{jj'}
\right.
\no \\
& &
\left. \hspace{13ex}
	-\left[Z_1^{(1)}\right]_{ii'} \delta_{jj'}
	-\delta_{ii'} \left[Z_1^{(1)}\right]_{jj'} \right]
\left\langle \mathcal{R}_{i'j'} \right\rangle^{(0),\bare}
\no \\
& &
\hspace{-0.6em}
- 2 Z_{\psi,1}^{(1)} \frac{1}{\eps^2} \zbil{-1,(1)}{1}{ij}
\left\langle\oloc\right\rangle^{(0)}
-\frac{1}{\eps} \zbil{-1,(1)}{1}{ij}
\left\langle\oloc\right\rangle^{(1),\bare},
\no \\
& &
\label{bareR1}
\end{eqnarray}
\end{subequations}
where we have used the notation \eq{ExpZSingle} and (\ref{qex}) for
the expanded $Z$-factors and matrix elements.  In
\eq{bareR} the symbol $\stackrel{\mathsf{div.}}{=}$ means that only
the divergent parts of the LHS and RHS need to be equal.

Now in the LO $\zbil{-1,(1)}{}{ij}$ is simply obtained from the
$1/\eps$ pieces $\zbil{-1,(1)}{1}{ij}$ of
$\left\langle\mathcal{R}_{ij}\right\rangle^{(0)}$.  These terms are
calculated by the evaluation of the diagrams in \fig{fig:double-cc-lo}
and \fig{fig:double-peng-lo} for $i=+,-$ and $j=1,2$ and $j=3,\ldots,6$
respectively.

For the NLO one has to know the $O(\as)$ corrections to
$\left\langle\oloc\right\rangle$, which are related to the matrix
elements of $\oll$ through the definition {\eq{q7bare}}:
\begin{eqnarray}
\left\langle\oloc\right\rangle^{(1),\bare} &=&
\frac{\mc}{g^2 \mu^{2\eps}} \left(2 Z_{m,1}^{(1)} + \beta_0\right)
\frac{1}{\eps} \left\langle\oll\right\rangle^{(0)}
+\frac{\mc}{g^2 \mu^{2\eps}} \frac{1}{\eps}
\left\langle\oll\right\rangle^{(1),\bare}.
\label{meq7b1}
\end{eqnarray}
Now the divergent parts of the two-loop diagrams in
{\fig{fig:double-cc-nlo}} and {\fig{fig:double-peng-nlo}} including
the corresponding subloop counterterm diagrams yield the terms in the
first two lines of {\eq{bareR1}} and the last term in \eq{meq7b1}.
\begin{nfigure}
\centerline{\epsfxsize=\textwidth \epsffile{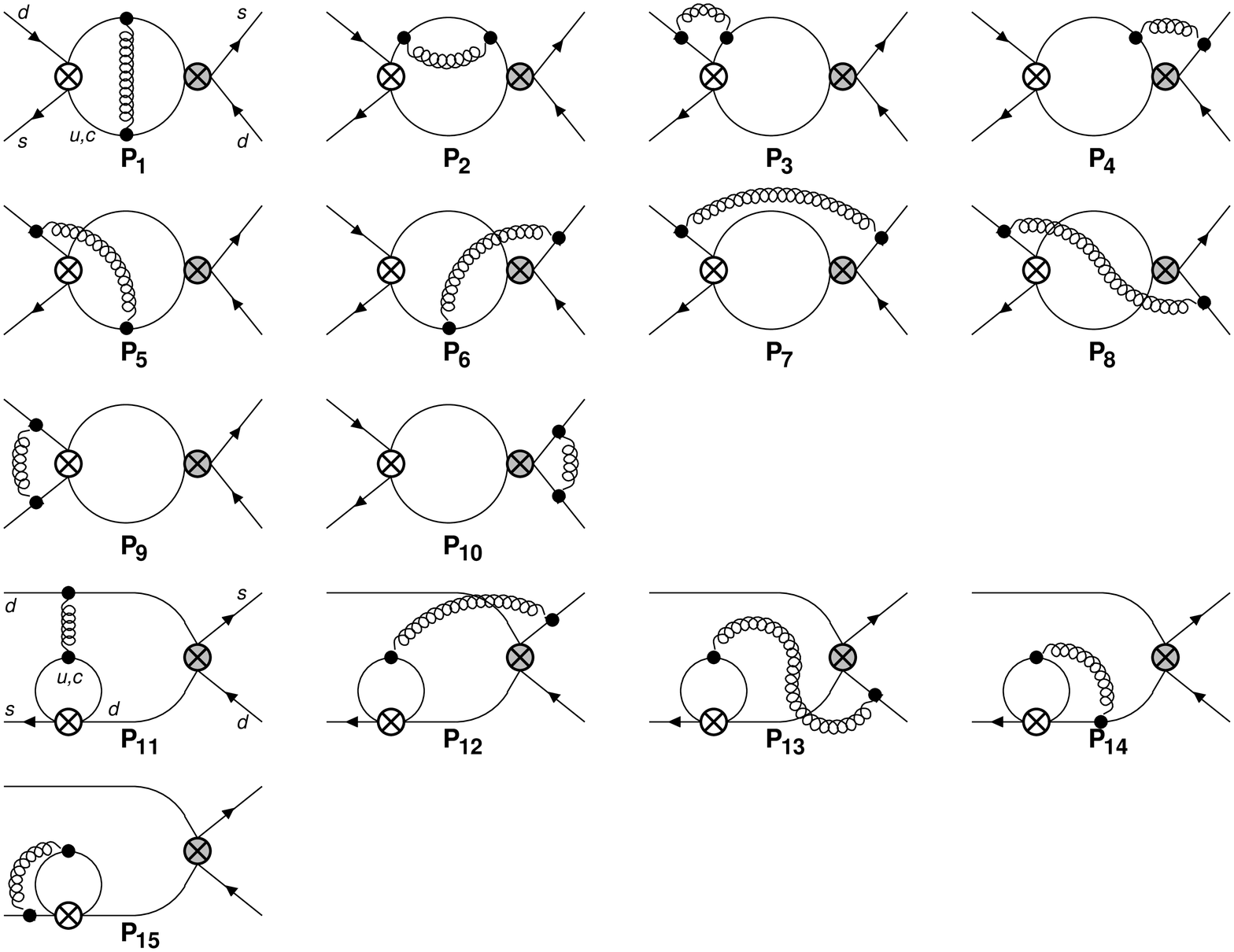}}
\caption[nothing]{\slshape
The classes of diagrams yielding the $O(\as)$ contribution to
$\mathcal{R}_{ij}$, $i=+,-$, $j=3,\ldots,6$ in \eq{GreenMixDoubleR} in
the effective five- and four-flavour theory, the remaining diagrams
are obtained by mirror reflections.  The white crosses denote
insertions of {\dsone}\ current-current operators, the shaded ones
insertion of {\dsone}\ four-quark penguin operators.  Additional QCD
counterterms have to be included.  The divergences of the diagrams
$\mathsf{P_i}$, $i=1,\ldots,15$ are summarized in {\apx{apx:int}}
(\eq{div-p} and \tab{tab:div-p}).  The diagrams $\mathsf{P_i}$ belong
to two different species: In $\mathsf{P_{1}}$ -- $\mathsf{P_{10}}$ the
penguin operator contributes through its up-type quark foot (as in the
LO diagram $\mathsf{P_0}$ of {\fig{fig:double-peng-lo}}), while in
$\mathsf{P_{11}}$ -- $\mathsf{P_{15}}$ its down-type quark foot is
involved --- a new feature of the NLO calculation.  Diagrams vanishing
identically are not displayed.}
\label{fig:double-peng-nlo}
\end{nfigure}\clearpage
The remaining divergences therefore correspond to
\begin{eqnarray}
- \frac{1}{\eps^2} \left[\zbil{-1,(2)}{2}{ij}
+\left(2Z^{(1)}_{m,1}+\beta_0+2Z^{(1)}_{\psi,1}\right)
\zbil{-1,(1)}{1}{ij} \right]
- \frac{1}{\eps} \zbil{-1,(2)}{1}{ij}.
\label{DivZbi1}
\end{eqnarray}
For clarity counterterms proportional to unphysical operators have
been omitted in (\ref{bareR}-\ref{DivZbi1}).

Hence we can simply read off $\zbil{-1,(2)}{1}{ij}$ from the $1/\eps$
divergences of the diagrams of \fig{fig:double-cc-nlo} and
{\fig{fig:double-peng-nlo}} after the inclusion of subloop
counterterms.

The diagrams $\mathsf{P_i}$ in \fig{fig:double-peng-nlo} appear in two
different species: For $i=1,\ldots,10$ the penguin operator contributes
via its up-type quark foot, i.e.\ through the couplings
$(\ov{s}d)_{V-A} (\ov{c}c)_{V\pm A}$ and $(\ov{s}d)_{V-A}
(\ov{u}u)_{V\pm A}$ as in the LO diagram $\mathsf{P_0}$ of
\fig{fig:double-peng-lo}.  In contrast in
$\mathsf{P_{11}},\ldots,\mathsf{P_{15}}$ the down-type quark foot of
the penguin contributes, i.e.\ the couplings $(\ov{s}d)_{V-A}
(\ov{s}s)_{V\pm A}$ and $(\ov{s}d)_{V-A} (\ov{d}d)_{V\pm A}$.  
Naively one would not expect $\mathsf{P_{11}},\ldots,\mathsf{P_{15}}$
to be proportional to $\mc$, since the subdiagram involving $m_c$ is
proportional to 
\begin{eqnarray}
&& \ov{s} \g_{\mu} L d \cdot \left( k^{\mu} k^{\nu} - k^2 g^{\mu \nu} 
    \right) f\left( \frac{k^2}{\mc} \right) \label{trv}
\end{eqnarray}
i.e.\ transverse with respect to the virtual gluon momentum $k$.  If
one expanded $f\left( \frac{k^2}{\mc} \right)$ around $D=4$, one would
only find a logarithmic dependence on $\mc$. Yet the second loop
integration over $k$ is quadratically divergent yielding a result
proportional to $\mc$.  Nevertheless only $\mathsf{P_{12}}$ with
insertions of $Q_3$ or $Q_4$ has a non-vanishing divergent part.  The
other diagrams are finite.  Diagrams with insertions of $Q_5$ or $Q_6$
vanish altogether.  Of course there is no divergence proportional to
$\mc/\eps^2$ in $\mathsf{P_{12}}$, because the one-loop counterterm
diagrams vanish. The finiteness of $\mathsf{P_{13}}-\mathsf{P_{15}}$
is related to current conservation and ensures that these diagrams do 
not contribute to the NLO calculation of $ K \rightarrow \pi \nu
\ov{\nu}$ performed in \cite{bb}. We had to include the 1PR diagrams 
$\mathsf{P_{14}}$ and $\mathsf{P_{15}}$ into the consideration,
because the result of their 1PI subdiagram is proportional to 
$\Qfeom$ defined in \eq{defqeom}, but $\Qfeom$ has been dropped from
the operator basis. 
 
These unexpected contributions from transverse subdiagrams \eq{trv}
has another interesting consequence: Dimension-8 operators such 
as 
\begin{eqnarray}
\wt{Q}_9 &=& \ov{s} \g_{\mu} L D_\nu d \cdot \ov{s} 
\begin{array}[b] {@{}c}
\!\! \scriptstyle \leftarrow \\[-1.8ex] 
D^{\nu}
\end{array} \!\!\!
\g^{\mu} L d \no
\end{eqnarray}
mix into $\oloc$ and even into $m_b^2/m_c^2 \cdot \oloc$ via two-loop
diagrams containing a gluon self-energy subdiagram with a $c$- or
$b$-quark loop. Hence in an \emph{effective}\ field theory, where
gluons can appear in \emph{quadratically}\ divergent diagrams 
heavy degrees of freedom (here: the $b$- and $c$-quark) of the 
QCD-lagrangian do not decouple anymore. This distinguishes 
an effective theory with non-renormalizable interactions 
(here the four-fermion interactions) from renormalizable 
theories, in which the Appelquist-Carrazone theorem holds \cite{ac}. 
Yet in our case fortunately the GIM mechanism ensures that the bilocal
structures $\mathcal{O}_{ij}$ and $\mathcal{R}_{ij}$ of
\eq{GreenMixDouble} do not mix into $\wt{Q}_9$ and other 
physical dimension-8 operators apart from $\oloc$. Hence our \dstwo\
basis is complete.

Using \eq{AnomDoubleCoeff} we then obtain the elements of the
anomalous dimension tensor
\begin{subequations}
\label{ResAnomTens}
\begin{eqnarray}
\gbi{+}^{(0)} =
\left(
\begin{array}{r}
-16 \\ -8 \\ -32 \\ -16 \\ 32 \\ 16
\end{array}
\right),
\quad &&
\gbi{-}^{(0)} =
\left(
\begin{array}{r}
8 \\ 0 \\ 16 \\ 0 \\ -16 \\ 0
\end{array}
\right),
\label{ResAnomTens0}
\\
\gbi{+}^{(1)} =
\left(
\begin{array}{r}
-212 \\
-28 \\
-456 \\ 
-88 \\ 
\frac{1064}{3} \\
\frac{832}{3}
\end{array}
\right),
&&
\gbi{-}^{(1)} =
\left(
\begin{array}{r}
276 \\
-92 \\
520 \\ 
-216 \\ 
-\frac{1288}{3} \\
0
\end{array}
\right),
\label{ResAnomTens1}
\end{eqnarray}
\end{subequations}
where we have set $N=3$ for brevity.  As usual the NLO anomalous
dimension tensor depends on the renormalization scheme.  The result
{\eq{ResAnomTens1}} corresponds to the NDR scheme with the definition
of the evanescent operators corresponding to {\eq{StdEvasNum}}.

In \eq{ResAnomTens1} the diagram $\mathsf{P_{12}}$ involving the
down-type foot of the penguin operator contributes $-32$ to $\gbi{\pm
3}^{(1)}$ and $\gbi{\pm 4}^{(1)}$. The results of the individual
diagrams can be found in appendix~\ref{apx:int}.
Appendix~\ref{apx:zfact} contains $\gbi{\pm}$ for arbitrary $N$.

\subsubsection{Evanescent Scheme Dependence}\label{sect:evas}
In this section we illustrate some findings of \cite{hn2}.  In
\cite{hn2} the transformation rule between two anomalous dimension
tensors calculated with two different definitions of the evanescent
operators in \eq{StdEvas} has been derived.

In our two-loop calculation we have kept $a_1,a_2, \wt{a}_1$ and
$\wt{b}_1$ in \eq{StdEvas} arbitrary yielding
\begin{eqnarray}
\!\!\!
\gbi{+}^{(1)} =
\left(
\begin{array}{@{}*{4}{r@{}}}
-\frac{188}{3} & -\frac{74}{3} a_1 & & + \frac{130}{3} \tilde{a}_1 \\
-\frac{100}{3} & -\frac{34}{3} a_1 & & + \frac{32}{3} \tilde{a}_1 \\
-\frac{1816}{3} & - \frac{88}{3} a_1 & & + \frac{32}{3} \tilde{a}_1 \\
-\frac{680}{3} & - \frac{80}{3} a_1 & & + \frac{28}{3} \tilde{a}_1 \\
\frac{1576}{3} & + \frac{80}{3} a_1 & + \frac{8}{3} a_2 &
	- \frac{32}{3} \tilde{a}_1 \\
\frac{1664}{3} & + \frac{28}{3} a_1 & + \frac{52}{3} a_2 &
	- \frac{28}{3} \tilde{a}_1
\end{array}
\right),
&& \!\!\!\!
\gbi{-}^{(1)} =
\left(
\begin{array}{@{}*{4}{r@{}}}
\frac{124}{3} &+\frac{22}{3} a_1 & & - \frac{110}{3} \tilde{a}_1 \\
-12 &+6 a_1 & & + 4 \tilde{a}_1 \\
\frac{1496}{3} & + \frac{8}{3} a_1 & & - \frac{16}{3} \tilde{a}_1 \\
-120 & + 8 a_1 & & + 4 \tilde{a}_1 \\
\frac{1160}{3} & - \frac{16}{3} a_1 & + \frac{8}{3} a_2 &
	+ \frac{16}{3} \tilde{a}_1 \\
-128 & - 4 a_1 & - 4 a_2 & - 4 \tilde{a}_1
\end{array}
\right) \label{ResAnomTens1a}        .
\end{eqnarray}

Let us first look at the dependence of $\gbi{\pm}^{(1)}$ on $a_1$ and
$a_2$ parameterizing the \dsone\ evanescent operators in
(\ref{StdEvas1-12}-\ref{StdEvas1-56}): In our case the corresponding
formula (cf.\ Eq.\ (50) of \cite{hn2}) reads
\begin{eqnarray}
\hspace{-1.2em}
\gbi{\pm j}^{(1)} (a^\prime)-\gbi{\pm j}^{(1)} (a) \!\! &=& \! \!\!
\sum_{k=+,-} \!\! \left[ \overline{D} \cdot (a^\prime -a) \right]_{\pm,k}
\gbi{kj}^{(0)} + \!
\sum_{j^\prime=1}^{6}
      \left[ \overline{D} \cdot (a^\prime -a) \right]_{j j^\prime}
      \gbi{\pm j^\prime}^{(0)} . \label{gbitrafo}
\end{eqnarray}
Here $\overline{D}$ is a 6$\times$6 diagonal matrix with
$\overline{D}_{im} = \left[Z_1^{(1)} \right]_{i,E_{1i}} \delta_{im} $.
It involves the evanescent part $ -\as/(4 \pi) \cdot \left[Z_1^{(1)}
\right]_{i,E_{1i}} E_{1} \left[ Q_i \right]$
of the one-loop counterterm to $Q_i$ needed to render the one-loop
diagrams of \fig{fig:ds1-cc-nlo} finite.  The normalization in
\eq{StdEvas} is chosen such that $\overline{D}$ is the unit matrix.
$a$ and $a^\prime$ are 6$\times$6 matrices defining the evanescent
operators $E_1 \left[Q_i\right]$ (cf.\ Eq.\ (5) of
\cite{hn2}). $a$ is easily obtained in terms of $a_1$ and $a_2$
from the colour factors in \eq{StdEvas} and \eq{DefColourEva}:
\begin{eqnarray}
\overline{D} \cdot a &=& a \; = \;
\left(
\begin{array}{cccccc}
\frac{7}{12} a_1 & \frac{a_1}{4} &  0 & 0 & 0 & 0 \\
\frac{a_1}{2}    & - \frac{a_1}{6} &  0 & 0 & 0 & 0 \\
0 & 0 & - \frac{a_1}{6}    & \frac{a_1}{2}  &  0 & 0 \\
0 & 0 & \frac{a_1}{4}    & \frac{7}{12} a_1 &  0 & 0 \\
0 & 0 & 0 & 0 & - \frac{a_2}{6}    & \frac{a_2}{2}   \\
0 & 0 & 0 & 0 &   \frac{a_2}{4}    & \frac{7}{12} a_2
\end{array}
\right) \label{amatrix} ,
\end{eqnarray}
where we have chosen $N=3$ for simplicity.
For the first term in \eq{gbitrafo} we further need the
current-current part of $a$ in the basis $(Q_+,Q_-)$:
\begin{eqnarray}
a_{++} &=& \frac{7}{12} a_1, \qquad a_{+-} \; = \; - \frac{a_1}{2}
\nonumber\\
a_{-+} &=& - \frac{a_1}{4},  \qquad
a_{--} \; = \; - \frac{a_1}{6} .\label{apm}
\end{eqnarray}
Here we have tacitly corrected an error in the example at the end of
sect.~4 of \cite{hn2}.  Inserting \eq{amatrix} and \eq{apm} into
\eq{gbitrafo} with $a^\prime=0$ correctly reproduces the dependence of
$\gbi{\pm}^{(1)}$ on $a_1$ and $a_2$ in \eq{ResAnomTens1a} found by our
explicit two-loop calculation.

The dependence on $\wt{a}_1$ is more interesting to study, because it
reveals some of the subtleties of the Fierz transformation in dimensional
regularization:  In addition to $\oloc$ we need its Fierz transform
\begin{eqnarray}
\wt{Q}_8
&=& \frac{\mc}{g^2 \mu^{2 \eps}} \cdot
     \ov{s}_i \g_\mu L d_j \cdot \ov{s}_j \g^\mu L d_i
\label{defq8}
\end{eqnarray}
with $i,j$ being colour indices. $\oloc$ is the Fierz transform of
$\wt{Q}_8$ for $D=4$, hence their difference is evanescent.  The
one-loop counterterm diagrams involved in the calculation of
\eq{ResAnomTens1a} have been accounted for diagram-by-diagram. This
effectively corresponds to keeping both $\oloc$ and $\wt{Q}_8$ in the
operator basis and prevents the incorrect use of Fierz symmetry in
$D$-dimensional expressions. For the scheme transformation formula we
therefore need the one-loop renormalization constants in the basis
$(\oloc,\wt{Q}_8)$:
\begin{subequations}
\label{nf}
\begin{eqnarray}
\left(  \zbil{-1,(1)}{1}{+j}^\mathrm{nF}    \right)
&=& (0,-2,-16,-4,16,4),
\\
\left(  \left[\wt{Z}^{-1,(1)}_{1}\right]_{+j ,8}^\mathrm{nF} \right)
&=& (-8,-2, 0,-4, 0, 4)
\\
\left(  \zbil{-1,(1)}{1}{-j}^\mathrm{nF}    \right)
&=& (0,-2,8,4,-8,-4)
\\
\left(  \left[\wt{Z}^{-1,(1)}_{1}\right]_{-j ,8}^\mathrm{nF} \right)
&=& (4, 2, 0,-4, 0, 4).
\end{eqnarray}
\end{subequations}
Here ``nF'' means that no Fierz symmetry is used.  In the final step
to calculate the anomalous dimension tensor of the physical operators the
\dstwo\ operator basis is transformed from $(\oloc,\wt{Q}_8)$ to
$(\oloc, \wt{Q}_8-\oloc)$.  Thereby $\zbil{-1,(1)}{1}{\pm j}^\mathrm{nF}$
and $\left[\wt{Z}^{-1,(1)}_{1}\right]_{\pm j ,8}^\mathrm{nF} $ simply add
to $\zbil{-1,(1)}{1}{\pm j}=\gbi{\pm j}^{(0)}/2$.

Hence for $\gbi{\pm j}^{(1)}$ the transformation rule
\cite[Eq.~(51)]{hn2} reads
\begin{eqnarray}
\lefteqn{
\gbi{\pm j}^{(1)} (\wt{a}^\prime) - \gbi{\pm j}^{(1)} (\wt{a})\;=}
\nonumber \\
&&
\sum_{l,m=7}^8 \left\{
2 \, \left[\wt{Z}^{-1,(1)}_{1}\right]_{\pm j ,l}^\mathrm{nF} \cdot
 \left[ \wt{Z}_1^{-1,(1)} \right]_{l,\wt{E}_{1l}}
-2 \beta_0 \cdot  \left[\wt{Z}^{-1,(1)}_{1}\right]_{\pm j,\wt{E}_{1l}}
+ \gzero_{\pm} \left[\wt{Z}^{-1,(1)}_{1}\right]_{\pm j,\wt{E}_{1l}}
\nonumber \right. \\
&& \left.
\! \! \! + \sum_{j^\prime=1}^6 \gzero_{j j^\prime } \cdot
\left[\wt{Z}^{-1,(1)}_{1}\right]_{\pm j^\prime,\wt{E}_{1l}} \!
- \sum_{n=7}^8
  \left[\wt{Z}^{-1,(1)}_{1}\right]_{\pm j,\wt{E}_{1l}} \! \cdot 2 \,
  \left[\wt{Z}^{-1,(1)}_{1}\right]_{mn}^\mathrm{nF}
\right\} \, \left[ \wt{a}_{lm}^\prime \! -\wt{a}_{lm} \right].
\label{scheme2}
\end{eqnarray}
The extra sum over $m$ compared to \cite[Eq.~(51)]{hn2} adds the
contributions proportional to $\langle \oloc \rangle^{(0)}$ and
$\langle \wt{Q}_8 \rangle^{(0)}$ of the diagrams from which $\gbi{\pm
j}^{(1)}$ is calculated.  This corresponds to the transformation
$(\oloc, \wt{Q}_8-\oloc)\rightarrow (\oloc,\wt{Q}_8)$ described at the
end of the previous paragraph.

The evanescent operator $E_1[\wt{Q}_8]$ is obtained from
$E_1[\oloc]$ in \eq{StdEvas1-loc} by replacing $K_{12}$ with $K_{11}$.
The remaining ingredients of \eq{scheme2} are
\begin{subequations}
\label{ingr}
\begin{eqnarray}
 \left[ \wt{Z}_1^{-1,(1)} \right]_{7,\wt{E}_{17}} &=&
 \left[ \wt{Z}_1^{-1,(1)} \right]_{8,\wt{E}_{18}} \; = \; -1, \\
 \left(  \left[\wt{Z}^{-1,(1)}_{1}\right]_{+j ,\wt{E}_{17}} \right)
&=& (-\frac{9}{4}, \frac{3}{4}, 0,0,0,0),
\\
\left(  \left[\wt{Z}^{-1,(1)}_{1}\right]_{+j ,\wt{E}_{18}} \right)
&=& (-\frac{3}{2}, -\frac{3}{2}, 0,0,0,0) ,
\\
\left(  \left[\wt{Z}^{-1,(1)}_{1}\right]_{-j ,\wt{E}_{17}} \right)
&=& ( \frac{9}{8}, \frac{15}{8}, 0,0,0,0),
\\
\left(  \left[\wt{Z}^{-1,(1)}_{1}\right]_{-j ,\wt{E}_{18}} \right)
&=& ( \frac{3}{4}, -\frac{3}{4}, 0,0,0,0),
\end{eqnarray}
\end{subequations}
\begin{eqnarray}
\left(   \left[\wt{Z}^{-1,(1)}_{1}\right]_{mn}^\mathrm{nF} \right) &=&
\left(
\begin{array}{cc}
7 - \beta_0 & 3 \\
3 & 7 - \beta_0
\end{array}
\right), \qquad
\wt{a} \; = \;
\left(
\begin{array}{rl}
-\frac{\wt{a}_1}{6}  & \frac{\wt{a}_1}{2}  \\
\frac{\wt{a}_1}{4}   & \frac{7}{12} \wt{a}_1
\end{array}
\right) . \label{atw}
\end{eqnarray}
Inserting \eq{nf}, (\ref{ingr}-\ref{atw}) and \eq{ds1anom0} into
\eq{scheme2} with $\wt{a}^\prime=0$ indeed reproduces the correct
dependence on $\wt{a}_1$. Note that for $j \geq 3$ only the first term
in \eq{scheme2} is nonzero. From e.g.\ $\gbi{-4}^{(1)}$ it is easy to
see that one must distinguish $\oloc$ and $\wt{Q}_8$ to derive the
correct result.

The Wilson coefficient $\cloc$ depends on $\wt{a}_1$, too.  We find
for its initial coefficient:
\begin{eqnarray}
\cloc(\mu_{tW}) &=&
\frac{\as(\mu_{tW})}{4\pi} \left[
-8\ln\frac{\mu_{tW}}{M_W}+4F\left(x_t\left(\mu_{tW}\right)\right)
-\left(6+\tilde{a}_1\right)
\right] \label{clocsch}
\end{eqnarray}
Here the dependence on $\wt{a}_1$ enters through
\begin{eqnarray}
\langle \bilr{+2}  (\mu_{tW}) \rangle ^{(0)} &=&
\frac{\mc\left(\mu_{tW}\right)}{16 \pi^2}
\left[-8 \ln\left(m_c/\mu_{tW}\right)+\left(6 +\tilde{a}_1
\right)\right] \langle \oll \rangle^{(0)},
\label{rsch}
\end{eqnarray}
which coincides with \eq{rmtw} for $\wt{a}_1=-8$.
With the methods of \cite{hn2} one derives the general transformation
law:
\begin{eqnarray}
\lefteqn{
\cloc (\mu_{tW} , \wt{a}_1^\prime)-\cloc (\mu_{tW} , \wt{a}_1)
\; = } \nonumber \\
&&\sum_{k=+,-} \sum_{j=1}^6 \sum_{l,m=7}^8
\left[\wt{Z}^{-1,(1)}_{1}\right]_{kj ,\wt{E}_{1l}}
   C_k^{(0)} (\mu_{tW}) C_j^{(0)} (\mu_{tW})
   \left[ \wt{a}_{lm}^\prime - \wt{a}_{lm}   \right] .
\label{cloctrafo}
\end{eqnarray}
With the leading order Wilson coefficients $C_k^{(0)} (\mu_{tW})$ from
\eq{ds1initial} and \eq{ds1diaginit} one easily reproduces
\eq{clocsch} from \eq{cloctrafo}.

Finally we discuss the dependence on $\tilde{b}_1$.  In \cite{hn2} it
has been proven that the NLO anomalous dimension tensor does not
depend on $\tilde{b}_1$.  We have verified this for our result
{\eq{ResAnomTens1a}}. Nevertheless the individual contributions in
(\ref{AnomDoubleCoeff2}) to the first two components of
$\gbi{\pm}^{(1)}$ depend on $\tilde{b}_1$.  As remarked in \cite{hn2}
the additive terms to $4 \zbil{-1,(2)}{1}{kn}$ in
(\ref{AnomDoubleCoeff2}) are automatically taken into account, if one
inserts the counterterms proportional to evanescent operators
with a factor of 1/2 rather than 1.  In the actual calculation we
have inserted them with a factor of $\lambda$. For $\lambda=1$ we have
obtained $4 \zbil{-1,(2)}{1}{\pm j}$, which has been found to depend on
$\wt{b}_1$. Setting $\lambda=1/2$ in our result yielding
$\gbi{\pm j}^{(1)}$ makes the coefficient of $\wt{b}_1$ vanish
(cf.~appendices \ref{apx:int} and \ref{apx:zfact}).

\subsection{Should One Sum $\mathbf{\ln m_t/M_W}$? }\label{sect:dort}
In the calculation of the initial conditions of the Wilson
coefficients in sect.~\ref{sect:match-tW} the top quark and the 
W-boson have been integrated out simultaneously.  This procedure is
sometimes criticized, because it neglects the RG evolution between 
the scales $\mu=m_t$ and $\mu=M_W$. In \cite{dpsr} this evolution 
has been investigated in a LO analysis, in which the top quark and the
W-boson are integrated out separately. While no effect has been 
found for $\eta_3$, the correction to $\eta_2$ is claimed to be of the
same size as the NLO correction calculated in \cite{bjw}. Let us
therefore look at $\wt{G}^t$ defined in \eq{smbox} in some detail:
The effect of the RG evolution between $\mu=m_t$ and $\mu=M_W$ is to 
sum $(\as \ln x_t)^n$, $n=0,1,\ldots$. The corresponding terms for 
$n=0$ and $n=1$ are also contained in the LO and NLO results of 
\eq{sxt} and \eq{h-1-t}. The smallness of both $\as (M_W)$ and 
$\ln x_t$ casts doubt on the necessity of this extra RG evolution.
And, more importantly, any RG summation of $\ln x_t$ is  accompanied 
by an OPE: 
\begin{eqnarray}
\wt{G}^t (x_t, \mu) &=& \lambda_t^2 \frac{G_F^2}{2} \lt[  
      \sum_k \wt{C}_k ( m_t, \mu ) \cdot 
   \langle \wt{O}_k \rangle (M_W, \mu) \rt. \nn
&& \lt. \phantom{\lambda_t^2 \frac{G_F^2}{2}}
+ \sum_{j,k} C_j ( m_t, \mu ) C_k (
 m_t, \mu ) \cdot \langle \mathcal{S}_{jk} \rangle (M_W, \mu)
\rt], 
\label{opetop}
\end{eqnarray}
where $\mathcal{S}_{jk}$ are bilocal structures composed of two
$\dsone$ operators describing the coupling of an $\ov{s}$ and $d$
quark to two W-bosons. \eq{opetop} corresponds to an expansion of
$\wt{G}^t$ in inverse powers of $x_t$ with higher powers of $1/x_t$
corresponding to increasing dimensions of the operators. Hence the
price to pay for the summation of the small logarithm $\ln x_t$ to all
orders is the inclusion of just a finite number of terms in the
expansion of $\wt{G}^t$ in $1/x_t$. With the results \eq{sxt} and
\eq{h-1-t} we can check the convergence of this expansion. For
$x_t=4.6$ one finds that even the inclusion of the first seven
terms in
\begin{eqnarray}
S(x_t) \; = \; 
            \frac{x_t}{4} + \frac{3}{2} \ln x_t -
            \frac{9}{4} +  
	    \frac{9 \ln x_t}{2 x_t}  - \frac{15}{4 x_t}
            + \frac{9 \ln x_t}{x_t^2}  - 
            \frac{21}{4 x_t^2} + 
            O \lt( \frac{\ln x_t}{x_t^3} \rt)
&& \label{largext}
\end{eqnarray} 
corresponding to the inclusion of $\dstwo$ operators up to dimension
12 in a NLO calculation in \eq{opetop} results in a 10\% error in
$S(x_t)$. This error is larger than the size of the $ x_t \as^2 \ln^2
x_t$-term included by the RG summation between $m_t$ and
$M_W$. Moreover the expansion of $k^t$ of \eq{h-1-t} in $1/x_t$ shows
no convergence at all. Hence the RG evolution between $\mu=m_t$ and
$\mu=M_W$ is not only unnecessary, but $x_t$ is simply too small to 
allow for a meaningful OPE.  

Now in \cite{dpsr} the summation of $\ln x_t$ has been tried by the
methods of \cite{vys}, which first applies the OPE to the \dsone\
substructure of $\wt{G}^t$ and then circumvents the calculation of the
operator mixing into \dstwo\ operators by the extraction of the
relevant logarithms from the \dstwo\ loop diagrams.  Yet there are
some mistakes in the analysis of \cite{dpsr}: For example the operator
basis $(Q_{+}^{kl}, Q_{-}^{kl})$ has been used for the \dsone\
transitions, which is equivalent to shrinking the W-lines to a point
as in \fig{fig:double-cc-lo}. This is the appropriate method for the
case $m_t \ll M_W$. For $m_t \gg M_W$ one has to shrink the top quark
lines in \fig{box} instead. The expansion of $S(x)$ for large $x$ in
\eq{largext} and the one for small $x$ in \eq{ilgim} are obviously
different. Further the authors of \cite{dpsr} have not realized that
every power of $x_t$ in $\wt{G}^t$ requires different operators in
\eq{opetop} with different anomalous dimensions.  Hence the results of
\cite{dpsr} are incorrect.

%

\section[\dstwo\ Transitions at the Charm Threshold 
 and Below]{\dstwo\ Transitions at the Charm Threshold \\ 
 and Below}\label{sect:belowc}

In this section we will eliminate the charm quark as a dynamic degree
of freedom and describe the physics of the \dstwo\ transition with an
effective three-flavour lagrangian.  The necessary steps are as in the
preceding section:
\begin{enumerate}
\renewcommand{\labelenumi}{\roman{enumi})}
\item Match the effective four-flavour theory to the effective
three-flavour theory at the renormalization scale $\mu_c$.
\item Perform the RG running below $\mu_c$.
\end{enumerate}

\subsection{Matching to the Effective Three-Quark Theory}
\label{sect:match3}

After integrating out the charm quark all dependence on $m_c$
belongs to the Wilson coefficients.  This implies that the
term involving $\oloc$ in \eq{lags2} has to disappear from the
effective lagrangian, because $\oloc$ contains $m_c$ in its definition
{\eq{defqseven}}.  Further the \dsone\ operators are neglected in the
new effective lagrangian, because the matrix elements of double
insertions of these operators are at most proportional to $\ms$
rather than $\mc$.  We have already neglected such terms in all
preceding steps.

Therefore the new effective lagrangian to describe the physics below
$\mu_c$ reads:
\begin{eqnarray}
\leo &=&
-\frac{\gft}{16\pi^2} \left[
\lambda_c^2 \cll{c}\left(\mu\right)
+\lambda_t^2 \cll{t}\left(\mu\right)
+\lambda_c \lambda_t \cll{ct}\left(\mu\right)
\right] \zll^{-1}\left(\mu\right) \oll^\bare.
\no \\
& &
\label{lags2c}
\end{eqnarray}
This lagrangian already resembles $-H^{\dstwo}$ introduced in \eq{s2}.
For the matching we have to set the Green's function \eq{GreenMix} and
the one derived from \eq{lags2c} equal at the scale $\mu=\mu_c$.

Let us start the matching procedure with $\wt{G}^{ct}$: With
\eq{GreenMixHct} and \eq{lags2c} $\cloc(\mu_c)$ is calculated from
\begin{eqnarray}
\lefteqn{
\hspace{-8em}
\sum_{i=+,-} \sum_{j=1}^6 C_i \lt( \mu_c \rt) C_j \lt( \mu_c \rt) 
              \langle \bilr{ij} \rangle \lt( \mu_c \rt) 
+ \cloc \lt( \mu_c \rt) \langle \oloc \rangle \lt( \mu_c \rt)
\;\;  }
\no \\
&&  \hspace{2ex}   = \; 
\frac{1}{8 \pi^2} \cll{ct} \lt( \mu_c \rt) \langle \oll \rangle
   \lt( \mu_c \rt).
\label{matchct}
\end{eqnarray}
$\cloc(\mu_c)$ is already nonzero in the LO due to admixtures from
$C_2$.  Recalling the inverse power of $g^2$ in the definition 
\eq{defqseven} of $\oloc$ one identifies the LO in \eq{matchct} with 
the order $\as^{-1}$. Hence $\cll{ct}$ receives a contribution
\begin{eqnarray}
\cll{ct}\left(\mu_c\right) &=&
\frac{\mc\left(\mu_c\right)}{2} \frac{4\pi}{\as\left(\mu_c\right)}
\cloc\left(\mu_c\right)
\qquad\qquad \mbox{in LO}.
\label{matchCct7}
\end{eqnarray}

Since the double insertion
diagrams {\fig{fig:double-cc-lo}} and {\fig{fig:double-peng-lo}} as
well as their Wilson coefficients are of order $\as^0$, they first
contribute to order $\as^0$, i.e.\ in the NLO.  Hence \eq{matchCct7}
is fully sufficient for the LO matching.  For the NLO matching we
define coefficients $r_{ij,S2}$ by
\begin{eqnarray}
\left\langle\mathcal{R}_{ij}\left(\mu\right)\right\rangle^{(0)}
&=&
\frac{\mc\left(\mu\right)}{16\pi^2} \, 2 \, r_{ij,S2}\left(\mu\right)
\left\langle\oll\right\rangle^{(0)}.
\label{DefRmatch}
\end{eqnarray}
Then the NLO version of $\cll{ct}$ reads
\begin{eqnarray}
\cll{ct}\left(\mu_c\right) &=&
\mc\left(\mu_c\right) \left[\frac{1}{2}
\frac{4\pi}{\as\left(\mu_c\right)} \cloc\left(\mu_c\right)
+ \sum_{i=+,-} \sum_{j=1}^{6} r_{ij,S2}\left(\mu_c\right)
C_i\left(\mu_c\right) C_j\left(\mu_c\right) \right].
\no \\
&&
\label{cllNLOct}
\end{eqnarray}

The coefficients $r_{ij,S2}\left(\mu_c\right)$ in \eq{DefRmatch} are
given by the finite parts of the diagrams in {\fig{fig:double-cc-lo}}
and {\fig{fig:double-peng-lo}}.  We find:
\begin{eqnarray}
r_{ij,S2}\left(\mu_c\right)
&=&
\left\{
\begin{array}{l!{\hspace{1em}\mbox{for}}>{j=}l}
\displaystyle
\left[-4\ln\frac{m_c(\mu_c)}{\mu_c} -1\right]
\tau_{ij} & 1,2, \\[2.8ex]
\displaystyle
\left[-8\ln\frac{m_c(\mu_c)}{\mu_c} -4\right] \tau_{ij} & 3,4, \\[2.8ex]
\displaystyle
\left[8\ln\frac{m_c(\mu_c)}{\mu_c} +4\right] \tau_{ij} & 5,6,
\end{array}
\right.
\label{Rmatch}
\end{eqnarray}
where the $\tau_{ij}$'s denote the colour factors
\begin{eqnarray}
\begin{array}{l}
\displaystyle
\tau_{\pm1} = \tau_{\pm3} = \tau_{\pm5} = \frac{1 \pm N}{2}, \\
\displaystyle
\tau_{+2} = \tau_{+4} = \tau_{+6} = 1, \\
\displaystyle
\tau_{-2} = \tau_{-4} = \tau_{-6} = 0.
\end{array}
\label{DefTau}
\end{eqnarray}
Note that $r_{ij,S2}$ for $j=1,2$ depends on the definition of the
evanescent operator $E_1\left[\oloc\right]$ (cf.~\eq{rsch}). As usual 
\eq{Rmatch} only holds in the NDR scheme. 

If one switches off the RG summation by expanding $\as(\mu_{tW})$ and
$\as (\mu_b)$ contained in the LO Wilson coefficient $\cloc
(\mu_c)$ in \eq{matchCct7} around $\as (\mu_c)$, one finds
\begin{eqnarray}
 \cloc (\mu_c)&=& \frac{\as \lt( \mu_c \rt)}{4 \pi} 
     4 \, \mc \lt( \mu_c \rt) \ln \frac{\mu_{tW}^2}{\mu_c^2} , 
\label{explo}
\end{eqnarray}
so that the unfamiliar factor $1/\as (\mu_c)$ in \eq{matchCct7}
cancels. After inserting \eq{explo} into \eq{matchCct7} and the result
into \eq{lags2c} the thereby expanded $-\langle \leo \rangle $
reproduces the large logarithm of $S(x_c,x_t)$ in $i \wt{G}^{ct}$
(cf.\ \eq{sxcxt} and \eq{gct}). This logarithm appears as $\ln \lt(
\mu_c^2/\mu_{tW}^2 \rt) $. If one likewise expands the NLO coefficient
in \eq{cllNLOct}, one finds the NLO part $F(x_t)$ of \eq{sxcxt} and 
the two small logarithms $\ln \lt( \mu_{tW}^2/M_W^2 \rt) $ and 
$\ln \lt( \mc /\mu_c^2 \rt) $, which are needed to complete the 
large logarithm of the LO result to $\ln x_c$. This shows how the 
dependence on the matching scales $\mu_c$ and $\mu_{tW}$ cancels to 
the calculated order.

Let us now shortly discuss the matching for the two remaining cases,
i.e.\ the determination of $\cll{c}$ and $\cll{t}$ in \eq{lags2c}.
The latter case is particularly simple: The corresponding term in the
lagrangians \eq{lags2} and \eq{lags2c} is equivalent, the only effect
for $\cll{t}$ is the transition to the three-quark running of $\as$.
In the case of $\cll{c}$ only the \dsone\ operators contribute above
$\mu_c$, therefore we define coefficients $d_{ij,S2}$ to parametrize
the matching:
\begin{eqnarray}
\left\langle\mathcal{O}_{ij}\left(\mu\right)\right\rangle &=&
\frac{\mc\left(\mu\right)}{16\pi^2} \, 2 \, d_{ij,S2}\left(\mu\right)
\left\langle\oll\left(\mu\right)\right\rangle.
\label{DefDmatch}
\end{eqnarray}
Since there is no large logarithm in
$\left\langle\mathcal{O}_{ij}\left(\mu\right)\right\rangle^{(0)}$, the
LO matching is performed from the finite parts of the diagrams in
{\fig{fig:double-cc-lo}} and {\fig{loc}}.
In the NLO the finite parts of {\fig{fig:double-cc-nlo}} and
{\fig{fig:ds2-1}} are needed.  Expanding $d_{ij,S2}$ in the usual way
in $\as$ and calculating the required diagrams, we find 
\begin{eqnarray}
d_{ij,S2}^{(0)} &=& \tau_{ij} \qquad \qquad \mbox{in LO}
\label{DmatchLO}
\end{eqnarray}
and 
\begin{eqnarray}
\!\!
d_{ij,S2}^{(1)}\left(\mu\right) \!\! &=& \!\! \left\{
\begin{array}{l}
\tau_{++}6\left(1-N\right)\ln\frac{m_c (\mu)}{\mu}
	+\frac{39-33N-9N^2+3N^3}{4N}
	+\pi^2 \frac{-6+6N+N^2-N^3}{12N} \\[0.3ex]
\hfill\mbox{for $(i,j) = (+,+)$~~} \\[0.3ex]
\tau_{+-}6\left(-1-N\right)\ln\frac{m_c(\mu)}{\mu}
	+\frac{13-13N+3N^2-3N^3}{4N}
	+\pi^2 \frac{-2+4N-3N^2+N^3}{12N} \\[0.3ex]
\hfill\mbox{for $(i,j) = (+,-),(-,+)$~~} \\[0.3ex]
\tau_{--}6\left(-3-N\right)\ln\frac{m_c(\mu)}{\mu}
	+\frac{-13+3N+7N^2+3N^3}{4N}
	+\pi^2 \frac{2-6N+5N^2-N^3}{12N} \\[0.3ex]
\hfill\mbox{for $(i,j) = (-,-)$~~}
\end{array}
\right.
\label{DmatchNLO}
\end{eqnarray}
in the NLO \cite{hn1}.
Here the color factors read
\begin{eqnarray}
\tau_{++} = \frac{N+3}{4}, \hspace{3em}
\tau_{+-} = \tau_{-+} = -\frac{N-1}{4}, \hspace{3em}
\tau_{--} = \frac{N-1}{4}.
\label{DefTau1}
\end{eqnarray}
With \eq{GreenMixHc} and (\ref{DefDmatch}-\ref{DmatchNLO}) the NLO
Wilson coefficient is found as 
\begin{eqnarray}
\cll{c} \lt( \mu_c \rt) &=& \mc (\mu_c) \, \sum_{i,j=+,-} 
       C_i \lt( \mu_c \rt) 
       C_j \lt( \mu_c \rt) \lt[ d_{ij,S2}^{(0)} + \frac{\as (\mu_c)}{4
       \pi} d_{ij,S2}^{(1)} \left(\mu_c \right) \rt] .
\label{cllNLOc}
\end{eqnarray} 

\subsection{RG in the Effective Three-Quark Theory}\label{sect:rg3}

The lagrangian \eq{lags2c} valid below the renormalization scale
$\mu_c$ only contains the single physical operator $\oll$.  The
evolution of the three Wilson coefficients $\cll{j}$, $j=c,t,ct$,
for $\mu\leq \mu_c$ is therefore equal and reads 
\begin{eqnarray}
\cll{j}\left(\mu\right) &=&
\cll{j}\left(\mu_c\right)
\left[\frac{\as\left(\mu_c\right)}{\as\left(\mu\right)}\right]^{d_{+}^{[3]}}
\left(1-J_{+}^{[3]}
	\frac{\as\left(\mu_c\right)-\as\left(\mu\right)}{4\pi}\right),
\label{EvolS2}
\end{eqnarray}
where $d_{+}^{[3]}$ and $J_{+}^{[3]}$ are the RG quantities for three
active flavours defined in \eq{defpm}.  The RG evolutions of $\oll$
and $Q_+$ are equal (see {\apx{apx:ds2anom}}, {\eq{DefAnomS2}}).

Finally we can express the NLO $\eta_i$'s in \eq{s2} in terms of the
coefficients:
\begin{subequations}
\label{ResEta}
\begin{eqnarray}
\!\!\!\!\!\!\!\eta_1 &=&
\frac{1}{\mc \left(\mu_c\right)} \cll{c}\left(\mu_c\right)
\left[\as\left(\mu_c\right)\right]^{d_{+}^{[3]}}
\left(1-J_{+}^{[3]}
	\frac{\as\left(\mu_c\right)}{4\pi}\right),
\label{ResEta1}
\\
\!\!\!\!\!\!\!\eta_2 &=&
\frac{1}{\mw S\left(x_t\left(\mu_c\right)\right)} 
             \cll{t}\left(\mu_c\right)
\left[\as\left(\mu_c\right)\right]^{d_{+}^{[3]}}
\left(1-J_{+}^{[3]}\frac{\as\left(\mu_c\right)}{4\pi}\right),
\label{ResEta2}
\\
\!\!\!\!\!\!\!\eta_3 &=& 
\frac{1}{2 \mw S\left(x_c\left(\mu_c\right),
                      x_t\left(\mu_{tW}\right)\right)} 
          \cll{ct}\left(\mu_c\right)
\left[\as\left(\mu_c\right)\right]^{d_{+}^{[3]}}
\left(1-J_{+}^{[3]}\frac{\as\left(\mu_c\right)}{4\pi}\right) \!.
\label{ResEta3}
\end{eqnarray}
\end{subequations}
The $\mu$-dependence present in \eq{EvolS2} is absorbed into
$b\left(\mu\right)$, which equals
\begin{eqnarray}
b\left(\mu\right) &=&
\left[\as\left(\mu \right)\right]^{-d_{+}^{[3]}}
\left(1+J_{+}^{[3]}\frac{\as\left(\mu\right)}{4\pi}\right)
\label{DefBmu}
\end{eqnarray}
in the NLO. 

The $\eta_i$'s defined in \eq{ResEta} are scheme independent except
that they depend on the definition of the quark masses.  We have
adopted the convention of
\cite{bjw,hn1} that the running masses in \eq{ResEta} are defined at
the scale at which they are integrated out, i.e.\
$m_c=m_c\left(\mu_c\right)$, $m_t=m_t\left(\mu_{tW}\right)$.  Whenever
the $\eta_i$'s are defined such that they multiply
$S\left(x_c\left(m_c\right)\right)$,
$S\left(x_t\left(m_t\right)\right)$ and
$S\left(x_c\left(m_c\right),x_t\left(m_t\right)\right)$ in the
effective hamiltonian \eq{s2}, we mark them with a star:
$\eta_i^{\star}$. For example 
\begin{eqnarray}
\eta_3^\star \; S\left(x_c\left(m_c\right), x_t\left(m_t\right) \right)
&=& \eta_3 \; S\left( x_c\left(\mu_c\right), x_t\left(\mu_{tW}\right) \right).
\label{DefEta3s}
\end{eqnarray}
We will further use
\begin{eqnarray}
m_i^\star=m_i\left(m_i\right)
&\hspace{2em}\mbox{and}\hspace{2em}&
x_i^\star = x_i (m_i)
\hspace{3em} \mbox{for}\quad i=c,t.
\label{mStar}
\end{eqnarray}
The result for $b(\mu)$ in \eq{DefBmu}, however, is scheme dependent
through $J_+^{[3]}$. The scheme and scale dependence of $b(\mu)$ must
cancel with that in the hadronic matrix element of $\oll$.
%
%

\section{The Final Result} 
\label{sect:final}

In this section we summarize the result and sketch our checks of our
NLO calculation of $\eta_3$.  Further we give an approximate formula
for quick implementations of $\eta_3$ in phenomenological programs.
We close the section with the NLO expressions for $\eta_1$ and
$\eta_2$.

\subsection{The Final Result for $\eta_3$ in the NLO}
\label{sect:final-eta3}

Combining \eq{ResEta3} and \eq{cllNLOct} we obtain
\begin{eqnarray}
\eta_3 &=&
\frac{x_c \left(\mu_c\right)}{2
S\left(x_c\left(\mu_c\right),x_t\left(\mu_{tW}\right)\right)}
\as\left(\mu_c\right)^{d_{+}^{[3]}}
\left(1 - J_{+}^{[3]}\frac{\as\left(\mu_c\right)}{4\pi}\right)
\no \\
&& \cdot
\left[
\frac{2\pi}{\as\left(\mu_c\right)} \cloc\left(\mu_c\right)
+\sum_{i=+,-} \sum_{j=1}^{6} r_{ij,S2}\left(\mu_c\right)
C_i\left(\mu_c\right) C_j\left(\mu_c\right)
\right].
\label{SummaryResEta3}
\end{eqnarray}
The Wilson coefficient functions at the renormalization scale $\mu_c$,
which are needed here, are obtained from those at the scale $\mu_{tW}$
by 
\begin{subequations}
\label{SummaryEvol}
\begin{eqnarray}
\cloc\left(\mu_c\right) &=&
C_+\left(\mu_c\right) D_7\left(\mu_c\right)
+ C_-\left(\mu_c\right) D_8\left(\mu_c\right),
\label{SummaryC7}
\\[1.2ex]
C_j \lt( \mu_c \rt) &=& 
   D_j \lt( \mu_c \rt) \qquad \qquad \qquad \qquad 
           \mbox{for $j=1,\ldots,6$}, \\[1.2ex]
C_{\pm}\left(\mu_c\right) &=&
\left[\frac{\as\left(\mu_b\right)}{\as\left(\mu_c\right)}\right]
	^{d_{\pm}^{[4]}}
\left[\frac{\as\left(\mu_{tW}\right)}{\as\left(\mu_b\right)}\right]
	^{d_{\pm}^{[5]}}
\left(1-J_{\pm}^{[4]}
	\frac{\as\left(\mu_b\right)-\as\left(\mu_c\right)}{4\pi}\right.
\no \\
&& \hspace{15ex}
\left. -J_{\pm}^{[5]}
	\frac{\as\left(\mu_{tW}\right)-\as\left(\mu_b\right)}{4\pi}\right)
C_{\pm}\left(\mu_{tW}\right),
\label{SummaryEvolPM}
\\[2.2ex]
\vec{D}\left(\mu_c\right) &=&
\left(1+\frac{\as\left(\mu_c\right)}{4\pi}\widehat{J}^{[4]}\right)
\exp\left[\widehat{d}^{[4]}\cdot
	\ln\frac{\as\left(\mu_b\right)}{\as\left(\mu_c\right)}\right]
\no \\
&& \cdot
\left(1+\frac{\as\left(\mu_b\right)}{4\pi}
	\left( \mbox{$\displaystyle\delta \widehat{r}$}^T\left(\mu_b\right)
              +\mbox{$\displaystyle\widehat{J}^{[5]}$}
              -\mbox{$\displaystyle\widehat{J}^{[4]}$}
         \right) \right)
\no \\
&& \cdot
\exp\left[\widehat{d}^{[5]}\cdot
	\ln\frac{\as\left(\mu_{tW}\right)}{\as\left(\mu_b\right)}\right]
\left(1-\frac{\as\left(\mu_{tW}\right)}{4\pi}\widehat{J}^{[5]}\right)
\cdot
\vec{D}\left(\mu_{tW}\right),
\label{SummaryEvol8}
\end{eqnarray}
\end{subequations}
and $\vec{D}(\mu_{tW})$ is obtained from the initial conditions
$C_{\pm}\left(\mu_{tW}\right)$, $\vec{C}\left(\mu_{tW}\right)$ and
$\cloc\left(\mu_{tW}\right)$ with the help of
\eq{RGdouble1matrix8vec}.  Further $\cloc$ is split into $\cloc^{\pm}$
according to {\eq{clocdec}}. The matrices
$\widehat{d}=\widehat{d}^{[f]}$ and $\widehat{J}=\widehat{J}^{[f]}$
encode the 8$\times$8 anomalous dimension matrix $\widehat{\gamma}$
defined in {\eq{RGdouble1matrix8vec}}:
\begin{eqnarray}
\widehat{d} = \frac{\mbox{$\displaystyle\widehat{\gamma}^{(0)}$}^T}{2 \beta_0},
\hspace{3em} &&
\widehat{J} + \left[\widehat{d}, \widehat{J}\right] =
-\frac{\mbox{$\displaystyle\widehat{\gamma}^{\left(1\right)}$}^T}{2 \beta_0}
+\frac{\beta_1}{\beta_0} \widehat{d}
\label{defdhat}
\end{eqnarray}
in analogy with {\eq{EvolSingleLO}} and {\eq{DefJ}}.  

We emphasize that one should consistently remove terms of order 
$\as^2$ in \eq{SummaryEvol} and terms of order 
$\as $ in \eq{SummaryResEta3}, because they do not belong to
the NLO. 

In {\tab{tab:initial}} we summarize the equations, in
which the initial conditions for the Wilson coefficients defined at
the renormalization scale $\mu_{tW}$ as well as the other ingredients
of {\eq{SummaryResEta3}} and {\eq{SummaryEvol}} can be found.

Finally $\eta_3^\star$ is obtained from $\eta_3$ in \eq{SummaryResEta3}
by 
\begin{eqnarray}
\eta_3^\star &=& \eta_3 \,
  \frac{S\left(x_c\left(\mu_c\right),x_t\left(\mu_{tW}\right)\right)}{
        S\left( x_c^\star,x_t^\star \right)} 
\label{SummaryResEta3*} ,
\end{eqnarray}
with $x_i^\star$ defined in \eq{mStar}.  The exact result for
$\eta_3^\star$ is independent of $\mu_c$, $\mu_b$ and $\mu_{tW}$. The
dependence of our NLO result on these scales will serve as an error
estimate in sect.~\ref{sect:num}.

\begin{ntable}
\begin{center}
\begin{tabular}{c*{6}{|c}}
$\vec{D}\left(\mu_{tW}\right)$ & $C_{\pm}\left(\mu_{tW}\right)$ &
$\vec{C}\left(\mu_{tW}\right)$ & $\cloc\left(\mu_{tW}\right)$ &
$r_{ij,S2}\left(\mu_{c}\right)$ & $S\left(x_c,x_t\right)$ &
$\widehat{d}$, $\widehat{J}$
\\
\hline
\eq{RGdouble1matrix8vec} & 
\eq{ds1diaginit}, \eq{ds1diaginitB} &
\eq{ds1initial}, \eq{ds1initials} &
\eq{clocInit}, \eq{deff} &
\eq{Rmatch}, \eq{DefTau}   &
\eq{sxcxt}, \eq{deff} &
\eq{defdhat}
\end{tabular}
\vspace{1ex}\par
\begin{tabular}{c*{8}{|c}}
$\widehat{\gamma}$ &
$\gbi{\pm}^{(0)}$, $\gbi{\pm}^{(1)}$ & 
$d_{\pm},J_{\pm}$ &
$\gamma^{(0)}_\pm$, $\gamma^{(1)}_\pm$ &
$\gamma^{(0)}$, $\gamma^{(1)}$ &
$\gloc^{(0)}$, $\gloc^{(1)}$ &
$\delta \widehat{r}$ &
$\delta r$ & 
$r^{[f]}$
\\
\hline
\eq{RGdouble1matrix8vec} &
\eq{ResAnomTens} &
\eq{defpm} &
\eq{anomQpm} &
\eq{ds1anom} &
\eq{AnomQ7exp} &
\eq{Defr8} &
\eq{defdr} &
\eq{ds1r} 
\end{tabular}
\vspace{1ex}\par
\begin{tabular}{c*{1}{|c}}
$\gamma_m^{(0)}$, $\gamma_m^{(1)}$ &
$\beta_0$, $\beta_1$
\\
\hline
\eq{MassAnom} &
\eq{QCDbetaNLO}
\end{tabular}
\end{center}
\caption{\slshape
The equations in which the ingredients of
{\protect\eq{SummaryResEta3}} and {\protect\eq{SummaryEvol}} are
defined.}
\label{tab:initial}
\end{ntable}

\subsection{Analytical Checks}
\label{sect:final-check}

We have performed several checks of our NLO result
in (\ref{SummaryResEta3}-\ref{SummaryResEta3*})
\begin{enumerate}
\renewcommand{\labelenumi}{\roman{enumi})}
\item 
The NLO anomalous dimension tensor $\gbi{\pm j}$ {\eq{ResAnomTens1}}
has been found independent of the infrared structure of the diagrams
in \fig{fig:double-cc-nlo} and {\fig{fig:double-peng-nlo}}, i.e.\ of
the small quark masses $m_s$, $m_d$ used as infrared regulators.
\item 
We have kept the gluon gauge parameter $\xi$ arbitrary. It has
vanished from $\gbi{\pm j}$ after adding the contributions of the
diagrams with their correct combinatorial weight.  Further we have
checked that $\xi$ vanishes from the $1/\eps^2$ terms in \eq{DivZbi1}
after subtracting the $\xi$-dependent term involving
$Z^{(1)}_{\psi,1}$. 
\item 
Another check has been provided by the well-known fact that the
$1/\eps^2$-part of the two-loop renormalization constant is related to
the one-loop $Z$-factors or equivalently to the LO anomalous
dimensions involved (see e.g.\ \cite{bw}).  We have confirmed the
corresponding relation for our case:
\begin{eqnarray}
\zbil{-1,(2)}{2}{\pm j} &=&
     \frac{1}{8}
       \left[ \gloc^{(0)} - 2 \beta_0 + \g_\pm^{(0)}  \right]
     \gbi{\pm j}^{(0)} + \frac{1}{8}  \sum_{j^\prime=1}^6 
    \g_{jj'}^{(0)}  \gbi{\pm j'}^{(0)}.
\end{eqnarray}
\item
$\ln (m_c/\mu)/\eps$-terms have disappeared from the sum of two-loop
diagrams and counterterm diagrams.
\item 
The dependences of the final result for $\eta_3^\star$ on the matching
scales $\mu_{tW}$, $\mu_b$ and $\mu_c$ cancel to order $\as$.
\item \label{item:FullCheck}
If one expands the final result in powers of $\as$, one recovers the
terms proportional to $\as^0 \ln^1 x_c$, $\as^0 \ln^0 x_c$, $\as^1
\ln^2 x_c$, $\as^1 \ln^1 x_c$ of the result without RG improvement in
\eq{gct} and {\eq{factgct}}.  The term proportional to $\as^1 \ln^0 x_c$
cannot be obtained, because it belongs to the NNLO, see
\tab{tab:rglogs}.  
The expansion of $\eta_3$ in terms of $\as (\mu_c)$ reveals how the 
coefficients of the leading and next-to-leading logarithms of $k^{ct}$
in \eq{h-1-ct} are related to the ingredients of the RG calculation:
The nonvanishing contributions to the LO term proportional to 
$x_c \ln^2 x_c$ are 
\begin{eqnarray}
\frac{1}{16} \g_m^{(0)} \gbi{+2}^{(0)} + 
\frac{1}{32} \sum_{j=1}^6 \lt( \gbi{+j}^{(0)}+\gbi{-j}^{(0)}\rt) \gzero_{2j}
+ \frac{1}{16} \g_{+}^{(0)} \gbi{+2}^{(0)} . \label{exp}
\end{eqnarray}
Note that only the second row of $\gzero$ appears here, the remaining
part only contributes to higher orders in $\as$.  Likewise the
coefficient of $x_c \ln x_c$ is found to involve the same LO
quantities as {\eq{exp}}, the NLO matching corrections $r_{\pm j,S2}
(\mu_c)$, $F(x_t)$, $B_{+}$, $\wt{E}(x_t)$ (see \eq{DefRmatch},
{\eq{deff}}, {\eq{ds1diaginit}} and \eq{ds1initial}) and the elements
$\gbi{\pm2}^{(1)}$ of the NLO anomalous dimension tensor. In addition
all terms related to the penguin operators sum to zero in the LO and
NLO part of $k^{ct}$, e.g.\ in \eq{exp} one finds
\begin{eqnarray}
\sum_{j=3}^6 \lt( \gbi{+j}^{(0)}+\gbi{-j}^{(0)}\rt) \gzero_{2j}=0.
\no
\end{eqnarray}
\item 
The initial condition for $\cloc$ in \eq{clocInit} as well as the
anomalous dimension tensor $\gbi{\pm j}$ in \eq{ResAnomTens1} depend
on the definition of the evanescent operators \eq{StdEvas}. We have
checked in sect.~\ref {sect:evas} that this dependence is in
accordance with the theorems of \cite{hn2}, so that the final result
is independent of the choice of the evanescent operators.
\end{enumerate}
We remark that all these checks are not sensitive to the results of 
diagrams $\mathsf{P_{11}}$-$\mathsf{P_{15}}$, which are gauge
independent and have no $1/\eps^2$-divergences.

\subsection{An Approximate Formula for $\eta_3$}
\label{sect:final-approx}

Since the numerical implementation of $\eta_3$ in \eq{SummaryResEta3} 
and {\eq{SummaryEvol}} is quite cumbersome for phenomenological
studies, we present a simple approximate formula for this
quantity.

Such an approximate formula is motivated by the following
observations, which are derived from the numerical study in
{\sect{sect:num}}:
\begin{enumerate}
\renewcommand{\labelenumi}{\roman{enumi})}
\item Variation of $\mu_b$ in the interval $\mu_c \leq \mu_b \leq
\mu_{tW}$ changes the result for $\eta_3$ on the permille level.
\item The contribution of the penguin operators $Q_3$, \ldots, $Q_6$
is of the order of 1\%.  This is so, because penguin effects
enter $G^{ct}$ in the order $\as^2$ rather than $\as$ 
(see vi) in sect.~\ref{sect:final-check}).
\end{enumerate}
We can therefore simply switch off the penguin operators and further
set $\mu_b \equiv \mu_{tW} = O(M_W,m_t)$, thereby neglecting any
effects from the effective five flavour theory.  This yields
\begin{eqnarray}
\eta_3 &=&
2\pi
\frac{x_c\left(\mu_c\right)}
	{S\left(x_c\left(\mu_c\right),x_t\left(\mu_{tW}\right)\right)}
\as\left(\mu_c\right)^{\frac{2}{9}} \cdot
\no \\
& &
\Biggl[ \frac{1}{\as\left(\mu_c\right)}
\left(-\frac{9}{7} A^{++} - \frac{6}{11} A^{+-} + \frac{3}{29} A^{--}
	+ \frac{3858}{2233} \tilde{A} \right)
\left(1-\frac{\as\left(\mu_c\right)}{4\pi} \frac{307}{162}\right)
\no \\
& &
+ \frac{1}{4\pi} \biggl[
\frac{262497}{17500} A^{++} - \frac{246}{625} A^{+-} +
\frac{1108657}{652500} A^{--} - \frac{277133}{25375} \tilde{A}
\no \\
& &\hspace{2em}
+ A \left(-\frac{21093}{4375} A^{++} + \frac{13331}{6875} A^{+-} -
\frac{20362}{18125} A^{--} - \frac{3462208}{2512125} \tilde{A} \right)
\no \\
& &\hspace{2em}
+ A \ln\frac{\mu_{tW}^2}{M_W^2} \left(-\frac{36}{7} A^{++} +
\frac{12}{11} A^{+-} - \frac{24}{29} A^{--} + \frac{10896}{2233}
\tilde{A} \right)
\no \\
& &\hspace{2em}
+ \left(-\frac{1}{2} - \ln\frac{m_c^2\left(\mu_c\right)}{\mu_c^2}
\right) \left(3 A^{++} - 2 A^{+-} + A^{--} \right)
\no \\
& &\hspace{2em}
+ A \tilde{A} \Bigl( -2 \ln\frac{\mu_{tW}^2}{M_W^2} - \frac{3}{2}
\frac{x_t\left(\mu_{tW}\right)}{1-x_t\left(\mu_{tW}\right)}
\no \\
& &\hspace{4em}
+ \left(2 - \frac{3}{2} \frac{x_t\left(\mu_{tW}\right)^2}
	{\left(1-x_t\left(\mu_{tW}\right)\right)^2} \right)
\ln x_t\left(\mu_{tW}\right) + 1 \Bigr)
\biggr] \Biggr]  ,
\label{eta3apx}
\end{eqnarray}
where
\begin{eqnarray}
& &
A = \frac{\as\left(\mu_{tW}\right)}{\as\left(\mu_c\right)},
\hspace{3em} \tilde{A} = A^{\frac{1}{5}},
\no \\
& & A^{++} = A^{\frac{12}{25}},
\hspace{3em} A^{+-} = A^{-\frac{6}{25}},
\hspace{3em} A^{--} = A^{-\frac{24}{25}}.
\label{eta3apxA}
\end{eqnarray}
We have kept the dependence on the scales $\mu_c$ and $\mu_{tW}$
because of their importance for our error estimate in
sect.~\ref{sect:num}. 

The accuracy of \eq{eta3apx} is 1\% with respect to combined
variations of $\mu_c$, $\mu_{tW}$, $m_c$ and $m_t$ in reasonable
intervals of the parameters.  For extremely high values of $\mu_{tW}$
and extremely low values of $\mu_c$ the precision reduces to something
like 2\%.

We can in the same way derive an approximate formula with $\mu_b
\equiv \mu_c = O(m_c)$, thereby performing the RG evolution in an
effective five-flavour theory, but this is considerably less accurate
than \eq{eta3apx} in the whole parameter space.

\subsection{$\eta_1$ and $\eta_2$}\label{sect:final-eta12}

Let us now shortly summarize the results for $\eta_1$ \cite{bjw} and
$\eta_2$ \cite{hn1}.  The former is obtained by combining
{\eq{ResEta1}}, {\eq{cllNLOc}} and {\eq{SummaryEvolPM}}.  The latter
is constructed by evolving the Wilson coefficient $\cll{t}(\mu_{tW})$
in {\eq{cllInit}} down to the scale $\mu_c$ and inserting this into
{\eq{ResEta2}}.  One finds
\begin{subequations}
\label{SummaryResEta12}
\begin{eqnarray}
\eta_1 &=&
\as\left(\mu_c\right)^{d_{+}^{[3]}}
\sum_{i=\pm} \sum_{j=\pm}
\left[\frac{\as\left(\mu_b\right)}{\as\left(\mu_c\right)}\right]
	^{d_{i}^{[4]}+d_{j}^{[4]}}
\left[\frac{\as\left(\mu_{tW}\right)}{\as\left(\mu_b\right)}\right]
	^{d_{i}^{[5]}+d_{j}^{[5]}}
\no \\
&& \cdot
\Biggl\{d_{ij,S2}^{(0)}
   \biggl[1
	+\frac{\as\left(\mu_c\right)}{4\pi}
	\left(J_i^{[4]}+J_j^{[4]}-J_{+}^{[3]}\right)
\no \\
&& \hspace{4em}
	+\frac{\as\left(\mu_b\right)}{4\pi}
	\left(J_i^{[5]}+J_j^{[5]}-J_i^{[4]}-J_j^{[4]}\right)
\no \\
&& \hspace{4em}
	+\frac{\as\left(\mu_{tW}\right)}{4\pi}
        \left(
		-J_i^{[5]}-J_j^{[5]}
		+\ln\frac{\mu_{tW}}{M_W}
                   \left(\gamma^{(0)}_i+\gamma^{(0)}_j\right)
                + B_i + B_j
        \right)
   \biggr]
\no \\
&& \hspace{1em}
   +d_{ij,S2}^{(1)}\left(\mu_c\right) \frac{\as\left(\mu_c\right)}{4\pi}
\Biggr\} ,
\label{SummaryResEta1}
\\
\eta_2 &=&
\as\left(\mu_c\right)^{d_{+}^{[3]}}
\left[\frac{\as\left(\mu_b\right)}{\as\left(\mu_c\right)}\right]
	^{d_{+}^{[4]}}
\left[\frac{\as\left(\mu_{tW}\right)}{\as\left(\mu_b\right)}\right]
	^{d_{+}^{[5]}}
\no \\
&& \cdot
\Biggl\{1
	+\frac{\as\left(\mu_c\right)}{4\pi}
	\left(J_{+}^{[4]}-J_{+}^{[3]}\right)
	+\frac{\as\left(\mu_b\right)}{4\pi}
	\left(J_{+}^{[5]}-J_{+}^{[4]}\right)
\no \\
&& \hspace{2em}
	+\frac{\as\left(\mu_{tW}\right)}{4\pi}
	\left(-J_{+}^{[5]}
		+\frac{k^t\left(x_t\left(\mu_{tW}\right),\mu_{tW}\right)}
                	{S\left(x_t\left(\mu_{tW}\right)\right)}
	\right)
\Biggr\} ,
\label{SummaryResEta2}
\end{eqnarray}
\end{subequations}
where terms not belonging to the LO and NLO have been consistently
removed.  From {\eq{SummaryResEta12}} it is immediately clear that the
masses enter as $m_c\left(\mu_c\right)$ and
$m_t\left(\mu_{tW}\right)$.  Table~\ref{tab:eta12in} refers to the
equations in which the various quantities entering
{\eq{SummaryResEta12}} are defined.

Finally $\eta_1^\star$ and $\eta_2^\star$ are defined analogously to
$\eta_3^\star$ in \eq{SummaryResEta3*}:
\begin{eqnarray}
\eta_1^\star = \eta_1 \, \frac{x_c (\mu_c)}{ x_c^\star }, \qquad && \qquad 
\eta_2^\star = \eta_2 \, 
\frac{S\left( x_t \left(\mu_{tW}\right) \right)}{S\lt( x_t^\star\rt)}.
\end{eqnarray}
\begin{ntable}
\begin{center}
\begin{tabular}{c*{5}{|c}}
$d_{\pm}$, $J_{\pm}$ &
$\gamma_{\pm}^{(0)}$, $\gamma_{\pm}^{(1)}$ &
$B_\pm$ &
$d_{ij,S2}^{(0)}$, $d_{ij,S2}^{(1)}$ &
$S\left(x\right)$ &
$k^t\left(x\right)$ 
\\
\hline
\eq{defpm} &
\eq{anomQpm} &
\eq{ds1initialB} &
(\ref{DmatchLO}-\ref{DefTau1}) &
\eq{sxt} &
\eq{h-1-t} 
\end{tabular}
\end{center}
\caption{\slshape The equations in which the ingredients of
$\eta_1$ and $\eta_2$ are defined.}
\label{tab:eta12in}
\end{ntable}

%

\section{Numerical Results}\label{sect:num}

This section is devoted to the numerical analysis of the results
derived in the preceding sections.  We will present the dependence
of $\eta_3$ and $\eta_3^\star$ on its various physical parameters and on
the renormalization scales at which particles are integrated out.  
Recall the relevant part of the effective low-energy hamiltonian from
{\eq{s2}}:
\begin{eqnarray}
H^{ct}\left(\mu\right) &=&
\frac{\gf^2}{16\pi^2} M_W^2 \, 2 \lambda_c \lambda_t \,
\eta_3 \,
S\left(x_c\left(\mu_c\right),x_t\left(\mu_{tW}\right)\right) \,
b\left(\mu\right) \oll\left(\mu\right)
\no \\
&=&
\frac{\gf^2}{16\pi^2} M_W^2 \, 2 \lambda_c \lambda_t \,
\eta_3^\star \,
S\left(x_c^\star,x_t^\star\right) \,
b\left(\mu\right) \oll\left(\mu\right).
\label{num-eff}
\end{eqnarray}
Here $\eta_3$ and $\eta_3^\star$ depend on the scales $\mu_{tW}$, $\mu_b$
and $\mu_c$.  Further they are functions of the masses $m_t$, $m_c$
and of $\laQCD$.  To establish a starting point let us pick a basic
set of input parameters
\begin{eqnarray}
\begin{array}{*{2}{l}}
m_c(m_c) = \mu_c = 1.3\gev,
&\laMSb = 0.31\gev \quad (\laQCD^\mathrm{LO} = 0.15\gev),
\\
\mu_b = 4.8\gev,
& M_W = 80\gev,
\\
\mu_{tW} = 130\gev,
& m_t(m_t) = 167\gev.
\end{array}
\label{InputParam}
\end{eqnarray}
In the following $\laQCD$ is always understood to be defined with
respect to four active flavours, the corresponding quantities in
effective three- and five flavour-theories are obtained by imposing
continuity on the coupling $\as$ at $\mu_c$ and 
$\mu_b$.\footnote{Threshold corrections appearing for $\mu_q\neq m_q$
are numerically negligible.}

The values of $\laQCD$ quoted in \eq{InputParam} require a comment:
The world average for $\as(M_Z)=0.117$ \cite{beth} corresponds to
$\laMSb=310\mev$ for $\mu_b=4.6\gev$ and to $\laMSb=315\mev$ for
$\mu_b=5.0\gev$.  The LO $\laQCD^{\mathrm{LO}}$, however, differs from
$\laMSb$ by an overall $\mu$-dependent factor.  If one equates the LO
coupling and the NLO \msb\ coupling constant at the scale
$\mu=M_Z=91\gev$, one finds that $\laMSb=310\mev$ corresponds to
$\laQCD^\mathrm{LO}=110\mev$.  If the matching relation is imposed at
the low scale $\mu=1.3\gev$, one finds $\laQCD^\mathrm{LO}=180\mev$.
This shows one contribution to the error bar in LO calculations.

The value for $\eta_3^\star$ corresponding to the set \eq{InputParam}
reads:
\begin{eqnarray}
{\eta_3^\star}^\mathrm{LO} = 0.365,
&\hspace{3em}&
{\eta_3^\star}^\mathrm{NLO} = 0.467.
\label{Num3s}
\end{eqnarray}
Hence the NLO calculation has enhanced $\eta_3^\star$ by 27\%.  From the
difference of 0.102 between the two values in \eq{Num3s} 0.022
originates from the change from the LO to the NLO running $\as$.  The
smallness of this contribution is of course caused by the adjustment
of $\laQCD^\mathrm{LO}$ to fit the NLO running coupling as described
in the previous paragraph.  The explicit $O(\as)$ corrections from the
NLO mixing and matching contribute 0.080.

Let us list the dominant sources of the enhancement: At the initial
scale $\mu_{tW}$ the magnitudes of the coefficients $C_2$, $C_+$ and
$C_-$ are of the order 1, while all other coefficients are almost
negligible. The RG evolution from $\mu_{tW}$ to $\mu_c$ enhances the
coefficient $C_-$ by roughly 75\% because of the negative sign of the
anomalous dimension of $Q_-$, while the coefficient $C_+$ is damped by
25\%.  Now the penguin coefficients $C_3$ to $C_6$ are still
negligible at $\mu=\mu_c$, only $C_1(\mu_c), C_2(\mu_c)$ and
$\cloc(\mu_c)$ are important.  In the matching at $\mu_c$ the
contribution of $C_+ (\mu_c) C_1(\mu_c)$ numerically cancels the one
of $C_- (\mu_c) C_1(\mu_c)$, because the RG damping of the former is
compensated by a larger colour factor $\tau_{+1}=2$ having the
opposite sign of $\tau_{-1}=-1$.  Finally $\cloc(\mu_c)\approx 0.7$
has become large due to the RG admixtures from $C_2$.  Hence only
$C_-(\mu_c) C_2(\mu_c)$, $C_+(\mu_c) C_2(\mu_c)$ and $\cloc(\mu_c)$
are important.  In the LO only $\cloc(\mu_c)$ enters $\eta_3$.  Keeping
only $\cloc(\mu_c)$ in the NLO expression, however, overestimates the
NLO enhancement by a factor of roughly 1.5, because $C_2(\mu_c)$
contributes with a negative sign to $\eta_3$ (for the standard
definition of the evanescent operators in \eq{StdEvasNum}).

Let us further quantify the influence of the penguin operators
$Q_{3,\ldots,6}$: If one neglects them completely, one obtains
${\eta_3^\star}^\mathrm{NLO,np}=0.472$ with the set in
{\eq{InputParam}}, i.e.\ their contribution is of the order of 1\%.
This strong suppression serves as a major motivation for the
derivation of the approximate formula without penguin effects in
{\sect{sect:final-approx}}.

\subsection{Scale Dependence of $\eta_3^\star$}\label{sect:num-scale}

Let us now analyze the scale dependences present in $\eta_3^\star$ in
detail.  In \eq{num-eff} $\eta_3$ multiplies
$S(x_c(\mu_c),x_t(\mu_{tW}))$, which is scale dependent.  Ideally
their product is scale independent.  Therefore $\eta_3^\star$ turns
out to be much more useful in the discussion of scale dependences,
because in {\eq{num-eff}} it multiplies $S(x_c^\star,x_t^\star)$ and
other quantities which are independent of $\mu_c$, $\mu_b$ and
$\mu_{tW}$.  So $\eta_3^\star$ should essentially behave flat with
respect to variations of the scales, any remaining dependence may
serve as a measure of the accuracy of the calculation.

We now have to discuss the scale dependence of $\eta_3^\star$
associated with the variation of the three scales $\mu_{tW}$, $\mu_b$
and $\mu_c$.  The one related to $\mu_b$ appears to be extremely mild.
This is due to the fact that no diagrams containing internal bottom
quarks contribute to the {\dstwo}\ process in order $\as$.  The only
places where $\mu_b$ enters are a) the running of $\as$, b) the NLO
matching condition of the {\dsone}\ penguin Wilson coefficients
{\eq{ds1r}}, and c) the anomalous dimensions of the {\dsone} penguins.
The latter two are strongly suppressed as we have seen already in a
preceding paragraph.

Because the $\mu_b$ dependence of $\eta_3^\star$ is so weak we could
even set $\mu_b \equiv \mu_c = O(m_c)$ or $\mu_b \equiv \mu_{tW} =
O(m_W,m_t)$, as we did for the approximate formula in
{\sect{sect:final-approx}}.  The error introduced by this is of the
order of 0.1--0.2\%.

Now let us turn to the more important cases, $\mu_{tW}$ and $\mu_c$.
First consider the variation of $\eta_3^\star$ with respect to $\mu_{tW}$,
which is displayed in \fig{fig:e3-mutW}.
\begin{nfigure}
\begin{minipage}[t]{\miniwidth}
\centerline{\epsfxsize=\miniwidthplot \epsffile{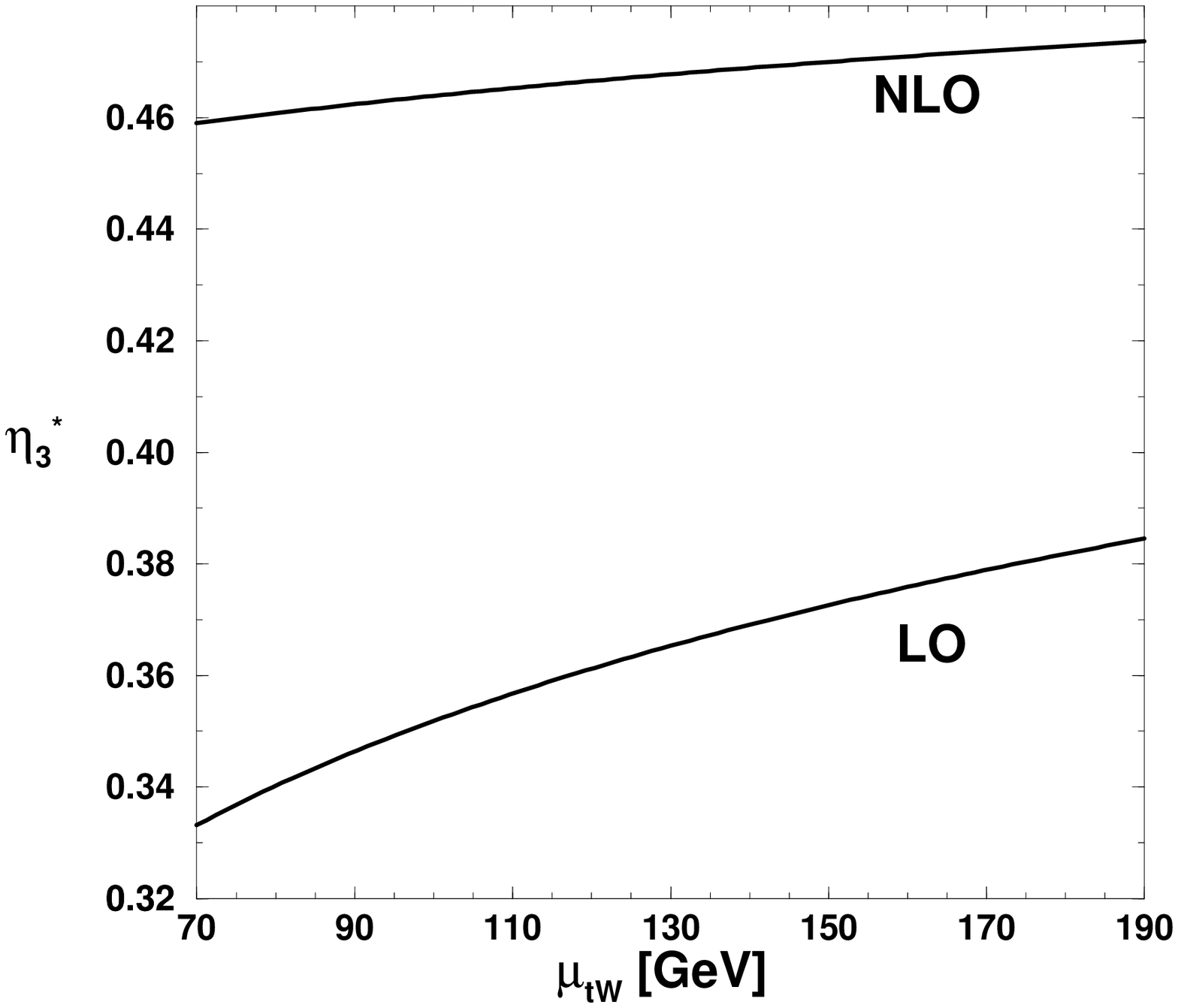}}
\caption{\slshape
The variation of $\eta_3^\star$ in LO and NLO with respect to the scale
$\mu_{tW}$, at which the initial condition is defined.  The other input
parameters are given in \eq{InputParam}.}
\label{fig:e3-mutW}
\end{minipage}
\hfill
\begin{minipage}[t]{\miniwidth}
\centerline{\epsfxsize=\miniwidthplot \epsffile{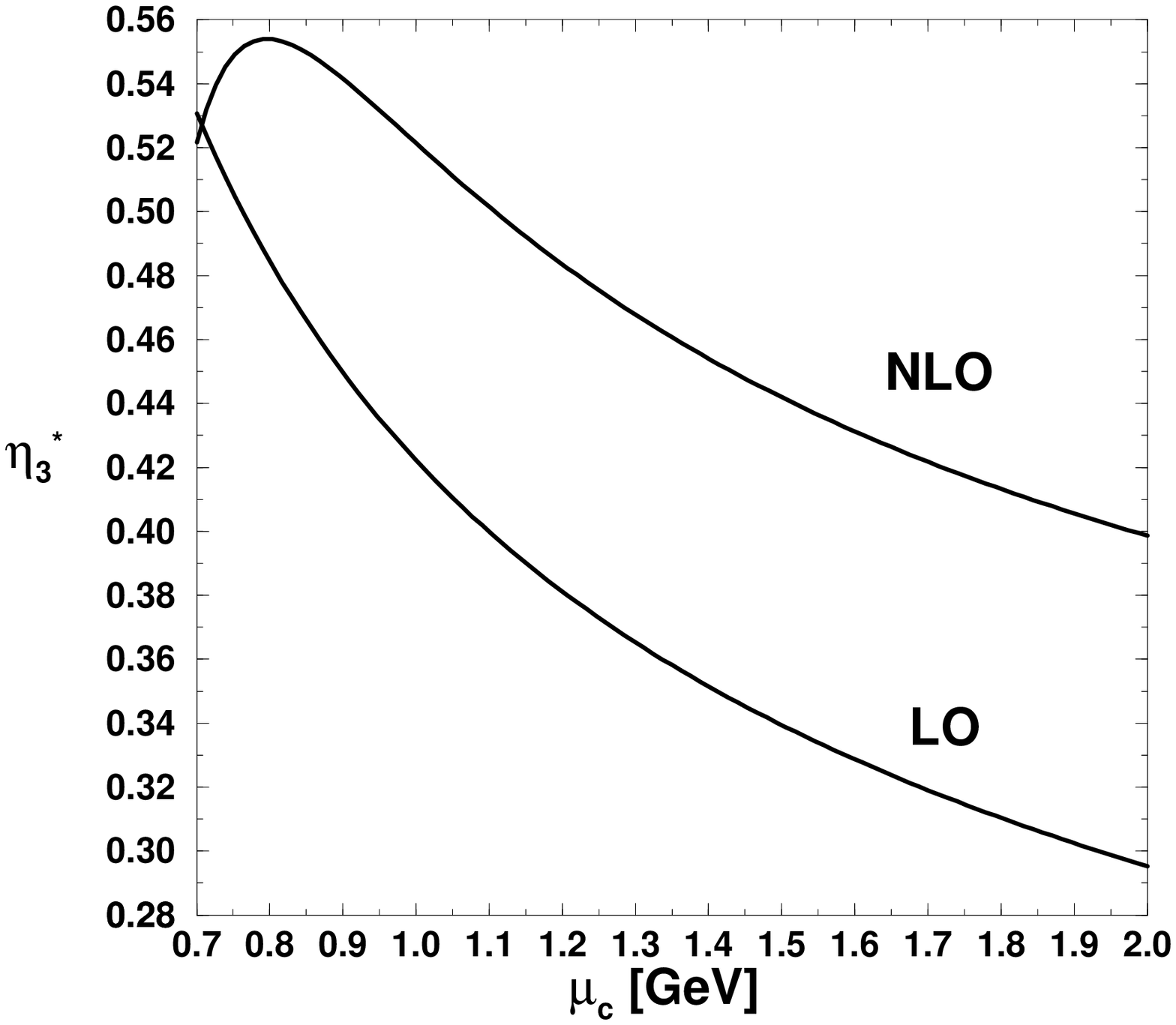}}
\caption{\slshape
The variation of $\eta_3^\star$ in LO and NLO with respect to the scale
$\mu_c$.  The range for the latter is taken unphysically large to
visualize the breakdown of perturbation theory.}
\label{fig:e3-muc}
\end{minipage}
\end{nfigure}
Since at $\mu_{tW}$ the top quark and the W-boson are integrated out
simultaneously, it is natural to choose the interval
$M_W\leq\mu_{tW}\leq m_t$ for the analysis.  In the LO result for
$\eta_3^\star$ we find a sizeable scale dependence of 12\%.  It is almost
totally removed in the NLO, where we obtain a variation of less than
3\% in this interval. This shows that it is very accurate to integrate
out the two heavy particles simultaneously. The strong improvement 
in the NLO is due to the smallness of $\ln x_t$ and vindicates 
our argumentation in sect.~\ref{sect:dort}.

The situation is not so nice in the case of the variation of $\mu_c$,
which is displayed in \fig{fig:e3-muc}.  We have intentionally extended
the range for $\mu_c$ to the unphysical low value of $0.7\gev$ to
visualize the breakdown of perturbation theory.  Varying $\mu_c$
within the interval $1.1\gev\leq\mu_c\leq1.6\gev$ yields
\begin{eqnarray}
0.33 \leq {\eta_3^\star}^\mathrm{LO} \leq 0.40,
&\hspace{3em}&
0.43 \leq {\eta_3^\star}^\mathrm{NLO} \leq 0.50 .
\label{num-muc}
\end{eqnarray}
This corresponds to a reduction of the scale dependence from 20\% to
14\%.  One reason for the poor improvement is the fact that the NLO
running of the mass is stronger than the LO one.

From \eq{num-muc} one realizes that the scale variation is not always
a good measure of the accuracy of the calculation: The central value
of ${\eta_3^\star}^\mathrm{NLO}$ does not lie in the range quoted for
${\eta_3^\star}^\mathrm{LO}$.  Yet in the NLO one may also judge the
contribution of the uncalculated $O(\as^2)$ terms by squaring the
calculated $O(\as)$ corrections.  This leads to the same interval as
quoted in \eq{num-muc}, so that we may consider the given interval as
a realistic estimate of ${\eta_3^\star}$.

\subsection{Dependence of $\eta_3^\star$ on Physical Quantities}
\label{sect:num-phys}

Let us now investigate the dependence of $\eta_3^\star$ on the physical
parameters.  From the smallness of the coefficient $\cloc$ at the
initial scale one expects $\eta_3^\star$ to be almost independent of
$m_t^\star=m_t(m_t)$.  This statement is confirmed numerically, see
{\fig{fig:e3-mt}}, allowing to treat $\eta_3^\star$ as $m_t$-independent
in phenomenological analyses.
\begin{nfigure}
\begin{minipage}[t]{\miniwidth}
\centerline{\epsfxsize=\miniwidthplot \epsffile{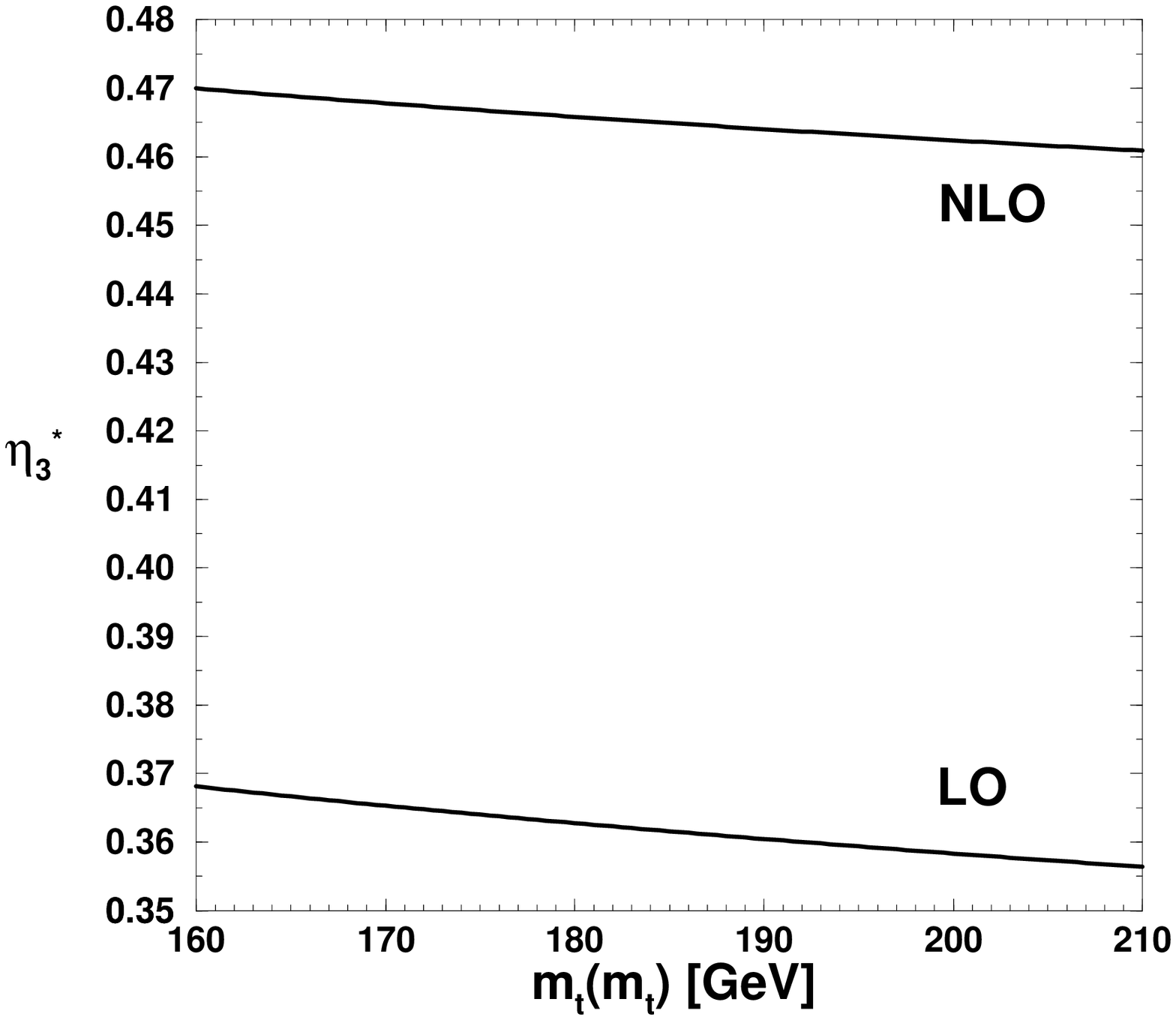}}
\caption{\slshape
The dependence of $\eta_3^\star$ on $m_t(m_t)$ in LO and NLO.  The other
input parameters are given in \eq{InputParam}.}
\label{fig:e3-mt}
\end{minipage}
\hfill
\begin{minipage}[t]{\miniwidth}
\centerline{\epsfxsize=\miniwidthplot \epsffile{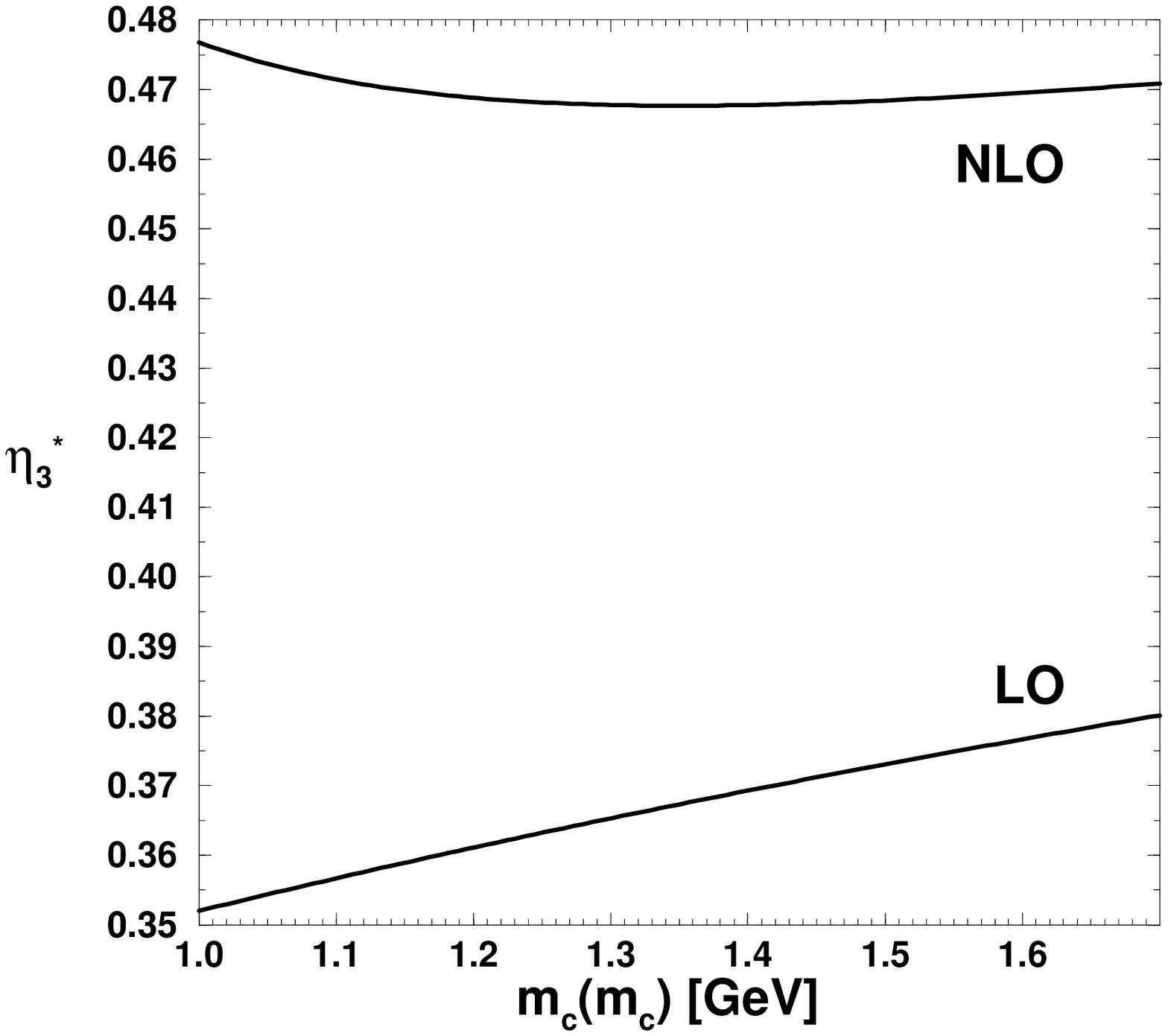}}
\caption{\slshape
The dependence of $\eta_3^\star$ on $m_c(m_c)$ in LO and NLO.}
\label{fig:e3-mc}
\end{minipage}
\end{nfigure}

The LO result for $\eta_3^\star$ depends on $m_c^\star=m_c(m_c)$ sizeably.
Yet this dependence is washed out nearly completely if one looks at
the NLO $\eta_3^\star$, see \fig{fig:e3-mc}.

$H^{\dstwo}$ in \eq{num-eff} and physical observables are entered by
the product $\eta_3^\star S(x_c^\star,x_t^\star)$.  It turns out to be
almost independent of $m_t$, but shows a very pronounced dependence on
$m_c^\star$, due to the dependence of $S(x_c^\star,x_t^\star)$ on $m_c^\star$.

We close this section by a look at the dependence of $\eta_3^\star$ on
$\laQCD$, which is plotted in \fig{fig:e3-laQCD}.  It also turns out
to be very moderate.
\begin{nfigure}
\begin{minipage}[t]{\miniwidth}
\centerline{\epsfxsize=\miniwidthplot \epsffile{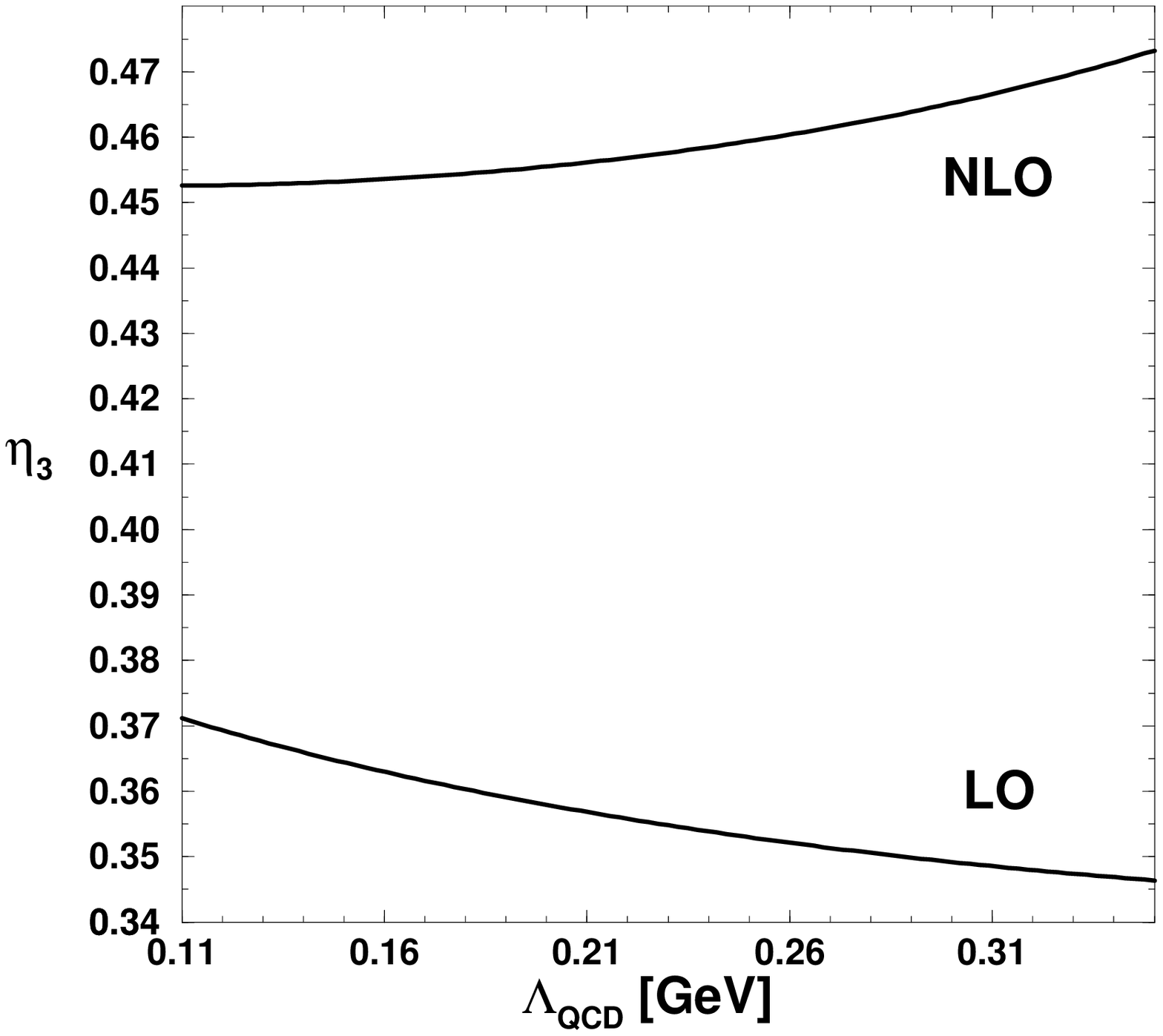}}
\caption{\slshape
The dependence of $\eta_3^\star$ on $\laQCD$ in LO and NLO. Actual values
for $\as (M_Z)$ correspond to $\laQCD^\mathrm{LO} \approx 0.15\gev$
and $\laMSb \approx 0.31\gev$.  The other input parameters are listed
in \eq{InputParam}.}
\label{fig:e3-laQCD}
\end{minipage}
\hfill
\begin{minipage}[t]{\miniwidth}
\centerline{\epsfxsize=\miniwidthplot \epsffile{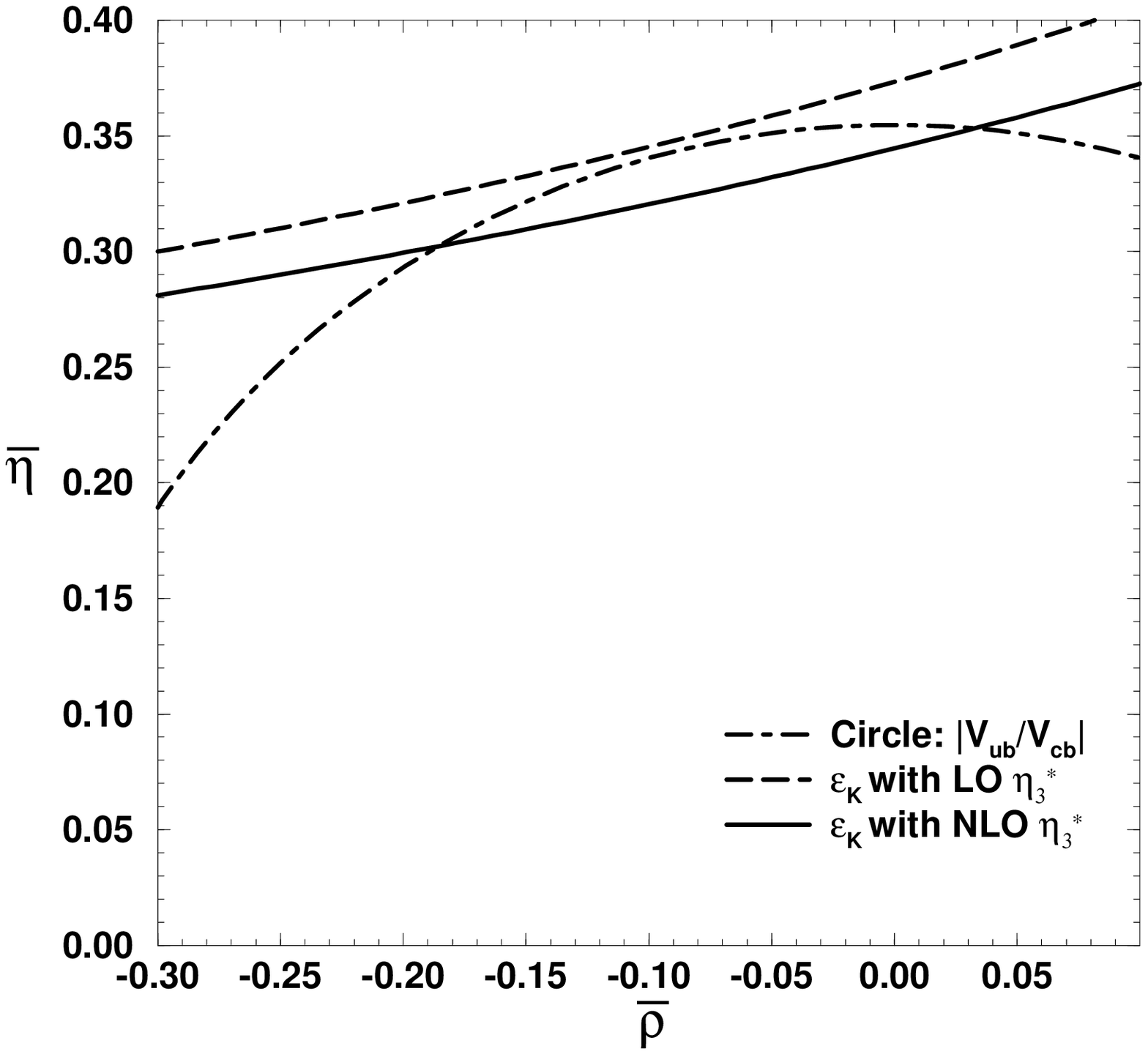}}
\caption{\slshape
The impact of the NLO calculation of $\eta_3^\star$ on the unitarity
triangle.  The parameters in the plot are $m_t(m_t)=165 \gev$,
$|V_{cb}|=0.041$, $|V_{ub}/V_{cb}|=0.08$, $m_c(m_c)=1.3 \gev$ and
$B_K=0.7$.  While the NLO result $\eta_3^\star=0.47$ permits two
solutions for $(\ov{\rho},\ov{\eta})$, there is no intersection, if
the old LO value $\eta_3^{\star \,\mathrm{LO}} =0.37$ is used. }
\label{fig:hyperbola}
\end{minipage}
\end{nfigure}

To summarize the findings of our numerical analysis: $\eta_3^\star$
behaves essentially flat with respect to all physical input parameters
and therefore may be treated as a constant in all phenomenological
applications.  The dominant uncertainty of our estimate of $\eta_3^\star$
is due to the scale variation with respect to $\mu_c$ and we quote as
our final result:
\begin{eqnarray}
{\eta_3^\star}^\mathrm{NLO} &=& 0.47
\raisebox{-0.5ex}{\shortstack[l]{$\scriptstyle+0.03$\\$\scriptstyle-0.04$}}.
\label{num-final}
\end{eqnarray}

\subsection{Results for $\eta_1^\star$ and $\eta_2^\star$}
\label{sect:num-eta12}

The only coefficient which is sensitive to the physical input
parameters is $\eta_1^\star$.  We update the result of \cite{hn1} to
actual values of $\as(M_Z)$ and $m_c^\star$ in Table~\ref{tab:eta12}.
\begin{ntable}
\begin{center}
\begin{tabular}{r|*{4}{r@{}l}}
$m_c^\star$
	&\multicolumn{2}{c}{1.20}
	&\multicolumn{2}{c}{1.25}
	&\multicolumn{2}{c}{1.30}
	&\multicolumn{2}{c}{1.35}
\\
$\as\left(M_Z\right)$ & \\
\hline
 0.111
	& 1.104&\errorpm{0.144}{0.135}
	& 1.070&\errorpm{0.131}{0.123}
	& 1.041&\errorpm{0.120}{0.115}
	& 1.014&\errorpm{0.111}{0.106} \\
 0.114	  	      
	& 1.243&\errorpm{0.211}{0.186}
	& 1.197&\errorpm{0.190}{0.169}
	& 1.157&\errorpm{0.172}{0.156}
	& 1.121&\errorpm{0.157}{0.143} \\
 0.117	  	      
	& 1.431&\errorpm{0.322}{0.264}
	& 1.367&\errorpm{0.284}{0.238}
	& 1.311&\errorpm{0.254}{0.217}
	& 1.261&\errorpm{0.229}{0.197} \\
 0.120	  	      
	& 1.695&\errorpm{0.512}{0.388}
	& 1.601&\errorpm{0.443}{0.345}
	& 1.520&\errorpm{0.389}{0.309}
	& 1.450&\errorpm{0.344}{0.279} \\
 0.123	  	      
	& 2.083&\errorpm{0.851}{0.594}
	& 1.937&\errorpm{0.723}{0.517}
	& 1.815&\errorpm{0.623}{0.455}
	& 1.712&\errorpm{0.541}{0.405}
\end{tabular}
\end{center}
\caption{\slshape
Numerical results for $\eta_1^\star$ in the NLO. The remaining input
parameters are as in {\protect\eq{InputParam}} except for
$\mu_{tW}=M_W$.  The errors are estimated by varying $\mu_c$ in the
interval corresponding to $-0.2 \leq \ln (\mu_c/m_c^\star) \leq 0.2$.
The accuracy of the NLO result is best in the upper right corner and
worst in the lower left corner of the table.}
\label{tab:eta12}
\end{ntable}

Finally the result for $\eta_2^\star$ reads \cite{bjw}
\begin{eqnarray}
\eta_2^\star &=& 0.573\errorpm{0.003}{0.010}
\label{num-final2}
\end{eqnarray}
for $80\gev \leq \mu_{tW} \leq 180\gev$ and $1.1\gev \leq \mu_c \leq
1.6\gev$.  The remaining input parameters are taken from
{\eq{InputParam}}, they affect the result marginally.

\subsection{Impact on the Phenomenology of $\eps_K$}\label{sect:impact}
Today's key quantity to determine the shape of the unitarity 
triangle is the well-measured value for $\epsilon_K$ \cite{cplear,a}.
With our result for $\eta_3$ one can now perform this analysis 
with NLO precision. This has been done first in  \cite{hn3}, where
also the \kkmd\ has been investigated.

Hence we only briefly discuss the impact of our new result for
$\eta_3$ on the unitarity triangle here. A useful tool to parametrize
the CKM elements are the improved Wolfenstein parameters
$\lambda,A,\ov{\rho},\ov{\eta}$, for a precise definition we refer to
\cite{blo,hn3}.

The constraint from $\eps_K$ reads \cite{hn3}
\begin{eqnarray}
5.3 \cdot 10^{-4}
&=&
B_K A^2 \overline{\eta} \left[
\left(1-\overline{\rho}\right) A^2 \lambda^4
	\eta_2^\star S\left(x_t^\star\right)
+ \eta_3^\star S\left(x_c^\star,x_t^\star\right)
- \eta_1^\star x_c^\star
\right].
\label{epsK}
\end{eqnarray}
Here $B_K$ has been defined in \eq{bk}.  The three terms on the RHS of
\eq{epsK} contribute roughly 75\%, 37\% and --12\%, i.e.\
$\eta_2^\star$ is most important and $\eta_1^\star$, which contains
the largest uncertainties, is least important.  If we look at the
impact of the NLO corrections, however, we find that the one to
$\eta_3^\star$ affects the RHS of \eq{epsK} most, because the NLO
calculation changes $\eta_2^\star$ by only --3\%, while $\eta_3^\star$
and $\eta_1^\star$ are shifted by 26\% and 77\% for $\as(M_Z)=0.117$.
This feature is due to the low scales affecting $\eta_1^\star$ and
$\eta_3^\star$.  The NLO correction to $\eta_3^\star$ has the same
effect on the RHS of \eq{epsK} as a shift in the non-perturbative
parameter $B_K$ from $0.75$ to $0.82$.  Hence the uncertainty of the
LO result for $\eta_3$ had influenced the phenomenology in a similar
way as the present hadronic uncertainty.

Finally we mention a special feature of the analysis of $\epsilon_K$:
{\eq{epsK}} defines a hyperbola in the $(\ov{\rho},\ov{\eta})$-plane 
as a function of $B_K,m_t$ and the magnitude of the CKM element
$V_{cb}$.  Its intersection points with a circle defined by the fourth
key parameter $|V_{ub}/V_{cb}|$ are the allowed solutions for the top
corner $(\ov{\rho},\ov{\eta})$ of the unitarity triangle.  If these
four input parameters are too low, one cannot find a solution.  Hence
\eq{epsK} encodes a ``new physics borderline'' in the parameter space
(see \cite{hn3}): If future determinations pin down the input
parameters to too low values, the Standard Model will be unable 
to explain the observed CP violation in \kkm.  Todays central values
for $|V_{cb}|,|V_{ub}/V_{cb}|, m_t$ and $B_K$ are very close to this
line.  Now the NLO shift to $\eta_3^\star$ has \emph{enlarged} the
allowed range for the input parameters and thereby vindicated the 
CKM mechanism of CP violation.  This can be seen in
{\fig{fig:hyperbola}}, where the hyperbola \eq{epsK} has been
displayed for the LO and NLO values of $\eta_3^\star$.

%
%

\section{Conclusions}\label{sect:concl}
We have calculated the QCD short distance coefficient $\eta_3$ of the
low energy \dstwo -hamiltonian in the next-to-leading order (NLO) of
renormalization group improved perturbation theory. The NLO
calculation of the other two coefficients $\eta_1$ and $\eta_2$ has
been repeated. 
The three coefficients read
\begin{eqnarray}
\eta_1^{\star} &=& 1.31\errorpm{0.25}{0.22}\,,
\quad \quad 
\eta_2^{\star} \; = \; 0.57\errorpm{0.01}{0.01}\,,
\quad \quad 
\eta_3^{\star} \; = \;  0.47\errorpm{0.03}{0.04}\,.
\label{coneta}
\end{eqnarray} 
The coefficients are scheme independent except that they depend on the
definition of the quark masses in $H^\dstwo$. The results in 
\eq{coneta} correspond to $\ov{\rm MS}$-masses $m_c(m_c)$ and 
$m_t(m_t)$ as indicated by the superscript ``$\star$''.  Only
$\eta_1^\star$ is sensitive to the quark masses and $\as$, the quoted
value corresponds to $m_c(m_c)=1.3
\gev$ and $\as (M_Z) =0.117 \gev$.  The NLO values in \eq{coneta} have to
be compared with the old LO results:
\begin{eqnarray}
\eta_1^{\star \,\mathrm{LO} } & \approx & 0.74 \, ,
\quad \quad
\eta_2^{\star \,\mathrm{LO} } \approx  0.59 \, , 
\quad \quad
\eta_3^{\star \,\mathrm{LO} } \approx  0.37 \, .
\label{old}
\end{eqnarray} 
The large differences between \eq{old} and \eq{coneta} illustrate the
big uncertainty of the crude leading log approximation. Conceptually
the correct definition of quark masses and the consistent use of the
QCD scale parameter $\laMSb$ can only be addressed in quantities
calculated beyond the LO. Especially the dependence of $\eta_3 \cdot
S(x_c,x_t)$ on $m_t$ enters in the NLO. The dependence of
$\eta_3^{\star}$ on renormalization scales and physical quantities has
been discussed in detail.

In the standard analysis of the unitarity triangle $\epsilon_K$ and
the \bbm\ parameter $x_d$ play pivotal roles. Since the QCD correction
factor of the $\mathrm{|\Delta B| \!=\!2}$-hamiltonian is also known
in the NLO \cite{bjw}, this analysis is now completely possible with
NLO accuracy \cite{hn3}. The shift in $\eta_3$ between \eq{old} and
\eq{coneta} has a larger impact on $\epsilon_K$ than the NLO
corrections to $\eta_1$ and $\eta_2$. The use of $\eta_3^{\star
\,\mathrm{LO} }$ imposes an error onto $\epsilon_K$ which is comparable
to the present hadronic uncertainty in $B_K$. The NLO shift to
$\eta_3^\star$ has enlarged the range for
$(|V_{cb}|,|V_{ub}/V_{cb}|,m_t,B_K)$ complying with the Standard Model
explanation of $\epsilon_K$.

We have further explained the construction of the effective
lagrangians needed to achieve {\eq{coneta}}. Here we have described
the elimination of unphysical operators with emphasis on the
discussion of evanescent operators, which impose a new type of scheme
dependence on the NLO anomalous dimension tensor and the Wilson
coefficients.  We have shown that this scheme dependence complies with
the results of \cite{hn2} implying its cancellation in $\eta_3$.
Further a compact solution for the renormalization group equation for
Green's functions with two operator insertions has been derived. 
Finally the results of the two-loop diagrams constituting the NLO
anomalous dimension tensor have been listed.

\section*{Acknowledgements}
The authors thank Andrzej Buras for suggesting the topic and permanent
encouragement. We have enjoyed fruitful discussions with him, Gerhard
Buchalla and Miko{\l}aj Misiak.  We thank Hubert Simma for his
thorough explanation of the role of the equation of motion in field
theory. We are grateful to Matthias Jamin and Markus Lautenbacher for
providing us with their computer programs on \dsone\ Wilson
coefficients.  We thank Andrzej Buras and Fred Jegerlehner for
carefully reading the manuscript.  S.H. is grateful to Fred
Jegerlehner for interesting discussions and thanks Andrzej Buras for
several opportunities to visit the TUM to complete this work.

\appendix
%
%

\section{Results of the Two-loop Diagrams}\label{apx:int}
Here we list the results for the two-loop diagrams of
{\fig{fig:double-cc-nlo}} and {\fig{fig:double-peng-nlo}}:
\begin{subequations}
\label{div}
\begin{eqnarray}
D_k^{rs} &=& i \frac{g^2}{\left( 4 \pi \right)^4} \mu^{-2 \eps} 
        \cdot	\gamma_\mu L \otimes \gamma^\mu L \cdot 
        \mathcal{C}_k^{rs} 
     \left[ \frac{d^{(k)}_2}{\eps^2} + \frac{d^{(k)}_1}{\eps } 
     +O \left( \eps^0 \right) \right] , 
\label{div-d} 
\\
&& \qquad \qquad  
k=1,\ldots,7, \nonumber \\
P_{k\ell}^{rs} &=& i \frac{g^2}{\left( 4 \pi \right)^4} \mu^{-2 \eps} 
          \cdot	\gamma_\mu L \otimes \gamma^\mu L \cdot m_c^2 \cdot
        \mathcal{E}_k^{rs} 
     \left[ \frac{p^{(k)}_{2 \ell} }{\eps^2} + \frac{p^{(k)}_{1 \ell}}{\eps } 
     +O \left( \eps^0 \right) \right] , \label{div-p} \\ 
&& \qquad \quad
k=1,\ldots,15, \quad \ell=L,R.
\nonumber 
\end{eqnarray}
\end{subequations}
Here the one-loop counterterms have been included diagram-by-diagram.
The result is then proportional to the Dirac structure $\gamma_\mu L
\otimes \gamma^\mu L = (\ov{s}d)_{V-A} \otimes (\ov{s}d)_{V-A}$ apart
from terms involving evanescent Dirac structures, which are irrelevant
for the NLO calculation and are not displayed in {\eq{div}}.
$\mathcal{C}_k^{rs}$ and $\mathcal{E}_k^{rs}$ are colour factors
listed in \tab{tab:colorD} and \tab{tab:colorP}.  They depend on the
colour structure of the inserted operators $(Q_r,Q_s)$. $\ell=L,R$ is
the chirality of the $\left(\ov{q} q\right)_{V\mp A}$ foot of the
inserted penguin operator.  The factor $\mu^{-2 \eps}$ in \eq{div}
stems from the definition \eq{defqseven} of $\oloc$.

In \eq{div-p} an internal charm quark has been assumed, the
corresponding result with an up quark is zero.  The coefficients
$d^{(k)}_i,p^{(k)}_{i\ell}$ in the $\ov{\rm MS}$ scheme are listed in
Tables~\ref{tab:div-d} and \ref{tab:div-p} for an arbitrary gluon
gauge parameter $\xi$.  Note the footnote on p.~\pageref{foot}.  The
tabulated values for $d^{(k)}_1,p^{(k)}_{1\ell}$ refer to the NDR
scheme and the standard definition \eq{StdEvasNum} of the evanescent
operators, while $\wt{b}_1$ in \eq{StdEvas2-loc} is kept
arbitrary. The evanescent one-loop counterterms have been inserted
with a factor of $\lambda$.  For the NLO anomalous dimension tensor
$\gbi{\pm}^{(1)}$ in {\eq{ResAnomTens1}} we need the result with
$\lambda=1/2$.  Hence the independence of $\gbi{\pm}^{(1)}$ of
$\wt{b}_1$ can be easily verified from \tab{tab:div-d} thereby
illustrating a general feature proven in {\cite{hn2}}.  By comparing
those results for $D_k^{rs}$ and $P_{kL}^{rs}$ which correspond to
diagrams related by the Fierz transformation (such as $\mathsf{D_1}$
and $\mathsf{P_{7L}}$) one finds the same results for
$\lambda=1/2$. This is only true for the choice \eq{StdEvasNum}
adopted in \tab{tab:div-d} and \tab{tab:div-p}.  The Dirac algebra has
been calculated with the help of the computer package \textsc{Tracer}
\cite{jl2}.
 
\begin{ntable}
\newlength{\colwidth}
\setlength{\colwidth}{6em}
\begin{displaymath}
\begin{array}{r|*{4}{l}}
\left(Q_r,Q_s\right) &
	\left(Q_{+},Q_{1}\right) &
	\left(Q_{+},Q_{2}\right) &
	\left(Q_{-},Q_{1}\right) &
	\left(Q_{-},Q_{2}\right)
\\
k &
	\phantom{\hspace{\colwidth}} &
	\phantom{\hspace{\colwidth}} &
	\phantom{\hspace{\colwidth}} &
	\phantom{\hspace{\colwidth}}
\\
\hline
1 &
	\frac{-1+N^2}{4N} &
	\frac{-2+N+N^2}{4N} &
	\frac{-1+2N-N^2}{4N} &
	\frac{-1+N}{4}
\\
2 &
	\frac{-1-N+N^2+N^3}{4N} &
	\frac{-2+N+N^2}{4N} &
	\frac{-1+N+N^2-N^3}{4N} &
	\frac{1-N}{4}
\\
3 &
	\frac{-1+N^2}{4N} &
	\frac{-1+N}{2N} &
	\frac{-1+2N-N^2}{4N} &
	0
\\
4 &
	\frac{-1-N+N^2+N^3}{4N} &
	\frac{-2+N+N^2}{4N} &
	\frac{-1+N+N^2-N^3}{4N} &
	\frac{1-N}{4}
\\
5 &
	\frac{-1-N+N^2+N^3}{4N} &
	\frac{-1+N^2}{2N} &
	\frac{-1+N+N^2-N^3}{4N} &
	0
\\
6 &
	\frac{-1+N^2}{4N} &
	\frac{-2+N+N^2}{4N} &
	\frac{-1+N^2}{4N} &
	\frac{-1+N}{4}
\\
7 &
	\frac{-1+N^2}{4N} &
	\frac{-1+N}{2N} &
	\frac{-1+N^2}{4N} &
	0
\end{array}
\end{displaymath}
\begin{displaymath}
\begin{array}{r|*{4}{l}}
\left(Q_r,Q_s\right) &
	\left(Q_{1},Q_{+}\right) &
	\left(Q_{2},Q_{+}\right) &
	\left(Q_{1},Q_{-}\right) &
	\left(Q_{2},Q_{-}\right)
\\
k &
	\phantom{\hspace{\colwidth}} &
	\phantom{\hspace{\colwidth}} &
	\phantom{\hspace{\colwidth}} &
	\phantom{\hspace{\colwidth}}
\\
\hline
6 &
	\frac{N-1}{4N} &
	\frac{N^2-1}{2N} &
	\frac{N-1}{4N} &
	0 
\\
7 &
	\frac{N-1}{4N} &
	\frac{N^2+N-2}{4N} &
	\frac{N-1}{4N} &
	\frac{1-N}{4} 
\end{array}
\end{displaymath}
\caption{\slshape
The colour factors $\mathcal{C}_k^{rs}$ of the diagrams in
{\fig{fig:double-cc-nlo}}. $Q_r$ ($Q_s$) is the left (right) operator
in the diagrams. For $k\leq 5$ one has
$\mathcal{C}_k^{rs}=\mathcal{C}_k^{sr}$.}  
\label{tab:colorD}
\end{ntable}
\begin{ntable}
\setlength{\colwidth}{6em}
\begin{displaymath}
\begin{array}{r|*{4}{l}}
\left(Q_r,Q_s\right) &
	\left(Q_{+},Q_{3,5}\right) &
	\left(Q_{+},Q_{4,6}\right) &
	\left(Q_{-},Q_{3,5}\right) &
	\left(Q_{-},Q_{4,6}\right)
\\
k &
	\phantom{\hspace{\colwidth}} &
	\phantom{\hspace{\colwidth}} &
	\phantom{\hspace{\colwidth}} &
	\phantom{\hspace{\colwidth}}
\\
\hline
1 &
	\frac{-1-N+N^2+N^3}{4N} &
	\frac{-2+N+N^2}{4N} &
	\frac{-1+N+N^2-N^3}{4N} &
	\frac{1-N}{4}
\\
2 &
	\frac{-1-N+N^2+N^3}{4N} &
	\frac{-1+N^2}{2N} &
	\frac{-1+N+N^2-N^3}{4N} &
	0
\\
3 &
	\frac{-1+N^2}{4N} &
	\frac{-2+N+N^2}{4N} &
	\frac{-1+N^2}{4N} &
	\frac{-1+N}{4}
\\
4 &
	\frac{-1+N}{4N} &
	\frac{-1+N^2}{2N} &
	\frac{-1+N}{4N} &
	0
\\
5 &
	\frac{-1+N^2}{4N} &
	\frac{-1+N}{2N} &
	\frac{-1+N^2}{4N} &
	0
\\
6 &
	\frac{-1+N}{4N} &
	\frac{-2+N+N^2}{4N} &
	\frac{-1+N}{4N} &
	\frac{1-N}{4}
\\
7 &
	\frac{-1+N^2}{4N} &
	\frac{-2+N+N^2}{4N} &
	\frac{-1+2N-N^2}{4N} &
	\frac{-1+N}{4}
\\
8 &
	\frac{-1+N^2}{4N} &
	\frac{-1+N}{2N} &
	\frac{-1+2N-N^2}{4N} &
	0
\\
9, 10 &
	\frac{-1-N+N^2+N^3}{4N} &
	\frac{-2+N+N^2}{4N} &
	\frac{-1+N+N^2-N^3}{4N} &
	\frac{1-N}{4}
\\
12 &
	\frac{-1+N}{4N} &
	\frac{-1+N}{4N} &
	\frac{-1+N}{4N} &
	\frac{-1+N}{4N} 
\end{array}
\end{displaymath}
\caption{\slshape
The colour factors $\mathcal{E}_k^{rs}$ of the diagrams in
{\fig{fig:double-peng-nlo}}.  In $\mathcal{E}_k^{rs}$ the first index
$r$ refers to the current-current operator.}
\label{tab:colorP}
\end{ntable}
\begin{ntable}
\begin{displaymath}
\begin{array}{r|*{2}{l}}
k & d^{(k)}_2 & d^{(k)}_1 \\
\hline
1 & \left(m_i^2+m_j^2\right) \left(-2\xi\right) &
	\left(m_i^2+m_j^2\right) \left(-6-3\xi+24 \lambda \right)
\\
2 & \left(m_i^2+m_j^2\right) \left(-4\xi\right) &
	\left(m_i^2+m_j^2\right) \frac{1}{2}\left(-\lambda+\frac{1}{2}\right)
	\tilde{b}_1
\\
3 & \left(m_i^2+m_j^2\right) \left(6+2\xi\right) &
	\left(m_i^2+m_j^2\right) \left(-1+3\xi+24\lambda 
	-\frac{1}{4} (\lambda-\frac{1}{2}) \tilde{b}_1\right)
\\
4 & \left(m_i^2+m_j^2\right) \left(-4\xi\right) &
	\left(m_i^2+m_j^2\right) \left(6\xi-\frac{1}{2}
	\left(\lambda-\frac{1}{2}\right) \tilde{b}_1\right)
\\
5 & -12m_i^2+2 \left(m_i^2+m_j^2\right)\xi &
	22 m_i^2 - 3\left(m_i^2+m_j^2\right)\xi
\\
6 & \left(m_i^2+m_j^2\right) \left(-2\xi\right) &
	\left(m_i^2+m_j^2\right) 3\xi
\\
7 & \left(m_i^2+m_j^2\right) \left(6+2\xi\right) &
	-11 m_i^2 - 23 m_j^2 - \left(m_i^2+m_j^2\right) \left(
	3 \xi + \frac{1}{4} (\lambda-\frac{1}{2}) \tilde{b}_1\right)
\end{array}
\end{displaymath}
\caption{\slshape
Divergent parts $d^{(k)}_s= d^{(k)}_s \left(m_i^2,m_j^2,\lambda 
 \right) $ of the diagrams $\mathsf{D_k}$ depicted in
 {\protect\fig{fig:double-cc-nlo}} according to \eq{div-d}.  Here $k$
 labels the diagram.  Two different internal up-type quark masses
 $m_i$ and $m_j$ are involved. In $\mathsf{D_5}$ and $\mathsf{D_7}$
 the mass corresponding to the upper quark line is $m_i$.  }
\label{tab:div-d}
\end{ntable}
\begin{ntable}
\begin{displaymath}
\begin{array}{r|*{2}{l}|*{2}{l}}
k & p^{(k)}_{2L} & p^{(k)}_{1L} & p^{(k)}_{2R} & p^{(k)}_{1R}
\\
\hline
1 & -8\xi & 12\xi & 8\xi & -12
\\
2 & -12+4\xi & 22-6\xi & 12-4\xi & -16
\\
3 & -4\xi & 6\xi & 4\xi & 6
\\
4 & -4\xi & -24+6\xi+48\lambda & 12+4\xi & -40+48\lambda
\\
5 & 12+4\xi & -34-6\xi & -12-4\xi & 28
\\
6 & 12+4\xi & -58-6\xi+48\lambda & -4\xi & -18+48\lambda
\\
7 & -4\xi & 12-6\xi & 4\xi & -6
\\
8 & 12+4\xi & 22+6\xi & -12-4\xi & -4
\\
9, 10 & -8\xi & 0 & 8\xi & 0
\\
11, 13, 14, 15 & 0 & 0 & 0 & 0
\\
12 & 0 & 24 & 0 & 0
\end{array}
\end{displaymath}
\caption{\slshape
Divergent parts $p^{(k)}_{i \ell}=p^{(k)}_{i \ell} \left( \lambda
\right)$ of the diagrams $\mathsf{P_{k}}$ depicted in
{\protect\fig{fig:double-peng-nlo}} according to \eq{div-p}.  Here $k$
labels the diagram and $\ell=L,R$ denotes the projection operator
present in the inserted penguin operator.}
\label{tab:div-p}
\end{ntable}

\section{Anomalous Dimension Tensor and $Z$-Factors}\label{apx:zfact}
Here we sketch the relation between the results \eq{div} for the
two-loop diagrams and $\gbi{\pm j}^{(1)}$ and $\zbil{(2)}{}{\pm
j}$. We further list the result for $\gbi{\pm j}^{(1)}$ for an
arbitrary number $N$ of colours. 

The $1/\eps$-terms of the  two-loop results combine to
\begin{eqnarray}
h^{rs} ( \lambda )
\hspace{-1.2pt} &=& \hspace{-1.2pt}
\left\{
\begin{array}{>{\displaystyle}l}
\sum\limits_{k=1}^7 w_k
    \left( \mathcal{C}_k^{rs}+\mathcal{C}_k^{sr} \right) 
   \left( d^{(k)}_1 ( \mc, 0,\lambda ) + d^{(k)}_1 (0,\mc,\lambda)
     \right)
\mbox{~~for s=1,2,}
\\
\sum\limits_{k=1}^{15} v_k \, 2 \, 
         \mathcal{E}_k^{rs}  p^{(k)}_{1 \ell} \left( \lambda \right)
\qquad \qquad \qquad  \mbox{for s=3,\ldots,6,}
\end{array}
\right. \label{defhrs}
\end{eqnarray}
where $\ell=L$ for $s=3,4$, $\ell=R$ for $s=5,6$ and $r=+,-$.  Further
$w_k,v_k$ count the possible ways of left-right and up-down
reflections. On has $w_1=w_2=w_3=w_5=v_1=v_9=v_{10}=1/2$, $w_4=1/4$,
and the other weight factors equal unity.  Now the $1/\eps$-part of
the two-loop Z-factor is obtained by
\begin{eqnarray}
\zbil{-1,(2)}{1}{rs} &=& -h^{rs} (1) .
\end{eqnarray}
The NLO anomalous dimension tensor is obtained from \eq{defhrs} by
\begin{eqnarray}
\gbi{rs}^{(1)} &=& - 4 \cdot h^{rs}\!\left( \frac{1}{2} \right),
\end{eqnarray}
cf.~(\ref{AnomDoubleCoeff}) and sect.~\ref{sect:anom}.

For completeness we list the LO and NLO results in \eq{ResAnomTens}
for an arbitrary number $N$ of colours:
\begin{eqnarray}
\g_{+,7}^{(0)} \; = \;  
\lt( 
\begin{array}{c}
-4 (N+1) \\   -8 \\ -8 (N+1) \\ -16 \\ 8 (N+1) 
 \\  16 
\end{array} \rt) ,     &\quad & 
\g_{-,7}^{(0)} \; = \;  
\lt( 
\begin{array}{c}
4 (N-1) \\   0 \\ 8 (N-1) \\ 0 \\ -8 (N-1) 
 \\  0 
\end{array} \rt) ,     \no
\end{eqnarray}  
\begin{eqnarray}
\gbi{+}^{(1)} &=& 2 \, \frac{N-1}{N} \cdot 
\left(
\begin{array}{c}
 6 - 22 N - 11 N^2  \\
 12 - 11 N  \\ 
-2 \, \left( 6 + 22 N + 11 N^2 \right) \\ 
-22 N \\
2 \, \left( 4+ 10N + 11 N^2 \right) \\ 
16 \, \left( 1+4 N \right)
\end{array}
\right) ,
\nn
\gbi{-}^{(1)} &=& 2 \, \frac{N-1}{N} \cdot 
\left(
\begin{array}{c}
 6 + 34 N + 11 N^2 \\
-23 N \\ 
2 \, \left(-6 + 34 N + 11 N^2 \right) \\
-2 \, \left( 12 + 23 N \right) \\
-2 \, \left(-4 + 22 N + 11 N^2 \right)\\
0
\end{array}
\right). \no
\end{eqnarray}
\section{RG Quantities and Matching Corrections}\label{apx:rgquant}

In this appendix we collect various quantities needed for the RG
evolution.

The QCD beta function reads
\begin{eqnarray}
\beta\left(g\right) &=&
\frac{g^3}{16\pi^2} \cdot
\left[-\beta_0-\beta_1 \frac{g^2}{16\pi^2}-\cdots\right],
\label{QCDbeta}
\end{eqnarray}
where the coefficients in LO and NLO are given by
\begin{eqnarray}
\beta_0^{[f]} = \frac{11N-2f}{3},
&\hspace{3em}&
\beta_1^{[f]} = \frac{34}{3}N^2-\frac{10}{3}Nf-2C_F f.
\label{QCDbetaNLO}
\end{eqnarray}
Using this the NLO running QCD coupling constant 
$\as=g^2/4\pi$ equals
\begin{eqnarray}
\frac{\as\left(\mu\right)}{4\pi} &=&
\frac{1}{\beta_0 \ln\frac{\mu^2}{\laMSb^2}}
\left(1-\frac{\beta_1}{\beta_0^2}
\frac{\ln \ln \frac{\mu^2}{\laMSb^2}}{\ln\frac{\mu^2}{\laMSb^2}}
\right)
\label{alphas}
\end{eqnarray}
in the $\overline{\textrm{MS}}$ scheme. 
The LO version of {\eq{alphas}} is obtained by dropping the term
involving $\beta_1$ and replacing $\laMSb$ by $\laQCD^{\rm LO}$.

Further we need the anomalous dimensions for the masses and the
fermion fields.  We define the renormalization constants and anomalous
dimensions as follows:
\begin{eqnarray}
\begin{array}{r@{=}l@{,\hspace{3em}}r@{=}l}
m_\bare & Z_m m & \gamma_m &
	\frac{\mu}{Z_m} \frac{d Z_m}{d\mu}, \\
\psi_\bare & Z_\psi^{1/2} \psi &
	\gamma_\psi & \frac{\mu}{Z_\psi^{1/2}} \frac{d Z_\psi^{1/2}}{d\mu}.
\end{array}
\label{DefZAnomPsi}
\end{eqnarray}
As usual we expand $\gamma_m$ and $\gamma_\psi$ in powers of $\as$:
\begin{eqnarray}
\gamma_X &=&
\gamma_X^{(0)} \frac{g^2}{16\pi^2}
+\gamma_X^{(1)} \left(\frac{g^2}{16\pi^2}\right)^2
+\cdots,
\hspace{3em}X=m,\psi.
\label{AnomPsiPert}
\end{eqnarray}
We need the following coefficients of \eq{AnomPsiPert}:
\begin{subequations}
\begin{eqnarray}
\gamma^{(0)}_m &=& 6 C_F, 
\qquad \gamma^{(1)}_m \; = \; \gamma _m^{(1)[f]} \; = \; C_F \left( 3 C_F + 
                  \frac{97}{3} N - \frac{10}{3} f \right),
\label{MassAnom}
\\
\gamma^{(0)}_{\psi} &=& \xi C_F.
\label{FermionAnom}
\end{eqnarray}
\end{subequations}
To transform masses between different scales we use the NLO evolution
equation
\begin{eqnarray}
m\left(\mu\right) &=&
m\left(\mu_0\right)
\left[\frac{\as\left(\mu\right)}{\as\left(\mu_0\right)}\right]
	^{d_m}
\left(1+\frac{\as\left(\mu_0\right)-\as\left(\mu \right)}{4\pi} J_m \right),
\label{MassEvol}
\end{eqnarray}
where
\begin{eqnarray}
d_m = \frac{\gamma_m^{(0)}}{2\beta_0}
&\hspace{3em}\mbox{and}\hspace{3em}&
J_m = -\frac{\gamma_m^{(1)}}{2\beta_0} + \frac{\beta_1}{\beta_0} d_m .
\label{MassCoeff}
\end{eqnarray}
The LO version of \eq{MassEvol} is obtained by dropping the $O(\as)$
term in the right bracket.

\subsection{\dsone\ Mixing and Matching Matrices}\label{apx:ds1mix}

Here we collect the important ingredients of the mixing and matching of
the \dsone\ operator basis in \eq{lags1} for $N=3$ \cite{bjlw}:
\begin{subequations}
\label{ds1anom}
\begin{eqnarray}
\gamma^{(0)} \hspace{-2.4pt}
&=& \hspace{-2.4pt}
\left(
\begin{array}{*{6}{c}}
-2 & 6 & 0 & 0 & 0 & 0 \\
6 & -2 & -\frac{2}{9} & \frac{2}{3} & -\frac{2}{9} & \frac{2}{3} \\
0 & 0 & -\frac{22}{9} & \frac{22}{3} & -\frac{4}{9} & \frac{4}{3} \\
0 & 0 & 6-\frac{2f}{9} & -2+\frac{2f}{3} & -\frac{2f}{9} & \frac{2f}{3} \\
0 & 0 & 0 & 0 & 2 & -6 \\
0 & 0 & -\frac{2f}{9} & \frac{2f}{3} & -\frac{2f}{9} & -16+\frac{2f}{3}
\end{array}
\right) ,
\label{ds1anom0} \\[3ex]
\gamma^{(1)} \hspace{-2.4pt}
&=& \hspace{-2.4pt}
\left(
\begin{array}{*{6}{c@{\hspace{0.3pt}}}}
-\frac{21}{2}-\frac{2f}{9} & \frac{7}{2}+\frac{2f}{3}  & \frac{79}{9}
	& -\frac{7}{3} & \frac{65}{9} & -\frac{7}{3} \\
\frac{7}{2}+\frac{2f}{3} & -\frac{21}{2}-\frac{2f}{9} &
	-\frac{202}{243} & \frac{1354}{81} & -\frac{1192}{243} &
	\frac{904}{81} \\
0 & 0 & -\frac{5911}{486}+\frac{71f}{9} & \frac{5983}{162}+\frac{f}{3}
	& -\frac{2384}{243}-\frac{71f}{9} &
	\frac{1808}{81}-\frac{f}{3} \\
0 & 0 & \frac{379}{18}+\frac{56f}{243} & -\frac{91}{6}+\frac{808f}{81}
	& -\frac{130}{9}-\frac{502f}{243} &
	-\frac{14}{3}+\frac{646f}{81} \\
0 & 0 & -\frac{61f}{9} & -\frac{11f}{3} & \frac{71}{3}+\frac{61f}{9} &
	-99+\frac{11f}{3} \\
0 & 0 & -\frac{682f}{243} & \frac{106f}{81} &
	-\frac{225}{2}+\frac{1676f}{243} &
	-\frac{1343}{6}+\frac{1348f}{81}
\end{array}
\right) ,
\no \\
&&
\label{ds1anom1} \\
r &=& \left(
\begin{array}{*{6}{c}}
\frac{7}{3} & -7 & 0 & 0 & 0 & 0 \\
-7 & \frac{7}{3} & \frac{2}{27} & -\frac{2}{9} & \frac{2}{27} &
	-\frac{2}{9} \\
0 & 0 & \frac{67}{27} & -\frac{67}{9} & \frac{4}{27} & -\frac{4}{9} \\
0 & 0 & -7+\frac{5f}{27} & \frac{7}{3}-\frac{5f}{9} & \frac{5f}{27} &
	-\frac{5f}{9} \\
0 & 0 & 0 & 0 & -\frac{1}{3} & 1 \\
0 & 0 & \frac{5f}{27} & -\frac{5f}{9} & -3+\frac{5f}{27} &
	\frac{35}{3}-\frac{5f}{9}
\end{array}
\right) .
\label{ds1r}
\end{eqnarray}
\end{subequations}
Here $f$ denotes the number of active flavours.  We stress that
$\gamma^{(1)}$ and $r$ depend on the renormalization scheme.  The
results in \eq{ds1anom1} and \eq{ds1r} are specific to the NDR scheme
with the evanescent operators in (\ref{StdEvas1-12}-\ref{StdEvas1-56})
defined with $a_1=-8$ and $a_2=-16$. \eq{ds1r} is further specific to
the 't Hooft-Feynman gauge and the choice $\mu^2=p^2$ where $p$ is a
small external momentum used as an IR regulator. In a RG improved
Wilson coefficient like \eq{MatchWilsonS1} the dependence on the gauge
parameter and the IR regulator cancels.  Results for arbitrary $N$ may
be found in \cite{bjlw2}.

We further give the anomalous dimension of the current-current
subspace of the \dsone\ operators in the diagonal basis $Q_\pm$ for
the scheme specified above:
\begin{eqnarray}
\gamma_{\pm}^{(0)} = \pm 6 \frac{N \mp 1}{N},
\hspace{3em}
\gamma_{\pm}^{(1)} = \frac{N \mp 1}{2N} \left(-21 \pm \frac{57}{N} \mp
\frac{19}{3} N \pm \frac{4}{3} f\right).
\label{anomQpm}
\end{eqnarray}

\subsection{\dstwo\ Anomalous Dimensions}\label{apx:ds2anom}

Here we collect the important quantities for the mixing and matching of
the \dstwo\ operator basis in \eq{lags2} and \eq{lags2c}:
$\oll$ has the same Dirac structure as the \dsone\ operators $Q_\pm$.
Since it is Fierz self-conjugate, its anomalous dimension is equal to
the one of $Q_+$ in \eq{anomQpm}
\begin{eqnarray}
\gamma_{S2} &=& \gamma_{+}.
\label{DefAnomS2}
\end{eqnarray}
This result is specific to the definition of the evanescent operator
$E_1[\oll]$ with $\wt{a}_1=-8$ (cf.\ sect.~\ref{sect:unphysical}).
Using \eq{defqseven} the anomalous dimension of $\oloc$ can be related
to $\gamma_{+}$:
\begin{eqnarray}
\gloc\left(g\right) &=&
\gamma_{+}\left(g\right) + 2 \gamma_m\left(g\right) + \frac{2}{g}
\beta\left(g\right).
\label{DefAnomQ7}
\end{eqnarray}
We therefore get
\begin{eqnarray}
\gloc^{(k)} &=& \gamma_{+}^{(k)} + 2 \gamma_m^{(k)} - 2 \beta_k,
\hspace{3em}
k=0,1,\ldots
\label{AnomQ7exp}
\end{eqnarray}
%

\end{document}